\documentclass[11pt,a4paper]{article}

\usepackage{jheppub}

\usepackage{grffile}
\usepackage{graphicx}
\usepackage{amsmath,amsthm,amssymb,bm,amsfonts}
\usepackage{empheq}
\usepackage[detect-all]{siunitx}
\usepackage[utf8]{inputenc}

\renewcommand{\thefigure}{\arabic{figure}}

\usepackage{color}
\usepackage[makeroom]{cancel}
\usepackage[normalem]{ulem}
\definecolor{kgbcolor}{rgb}{0,0.1,0.7}
\definecolor{ascolor}{rgb}{1,0,1}

\newcommand\kgbout{\marginpar{\color{kgbcolor}$\clubsuit$}\bgroup\markoverwith{\color{kgbcolor}{\rule[0.4ex]{2pt}{0.8pt}}}\ULon}
\newcommand\asout{\marginpar{\color{ascolor}$\heartsuit$}\bgroup\markoverwith{\color{ascolor}{\rule[0.4ex]{2pt}{0.8pt}}}\ULon}

\def\bbold{{\mathbf{b}}}

\def\aprime{a^\prime}

\def\abis{a^{\prime\prime}}

\def\kperp{k_\perp}
\def\pperp{p_\perp}

\def\qperpb{\mathbf{q}_{\perp}}
\def\kperpone{k_{1\perp}}
\def\kperptwo{k_{2\perp}}

\newcommand{\be}{\begin{eqnarray}}
\newcommand{\ee}{\end{eqnarray}}

\newcommand{\bea}{\begin{align}}
\newcommand{\eea}{\end{align}}

\def\kperp{{k_\perp}}

\title{Numerical analysis of the unintegrated double gluon distribution}
\author[a]{Edgar Elias}
\author[b,c]{Krzysztof Golec-Biernat}
\author[a]{Anna M. Sta\'sto}

\affiliation[a]{Department of Physics, The Pennsylvania State University, University Park, PA 16802, U.S.A.}
\affiliation[b]{Institute of Nuclear Physics, Polish Academy of Sciences, 31-342 Cracow, Poland}
\affiliation[c]{Faculty of Mathematics and Natural Sciences, University of Rzesz\'ow,  35-959 Rzesz\'ow, Poland}

\abstract{ We present detailed  numerical analysis of the unintegrated  double gluon distribution which includes the dependence on the transverse momenta
of  partons.  The unintegrated double gluon distribution was obtained  following the Kimber-Martin-Ryskin method as a convolution of the perturbative gluon splitting function with the  collinear integrated double gluon distribution and  the Sudakov form factors.  We   analyze the dependence on the  transverse momenta,  longitudinal momentum fractions and  hard scales.  We find that the unintegrated gluon distribution factorizes into a product of two single unintegrated gluon distributions in the region of small values of $x$, provided the  splitting contribution is included and the momentum sum rule is satisfied.  
}


\begin{document}
\maketitle

\section{Introduction}
\label{sec:1}

One of the most intriguing features observed in QCD is the strong increase of the gluon density with the  decrease of the parton longitudinal momentum fraction $x$. This rise, originally discovered at HERA electron(positron)-proton collider  \cite{Abt:1993cb,Derrick:1993fta,Aid:1996au,Derrick:1996hn}, manifests itself as the  rise of the cross sections for variety of processes in hadronic collisions when the small values of longitudinal fractions of the partons participating in the  scattering are probed.  The other consequence of this increase in the gluon  density is the increase of the probability of multiparton interactions in hadron-hadron collisions. These are events when in one  encounter of hadrons, more than one elementary partonic scattering occurs. These type of processes were originally observed by the AFS collaboration at CERN \cite{Akesson:1986iv}, followed by the measurements at the Tevatron \cite{Abe:1997bp,Abe:1997xk,Abazov:2009gc} and more recently at the Large Hadron Collider (LHC) \cite{Aad:2013bjm,Chatrchyan:2013xxa,Aad:2014rua}.

The theoretical framework for the description of the multiparton interactions in QCD relies on the assumption of the factorization of the hard double parton interaction in the presence of sufficiently large scales, see for example \cite{Paver:1982yp,Diehl:2011yj}, and \cite{Treleani:2017zzl,Diehl:2017wew} for a recent overview. The case of double-parton scattering is  usually described within the framework of the standard collinear perturbative QCD by means of the double parton distribution functions (DPDFs) \cite{Shelest:1982dg,Zinovev:1982be,Ellis:1982cd,Bukhvostov:1985rn,Snigirev:2003cq,Korotkikh:2004bz,Gaunt:2009re,Blok:2010ge,Ceccopieri:2010kg,Diehl:2011tt,
Gaunt:2011xd,Ryskin:2011kk,Bartels:2011qi,Blok:2011bu,Diehl:2011yj,Luszczak:2011zp,
Manohar:2012jr,Ryskin:2012qx,Gaunt:2012dd,Blok:2013bpa,
vanHameren:2014ava, Maciula:2014pla, Snigirev:2014eua,  Golec-Biernat:2014nsa, Gaunt:2014rua, 
Harland-Lang:2014efa, Blok:2014rza, Maciula:2015vza}. The DPDF distributions satisfy DGLAP-type equations \cite{Kirschner:1979im,Shelest:1982dg,Zinovev:1982be,Snigirev:2003cq,Korotkikh:2004bz,Gaunt:2011xd}. Similar type of equations for double parton correlations were considered earlier \cite{Konishi:1978yx,Konishi:1979cb} in the context of  the multiparton correlation functions within jets. These equations contain  two terms: a  homogeneous term describing the independent evolution of the two chains of partons in the double parton distributions and a non-homogeneous term which originates from the perturbative splitting of two partons. The latter contribution is  driven by the single parton distribution functions.  In principle, some part of the splitting contribution can coincide with the higher order contribution to the single parton interaction, and thus one must perform special subtraction and matching in order to avoid  potential double counting when computing the double parton scattering cross section with this term \cite{Diehl:2017kgu}.

The collinear framework works well for sufficiently inclusive quantities, but can miss important information about the kinematics, see for example \cite{Collins:2005uv}.  The more detailed information about the kinematics of the process can be incorporated by using the unintegrated parton distribution functions (UPDFs) or transverse momentum dependent distributions (TMDs) \cite{Collins:2007ph,Collins:2011zzd,Collins:2011ca}. The unintegrated parton distribution functions in the case of the single scatterings are widely used in the literature, and naturally occur  in the high energy or small $x$ limit of QCD, see for example \cite{Catani:1990eg}.   It is thus of great phenomenological and theoretical interest to develop the formalism which includes the transverse momentum dependence in the context of the multiparton distributions.  There has been recently great theoretical effort to develop a consistent formalism for the double parton distribution functions which include transverse momentum dependence, see \cite{Buffing:2016qql,Buffing:2017mqm} and \cite{Golec-Biernat:2016vbt}. The framework presented in \cite{Buffing:2016qql,Buffing:2017mqm} is based on the extension of the TMD factorization to the double scattering case for colorless final states.

   In the present work we perform detailed numerical analysis of  the unintegrated double gluon distribution within the formalism which was proposed in our earlier work \cite{Golec-Biernat:2016vbt}. In that work, the  double parton distributions with the transverse momentum dependence were constructed from the integrated double parton distributions through the convolution with the Altarelli-Parisi splitting functions and by including the Sudakov form factors.   This is an extension of the formalism proposed  by Kimber, Martin and Ryskin \cite{Kimber:1999xc,Kimber:2001sc, Kimber:2000bg} for the case of the single parton distributions, which has been successfully applied to a wide range of processes which are sensitive to the transverse momentum dependence.

   We perform a thorough numerical analysis of the unintegrated double gluon distribution, in particular the dependence on the transverse momenta and  longitudinal momentum fractions of gluons, and also  on hard scales. In addition, we analyze  non-perturbative contributions
  in which at least one parton 
  is almost collinear with the parent hadron with a low, non-perturbative value of the transverse momentum. We observe that the distribution in the transverse momentum of the double gluon density shifts towards higher values for increasing values of hard scales and decreasing values of longitudinal momentum fraction $x$.    We also investigate the factorization of the unintegrated double gluon distribution into a product of two unintegrated single gluon distribution. We show that factorization holds for sufficiently low values of $x$ provided the splitting contribution  in the evolution equations for the integrated double gluon distribution is included and the momentum sum rule for this  distribution holds.

The outline of the paper is the following. In Sec.~\ref{sec:2a} we recall the evolution equations for the integrated double parton distribution functions. In Sec.~\ref{sec:3} we discuss the choice of the initial conditions and the issues related to the momentum sum rules.  In Sec.~\ref{sec:4} we present numerical results on various aspects of 
the dependence of the unintegrated double gluon distribution on kinematic variables while 
in Sec.~\ref{sec:5} we  discuss the factorization issues. Finally, in Sec.~\ref{sec:6} we present summary and outlook of our analysis. 

\section{Evolution equations for the double gluon distribution}
\label{sec:2a}
We shall start by recalling the evolution equations for the collinear double parton distribution functions. These equations have been first proposed in \cite{Konishi:1978yx,Konishi:1979cb}  in the context of jet physics for the perturbative QCD description of the jet  structure and later derived in \cite{Shelest:1982dg,Zinovev:1982be,Snigirev:2003cq} for the initial state double parton distribution functions.
We shall review first the evolution equations for the  integrated double parton distribution functions, following results of Ref.~\cite{Ceccopieri:2010kg}. We note that the DPDFs also depend  on the transverse momentum vector, $\qperpb$, which we set to zero.  This means that in the Fourier space one integrates over the relative position of the two partons  $\bbold$, see \cite{Diehl:2011tt,Buffing:2017mqm}.
In principle this dependence on $\qperpb$ could be reinstated in the presented formalism through the inclusion of the appropriate form factor, see for example \cite{Ryskin:2011kk}. Precise description of the transverse distribution of partons and the ensuing correlations between the partons in transverse plane is an outstanding and challenging problem \cite{Blok:2017alw,Strikman:2011cx,Rogers:2009ke,Rogers:2008ua}. For now,  we shall postpone this problem to a further study and consider distributions integrated over $\bbold$.  

For $\qperpb=0$, the DPDFs in the lowest order approximation are probabilities to find two partons with longitudinal momentum fractions $x_{1}$ and $x_2$, when probed at  two different scales $Q_{1}$ and $Q_2$   \cite{Diehl:2011yj}. 
The integrated double distributions are denoted from now on by
\be
D_{a_1a_2}\; \equiv \; D_{a_1a_2}(x_1,x_2,Q_1,Q_2,\qperpb=0) \; ,
\ee
where $a_1$ and $a_2$ refer either  to  the quark flavor or gluon $g$.
 The solution to the two scale evolution of the DPDFs can be cast in the following form \cite{Ceccopieri:2010kg}
\begin{align}\nonumber
\label{eq:9}
D_{a_1a_2} &=
\sum_{\aprime,\abis}\bigg\{
\int_{x_1}^{1-x_2}\frac{dz_1}{z_1}\int_{x_2}^{1-z_1}\frac{dz_2}{z_2}\,
E_{a_1\aprime}\Big(\frac{x_1}{z_1},Q_1,Q_0\Big)
E_{a_2\abis}\Big(\frac{x_2}{z_2},Q_2,Q_0\Big) \\ \nonumber &\times D_{\aprime \abis}(z_1,z_2,Q_0,Q_0)
\\\nonumber
&+
\int^{Q_{\min}^2}_{Q_0^2}\frac{dQ^2_s}{Q^2_s}
\int_{x_1}^{1-x_2}\frac{dz_1}{z_1}
\int_{x_2}^{1-z_1}\frac{dz_2}{z_2}\,
E_{a_1\aprime}\Big(\frac{x_1}{z_1},Q_1,Q_s\Big)
E_{a_2\abis}\Big(\frac{x_2}{z_2},Q_2,Q_s\Big) 
\\ 
& \times D_{\aprime \abis}^{(sp)}(z_1,z_2,Q_s)\bigg\} \; ,
 \end{align}
where $Q_{\min}^2=\min\{Q_1^2,Q_2^2\}$, $Q_0$ is an initial scale for the evolution and the integration limits 
 take into account kinematic constraints $x_1,x_2>0$ and $x_1+x_2\le 1$.

\begin{figure}[t]
\begin{center}
\includegraphics[width=0.22\textwidth]{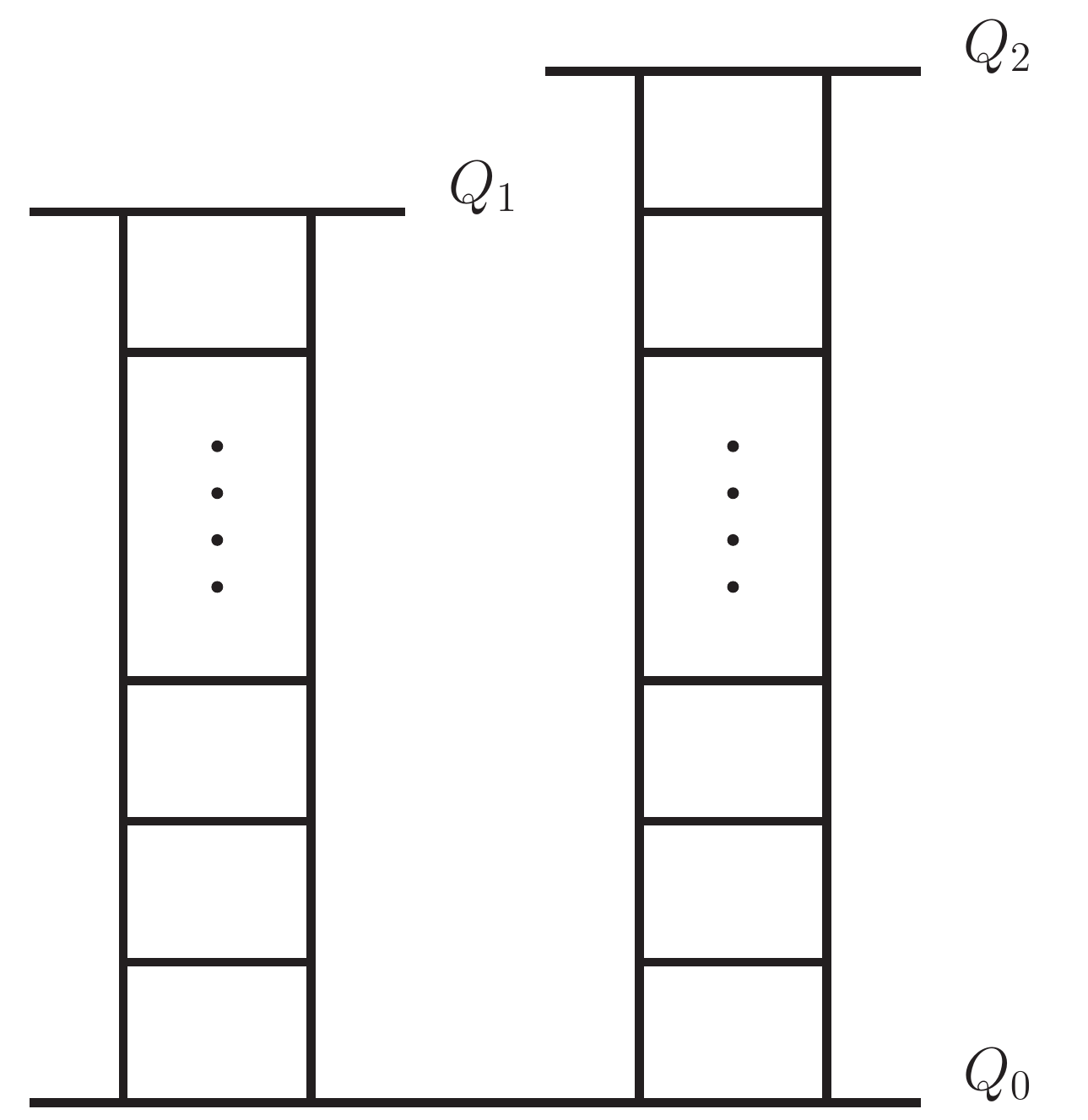}\hspace*{1.7cm}
\includegraphics[width=0.22\textwidth]{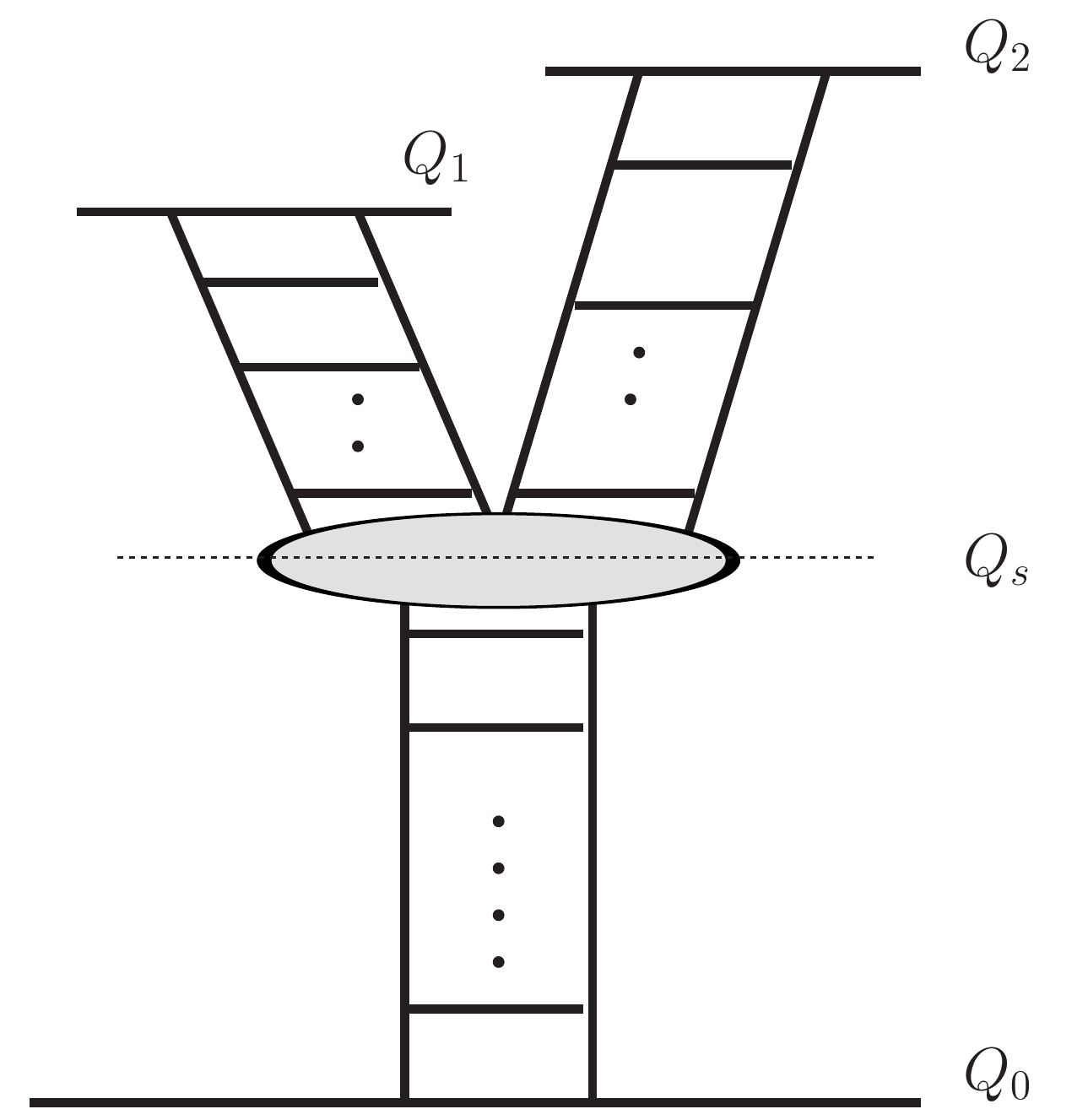}
\vspace*{0.5cm}
\caption{Schematic illustration of  the two contributions to the DPDFs (\ref{eq:9}). Left: homogeneous term; right: non-homogeneous term. It is understood that all the ladders are cut diagrams. $Q_0$ is the scale at which initial conditions are set.}
\label{fig:fig1a}
\end{center}
\end{figure}

 The functions $E_{ab}$ are   parton-to-parton evolution distributions which obey the DGLAP evolution equation, 
\begin{align}\nonumber
\label{eq:b12}
\frac{\partial}{\partial \ln \mu^2} E_{ab}(x,\mu,\mu_0) = \sum_{\aprime}\int_x^{1}\frac{dz}{z}\,P_{a\aprime}(z,\mu)\,
E_{\aprime b}\Big(\frac{x}{z},\mu,\mu_0\Big) 
\\
- E_{ab}(x,\mu,\mu_0)\,\sum_{\aprime}\int_0^{1}dz z P_{\aprime a}(z,\mu) \; ,
\end{align}
with the initial condition $E_{ab}(x,\mu_0,\mu_0)=\delta_{ab}\,\delta(1-x)$. 
The functions  $E_{ab}$ have the interpretation  of the  Green's functions for the DGLAP evolution, and using them we can construct the evolved single PDF as follows
\be
D_a(x,\mu) = \sum_b\, \int_{x}^1 \frac{dz}{z} \, E_{ab}\Big(\frac{x}{z},\mu,\mu_0\Big) \, D_b(z,\mu_0) \; .
\ee
Eq.~\eqref{eq:9}   is a sum of two contributions which are schematically shown in  Fig.~\ref{fig:fig1a}
\be
D_{a_1a_2} = D^{(h)}_{a_1a_2} + D^{(nh)}_{a_1a_2} \,.
\ee 
The first,  homogenous term, $D^{(h)}_{a_1a_2},$ is proportional to the double parton distribution and corresponds to the independent evolution of two partons  from the initial scale $Q_0$ to $Q_1$ and from $Q_0$ to $Q_2$. The second, non-homogeneous term, $ D^{(nh)}_{a_1a_2}$,
contains the distribution
\be
\label{eq:10}
D_{\aprime \abis}^{(sp)}(x_1,x_2,Q_s) =\frac{\alpha_s(Q_s)}{2\pi}\sum_{a}\frac{D_{a}(x_1+x_2,Q_s)}{x_1+x_2}\,
P_{a\to \aprime \abis}\bigg(\frac{x_1}{x_1+x_2}\bigg)\,,
\ee
which describes the splitting of the parton $a\to \aprime\abis$. 
Notice on the right hand side of Eq.~(\ref{eq:10}) the  single PDFs, $D_{a}$,
taken  at the splitting scale $Q_s$  along with the real emission leading order (LO) Altarelli-Parisi splitting functions,
 \be
\label{eq:1a}
P_{a\to \aprime \abis}(z) = P^{(0)}_{a\aprime}(z)\,.
\ee
In the LO approximation,  the flavor of the  second parton,  $\abis$, is uniquely determined from the splitting $a\to \aprime$. 
The single distributions $D_a$ are evaluated at $(x_1+x_2)$ due to  conservation of the parton longitudinal momentum in the evolution.

\section{Initial conditions for the evolution equations}
\label{sec:3}

Similarly to the case of the DGLAP evolution equations for single PDFs, the  evolution equations for the double parton distributions need to be supplemented by the appropriate non-perturbative initial conditions at  a scale $Q_0$, usually  of the order of $1\;{\rm GeV}$.  The initial conditions for the single PDFs are usually parametrized by some flexible functional forms with many free parameters fixed by fits to  the experimental data. 
There are a few constraints that need to be obeyed by the initial conditions, and these are momentum and quark sum rules. For the double distributions, 
such sum rules  relate the integrals of the DPDFs to the single PDFs, thus placing important constraints on the allowed forms of the initial conditions.   This makes the construction of the initial conditions for the DPDFs much more complicated, see \cite{Broniowski:2013xba,Golec-Biernat:2014bva,Golec-Biernat:2015aza}  for various
proposals.

 For the case when $Q_1=Q_2=Q$, the  momentum sum rules for the DPDFs
 can be expressed in the following form
\be
\label{eq:sumrule}
\sum_{a_1}\int_0^{1-x_2} dx_1 \, x_1 D_{a_1 a_2}(x_1,x_2,Q,Q)=(1-x_2) D_{a_2}(x_2,Q)\;\;,
\ee
where $D_{a_1}(x_1,Q)$ is the single parton distribution function, known from global fits to hard scattering data. 
The above momentum sum rule can be interpreted in the following way: the ratio $D_{a_1a_2}/D_{a_1}$ can be viewed as the conditional probability for finding parton $a_1$ while the longitudinal momentum $x_2$ of the second parton $a_2$ is fixed. Therefore the total momentum carried by the remaining partons (except parton $a_2$) is equal to $(1-x_2)$. 
The  quark number sum rule   is of a similar form, which can be found in e.g. \cite{Gaunt:2009re}. 

 In this paper we restrict ourselves to the single channel with gluon only, but the entire analysis could be extended to include quarks as well.   In such a  case, we are only concerned about the momentum sum rule \eqref{eq:sumrule}
 with $a_1=a_2=g$. Assuming that the double gluon distribution is a symmetric function of parton momenta fractions,
$D_{gg}(x_1,x_2)=D_{gg}(x_2,x_1)$, it can easily be derived that the rule (\ref{eq:sumrule}) is also valid for the second variable.

In the following, we shall use the initial conditions suggested by the construction proposed in   \cite{Golec-Biernat:2015aza}  which leads to  the initial conditions for both the single and double gluon distribution satisfying the momentum sum rules. The construction is based on the observation that the appropriate set of functions for the initial conditions can be chosen from the set of Dirichlet distributions.
Taking advantage of a particular form  of the 
single gluon distribution at the initial scale $Q_0=1~{\rm GeV}$,  given in  \cite{Martin:2009iq},
\be
\label{eq:2.2}
D_g(x,Q_0)= \sum_{k=1}^N N^k_g\, x^{\alpha^k_g} \, (1-x)^{\beta_g^k}\,,
\ee
one can demonstrate  that the double gluon distribution
\be
\label{eq:dpdfinitial}
D_{gg}(x_1,x_2,Q_0,Q_0)=\sum_{k=1}^N \, N^k_g\frac{\Gamma(\beta^k_g+2)}{\Gamma(\alpha^k_g+2)\Gamma(\beta^k_g-\alpha^k_g)}\,
(x_1x_2)^{\alpha^k_g}\,(1-x_1-x_2)^{\beta^k_g-\alpha^k_g-1},
\ee
obeys the momentum sum rule (\ref{eq:sumrule}), see  \cite{Golec-Biernat:2015aza} for details. Note that, the initial double gluon distribution (\ref{eq:dpdfinitial}) is completely determined by the  parameters of the
single gluon distributions, $N^k_g, \alpha^k_g, \beta^k_g$, known from the global fits \cite{Martin:2009iq}. 
It is also worth emphasizing that the distribution \eqref{eq:dpdfinitial} is not a product
of two single gluon distributions, even for small values of parton momentum fractions.

The momentum sum rule is preserved by evolution equations for the double parton distribution \cite{Shelest:1982dg,Zinovev:1982be}.
Thus, the rule (\ref{eq:sumrule}) is independent of the hard scale $Q\equiv Q_1=Q_2$ at which the double gluon distribution is defined. 
 The distribution (\ref{eq:dpdfinitial}) will be used in the forthcoming discussion as the initial condition for the evolution equations at some initial scale $Q_0$ and we shall refer to it as the ${\rm GBLS}^3$ input.
For the comparison, we also use the distribution proposed in \cite{Gaunt:2009re} (referred to subsequently as the GS input),
\begin{equation}
\label{eq:gauntinput}
D_{gg}(x_1,x_2,Q_0,Q_0) \; = \; D_{g}(x_1,Q_0) \, D_{g}(x_1,Q_0) \, \frac{(1-x_1-x_2)^2}{(1-x_1)^2 \, (1-x_2)^2} \; .
\end{equation}
This form is approximately  factorizable into product of two single PDFs for small values of momentum fractions  and includes the correlating factor which guarantees smooth vanishing of the distribution when  $(x_1+x_2) \rightarrow 1$. This ansatz does not satisfy the momentum sum rule \eqref{eq:sumrule}, although we checked that  the violation is relatively small for almost all  values of $x$, that is below $5\%$ percent for $x<0.5$, and only becomes significant, up to $30\%$ level, for higher values of $x$.

\section{Unintegrated double gluon distribution}
\label{sec:4}

In this section we present the construction of the double  gluon distribution which is unintegrated over the transverse momenta.  The framework discussed in the current  work, proposed recently in \cite{Golec-Biernat:2016vbt}, allows to construct the unintegrated double parton distributions which additionally
depend on parton transverse momenta, $k_{1\perp}$ and $k_{2\perp}$.
In the following  analysis, we are only interested in the  unintegrated double gluon distribution, 
$f_{gg}(x_1,x_2,k_{1\perp},k_{2\perp},Q_1,Q_2)$, though similar investigations can be performed for the other distributions\footnote{
In this case though one would need to specify the appropriate initial conditions for the integrated DPDFs which satisfy the  momentum and quark sum rules  with known single PDFs, which  is a much more difficult  task.}. 

In the proposed construction \cite{Golec-Biernat:2016vbt}, which follows  the original proposal of Kimber-Martin-Ryskin for the single PDFs 
\cite{Kimber:1999xc,Kimber:2001sc}, 
the dependence
on the transverse momenta is obtained from unfolding the last step in the DGLAP-like evolution equations \eqref{eq:9} reduced to the gluonic sector.  
The integrated double gluon distribution is the sum of  two contributions described in the previous section and illustrated in Fig.~\ref{fig:fig1a}, 
\be
\label{eq:hnh}
D_{gg} = D_{gg}^{(h)}(x_1,x_2,Q_1,Q_2) + D_{gg}^{(nh)}(x_1,x_2,Q_1,Q_2) \; .
\ee

Similarly, the unintegrated double gluon distribution can also be written as a sum of the homogenous and non-homogeneous contributions,
\be
f_{gg}= f_{gg}^{(h)}(x_1,x_2,k_{1\perp},k_{2\perp},Q_1,Q_2)
+f_{gg}^{(nh)}(x_1,x_2,k_{1\perp},k_{2\perp},Q_1,Q_2) \; .
\ee

\subsection{Homogeneous contribution}

The form of the homogeneous contribution, $f_{gg}^{(h)}$, in the region where transverse momenta are perturbative,  $\kperpone, \kperptwo>Q_0$, is given by \cite{Golec-Biernat:2016vbt}
\begin{align}\nonumber
\label{eq:3.1}
f_{gg}^{(h)}(&x_1,x_2,k_{1\perp},k_{2\perp}, Q_1,Q_2)=
T_{g}(Q_1,\kperpone)\,T_{g}(Q_2,\kperptwo)\,\times
\\
&\times\,
\int_{\frac{x_1}{1-x_2}}^{1-\Delta_1}\frac{dz_1}{z_1}
P_{gg}(z_1,\kperpone)\,
\int_{\frac{x_2}{1-x_1/z_1}}^{1-\Delta_2}\frac{dz_2}{z_2}
P_{gg}(z_2,\kperptwo)\,
D_{gg}^{(h)}\Big(\frac{x_1}{z_1},\frac{x_2}{z_2},\kperpone,\kperptwo\Big).
\end{align}
In the above equation, $T_g$ is the well known Sudakov form factor which sums virtual gluon emissions,
\be
\label{eq:3.2}
T_g(Q,\kperp) =\exp\bigg\{
-\int_{k^2_{\perp}}^{Q^2} \frac{d\pperp^2}{\pperp^2}\int_0^{1-\Delta} dz z P_{gg}(z,\pperp)\bigg\}\,,
\ee
where $P_{gg}$ is the real part of  Altarelli-Parisi gluon-gluon splitting function in the leading order approximation,
\be
\label{eq:3.3}
P_{gg}(z,\pperp)= \frac{\alpha_s(\pperp)}{2\pi}\,2N_c\bigg\{\frac{z}{1-z}+\frac{1-z}{z}+z(1-z)\bigg\}\,.
\ee
The integration limits in Eq.~\eqref{eq:3.1}  impose the following condition for 
the  longitudinal momentum fractions
 \be
\label{eq:3.6}
0<\frac{x_1}{1-\Delta_1}+\frac{x_2}{1-\Delta_2}<1\,,
\ee 
which replaces the   condition for the integrated double parton distributions,  $0<x_1+x_2<1$.

The parameters $\Delta$ in Eqs.~\eqref{eq:3.1} and \eqref{eq:3.2}  regularize the integrals at $z=1$ and could depend on transverse momenta of partons,
$\kperpone$ and $\kperptwo$ for real emission and $\pperp$ for virtual one, as well as on the hard scales, $Q_1$ and $Q_2$.
Thus, Eq.~\eqref{eq:3.6}  introduces non-trivial correlations between transverse and longitudinal momenta among both partons.
We consider two options in our numerical studies for the kinematics of parton emissions. 
\begin{enumerate} 
\item \underline{Strong ordering (SO):}\\
The strong ordering  of transverse momenta of emitted partons, typical for DGLAP dynamics, leads to
\be
\label{eq:3.4}
\Delta_i=\frac{k_{i\perp }}{Q_i}\,,~~~~~~~~~i=1,2\,.
\ee
In this case, the transverse momenta are bounded from the above by  the hard scale,  $k_{i\perp} <Q_i$. 
\item \underline{Angular ordering (AO):}\\
The angular ordering   of parton emissions (AO) due to the coherence effects 
in spacelike cascades \cite{Ciafaloni:1987ur,Catani:1989sg,Catani:1989yc,Marchesini:1994wr} leads to another cutoff 
\cite{Kimber:2000bg,Kimber:2001sc} 
\be
\label{eq:3.5}
\Delta_i=\frac{k_{i\perp}}{k_{i\perp}+Q_i}\,,
\ee
In this case the transverse momenta are no longer bounded by the scales $Q_i$. 
\end{enumerate}

\begin{figure}[t]
\includegraphics[width=0.5\textwidth]{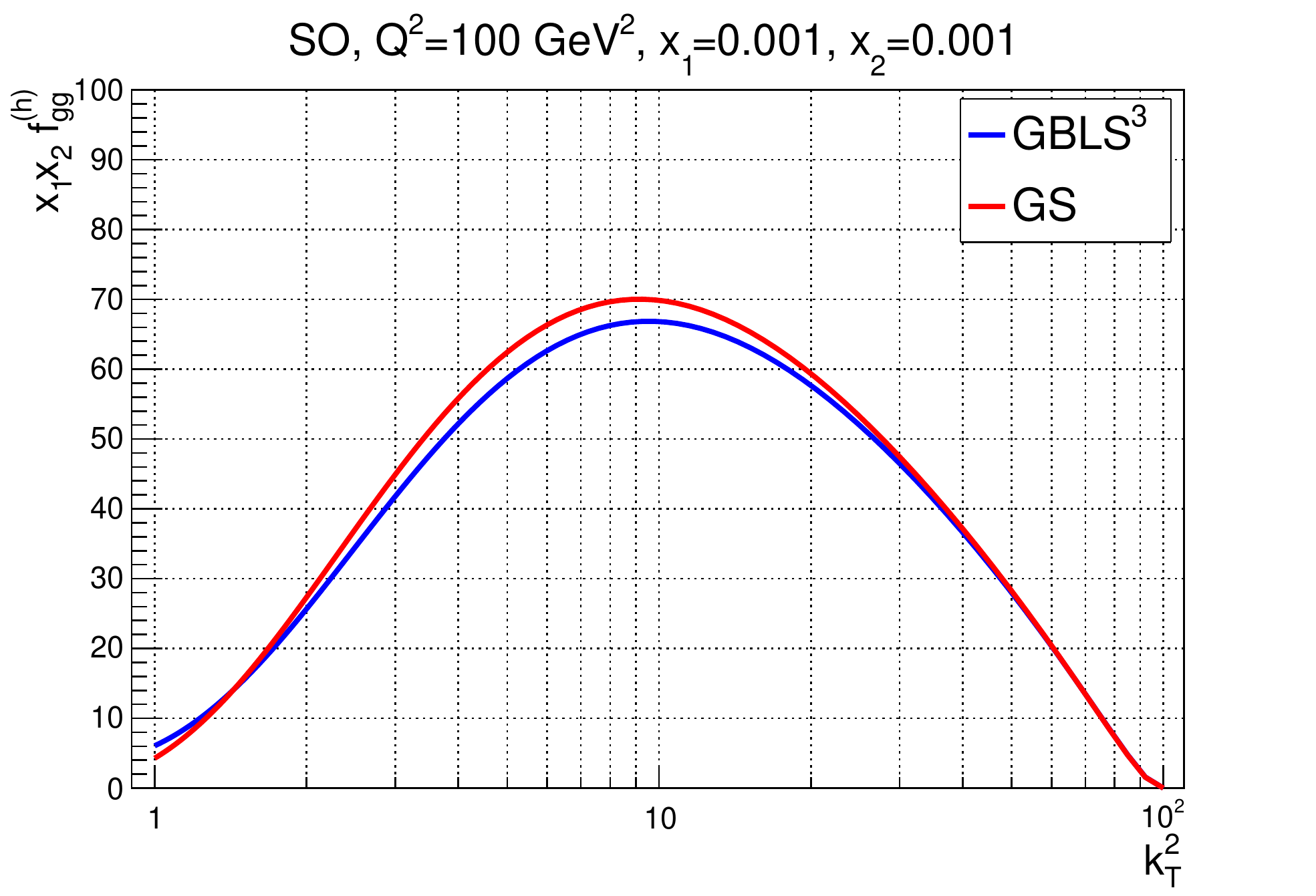}
\includegraphics[width=0.5\textwidth]{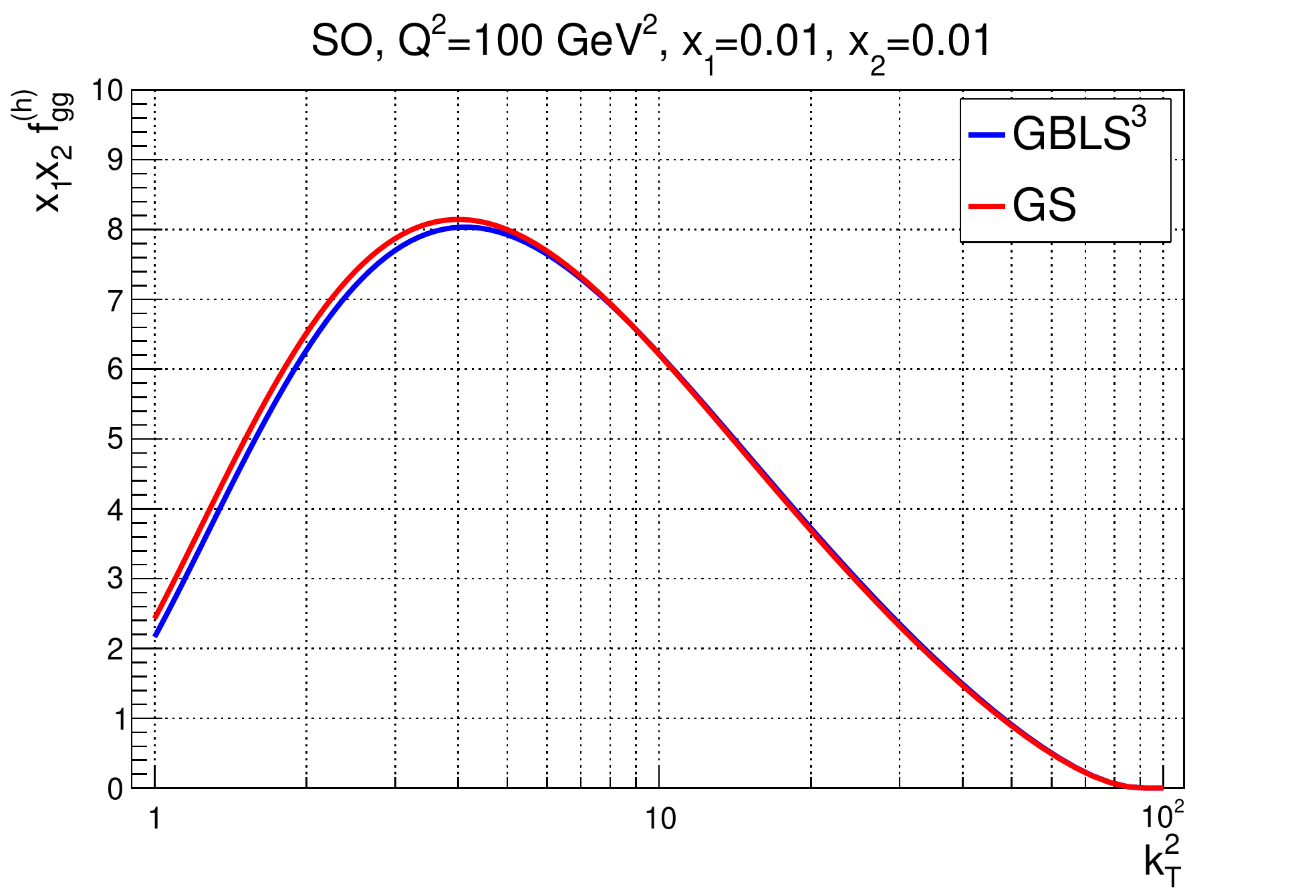}
\vspace*{1cm}
\includegraphics[width=0.5\textwidth]{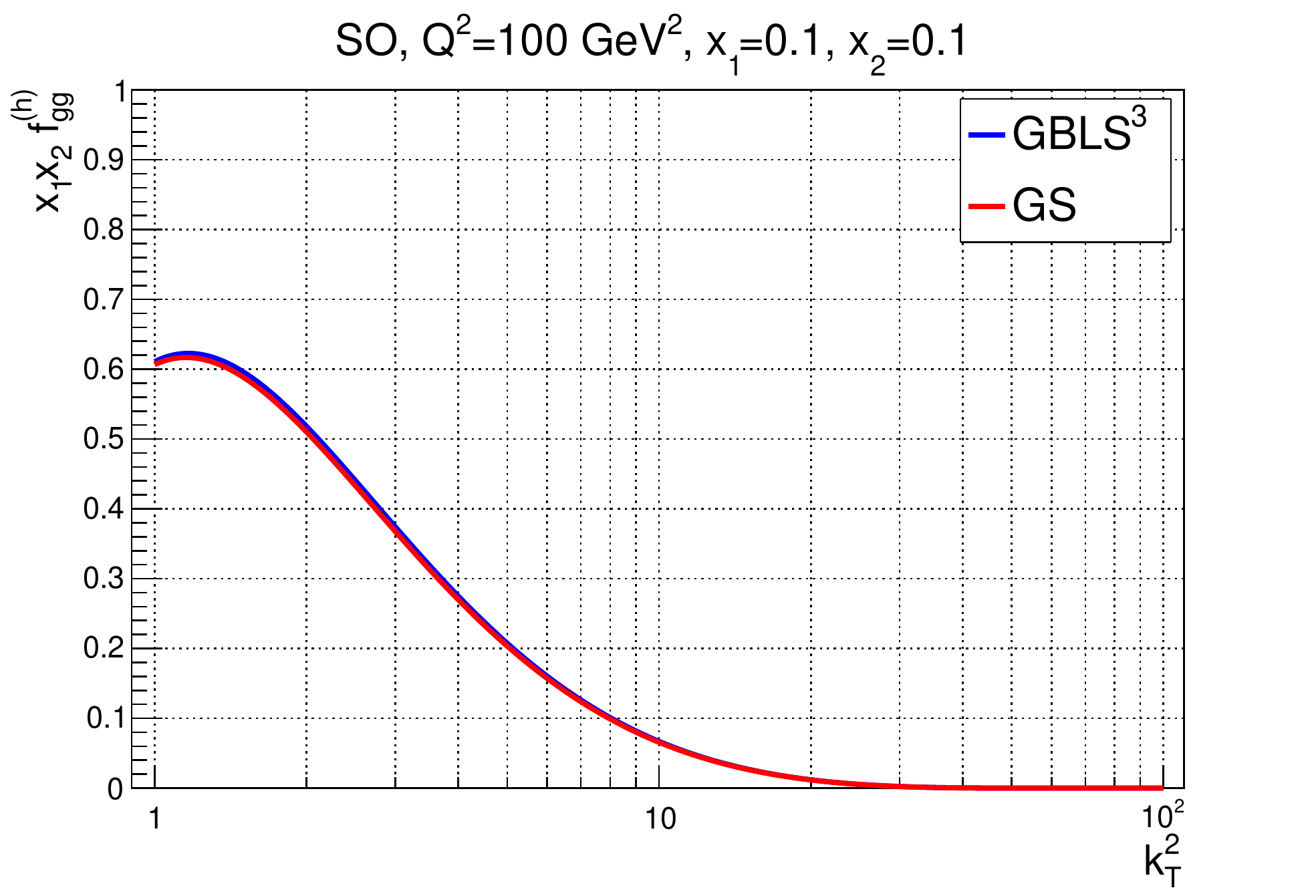}
\includegraphics[width=0.5\textwidth]{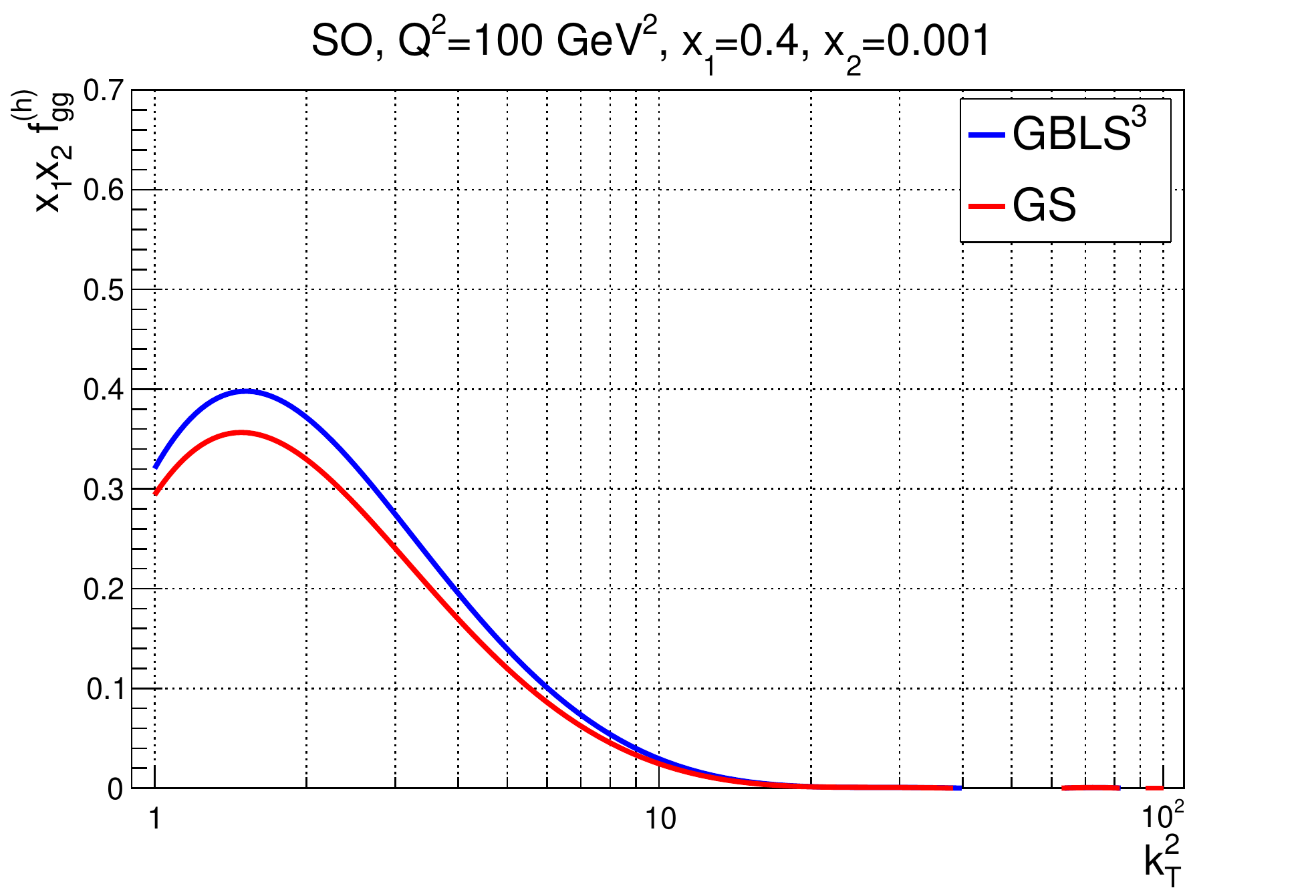}
\vskip -5mm
\caption{The distribution $x_1x_2 f_{gg}^{(h)}(x_1,x_2,\kperp,\kperp,Q,Q)$ from Eq.~\eqref{eq:3.1} plotted as a function of the transverse momentum $k^2_\perp \equiv\kperpone^2=\kperptwo^2$, for $Q^2=100\,{\rm GeV}^2$ and the indicated values of $(x_1,x_2)$,  starting from
the ${\rm GBLS}^3$  \eqref{eq:dpdfinitial}  (blue) and GS \eqref{eq:gauntinput} (red)  inputs. The cutoffs $\Delta_i$ were imposed according to the strong ordering condition \eqref{eq:3.4}. }
\label{fig:dpdf_kt_so2}
\end{figure}

\begin{figure}[t]
\includegraphics[width=0.5\textwidth]{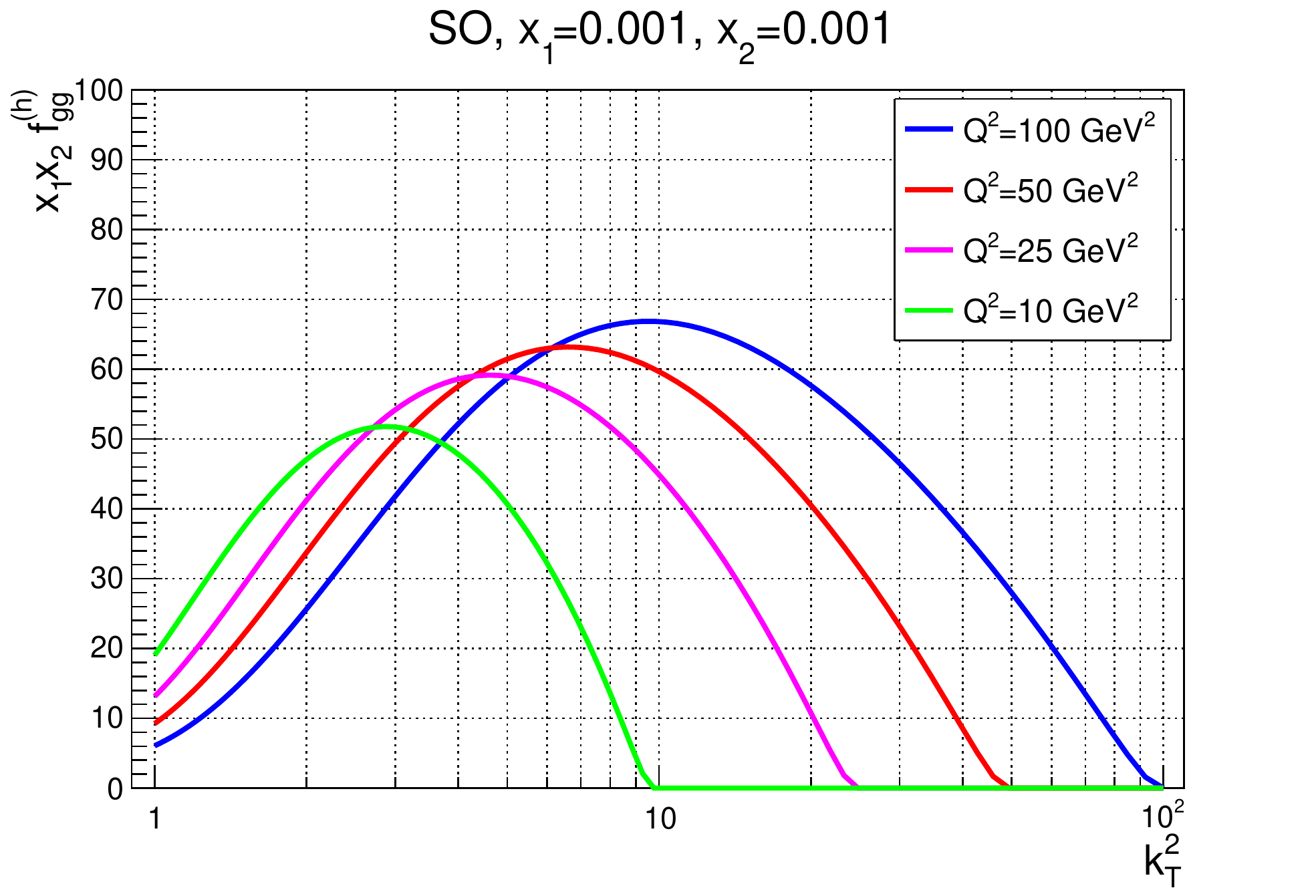}
\includegraphics[width=0.5\textwidth]{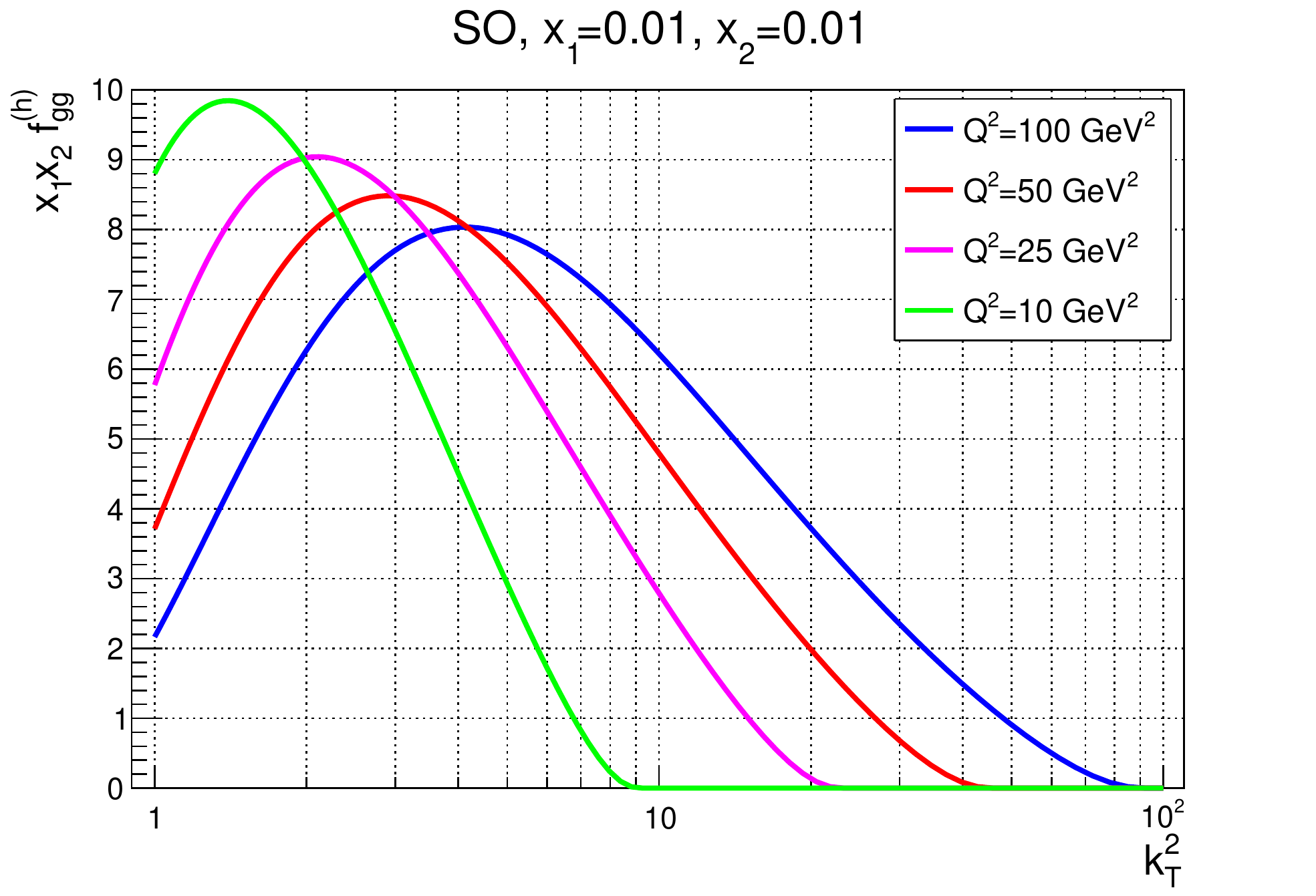}
\vspace*{1cm}
\includegraphics[width=0.5\textwidth]{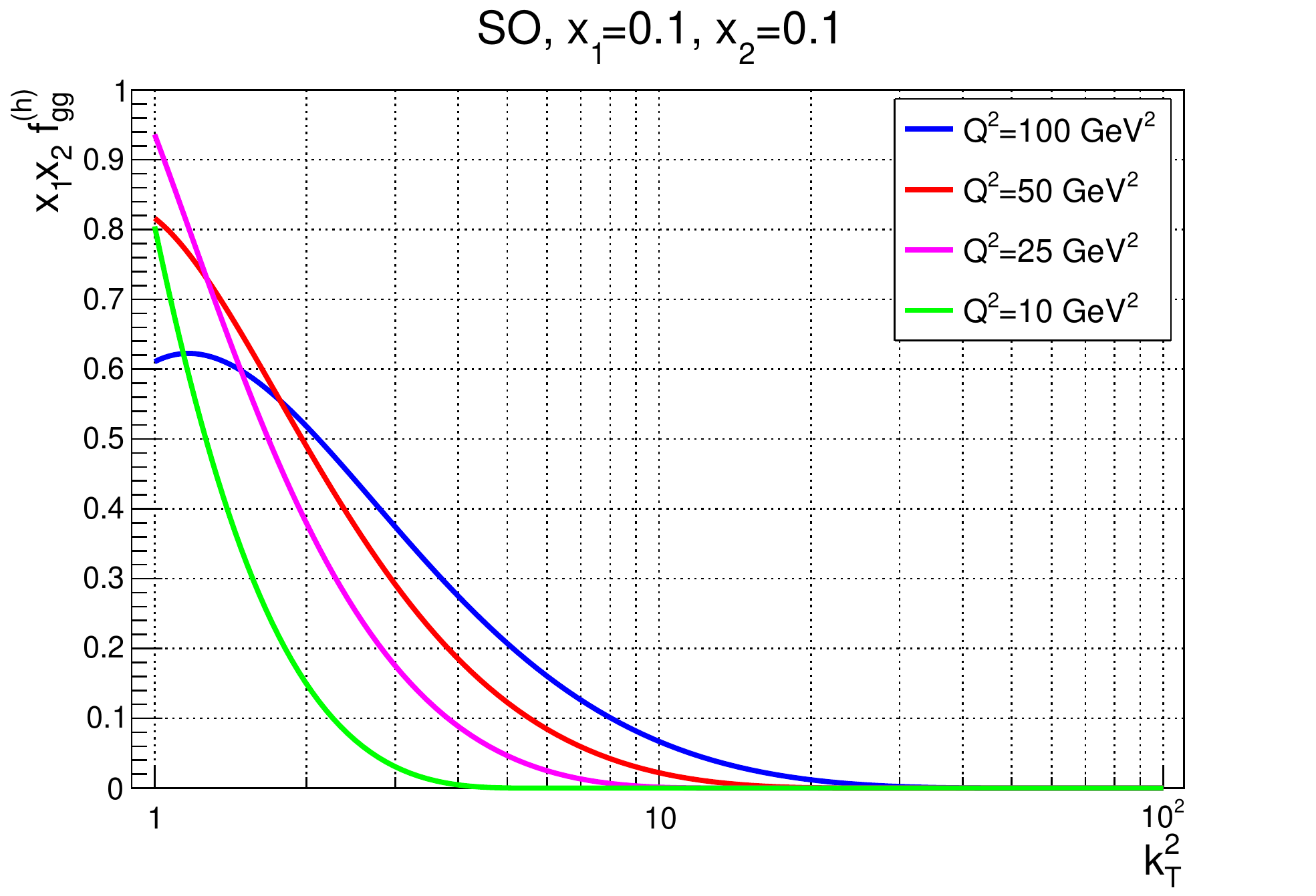}
\includegraphics[width=0.5\textwidth]{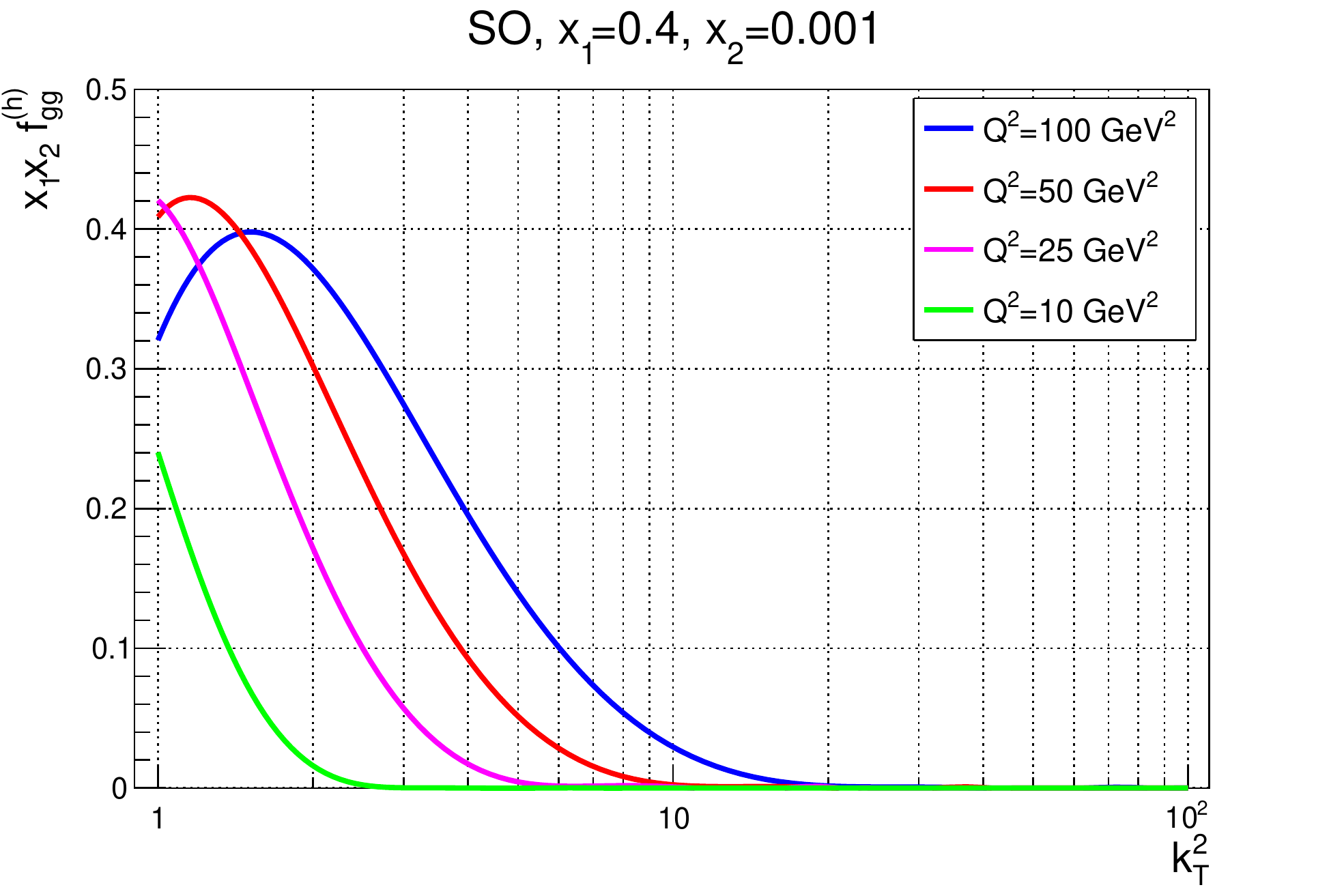}
\caption{The distribution $x_1x_2 f_{gg}^{(h)}(x_1,x_2,\kperp,\kperp,Q,Q)$ from  Eq.~\eqref{eq:3.1} as a function of the transverse momentum 
$k^2_\perp \equiv\kperpone^2=\kperptwo^2$,   for different values of $Q^2$ and the indicated values of $(x_1,x_2)$, 
starting from the ${\rm GBLS}^3$  input. 
The cutoffs $\Delta_i$ are imposed according to the condition \eqref{eq:3.4}.
}
\label{fig:dpdf_kt_q2depso}
\end{figure}

\begin{figure}[t]
\includegraphics[width=0.5\textwidth]{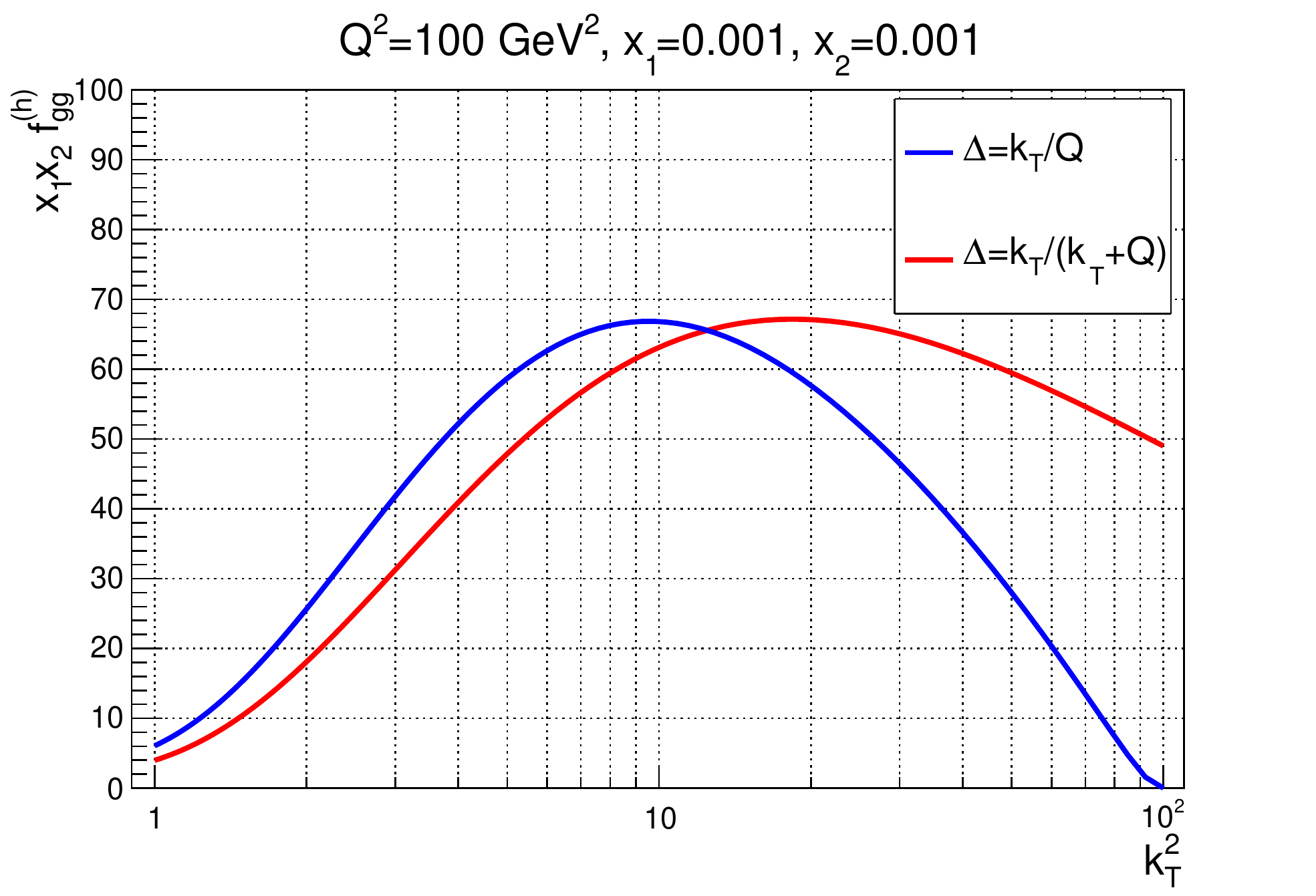}
\includegraphics[width=0.5\textwidth]{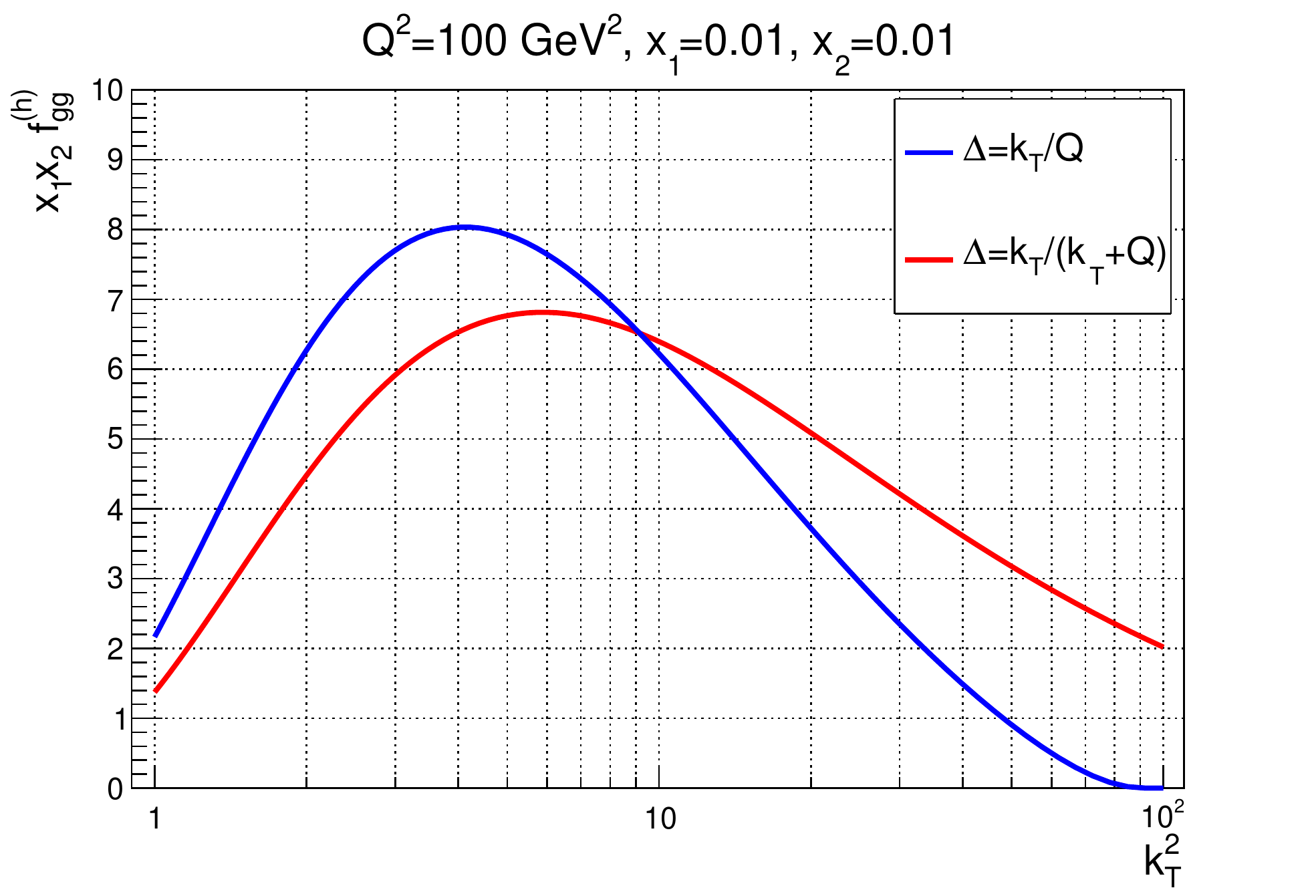}
\vspace*{1cm}
\includegraphics[width=0.5\textwidth]{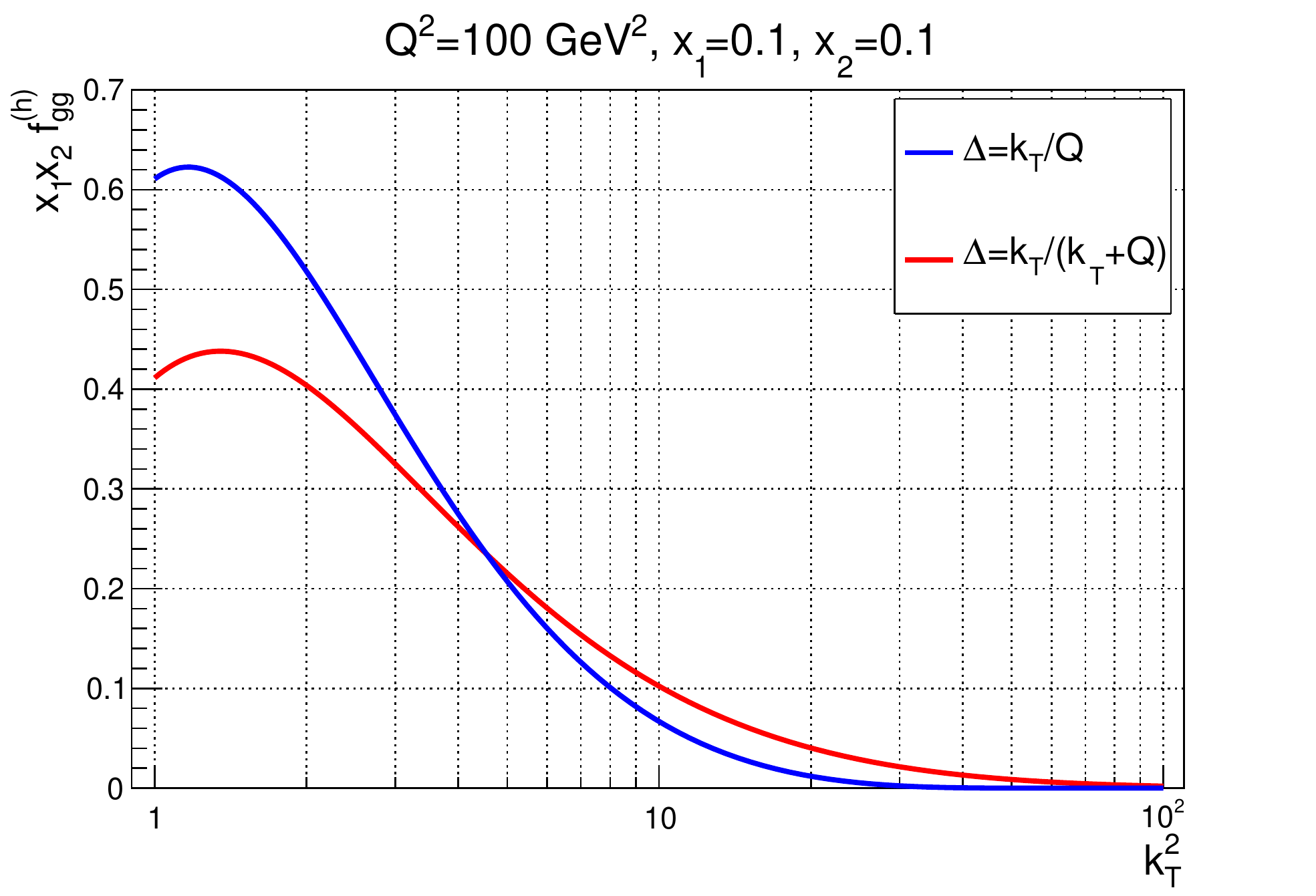}
\includegraphics[width=0.5\textwidth]{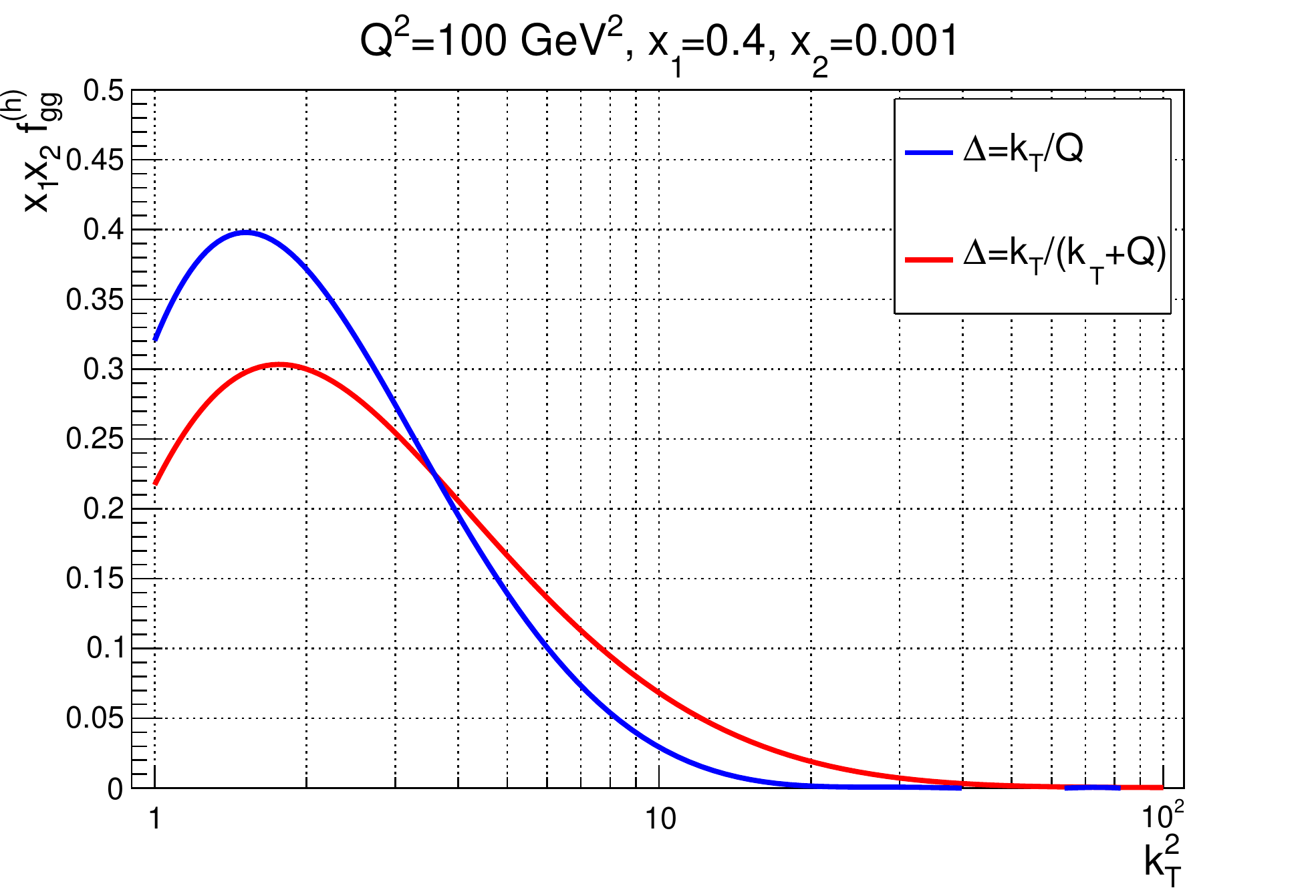}
\caption{The distribution $x_1x_2 f_{gg}^{(h)}(x_1,x_2,\kperp,\kperp,Q,Q)$ from Eq.~\eqref{eq:3.1} as a function of the transverse momentum 
$k^2_\perp \equiv\kperpone^2=\kperptwo^2$ in the strong ordering  \eqref{eq:3.4} (blue curve) and the angular ordering  \eqref{eq:3.5} (red curve) cases, 
for the  indicated values of $(x_1, x_2)$ and the ${\rm GBLS}^3$ input.
}
\label{fig:dpdf_kt_ao_gblsss}
\end{figure}

\begin{figure}[t]
\begin{center}
\includegraphics[width=0.49\textwidth]{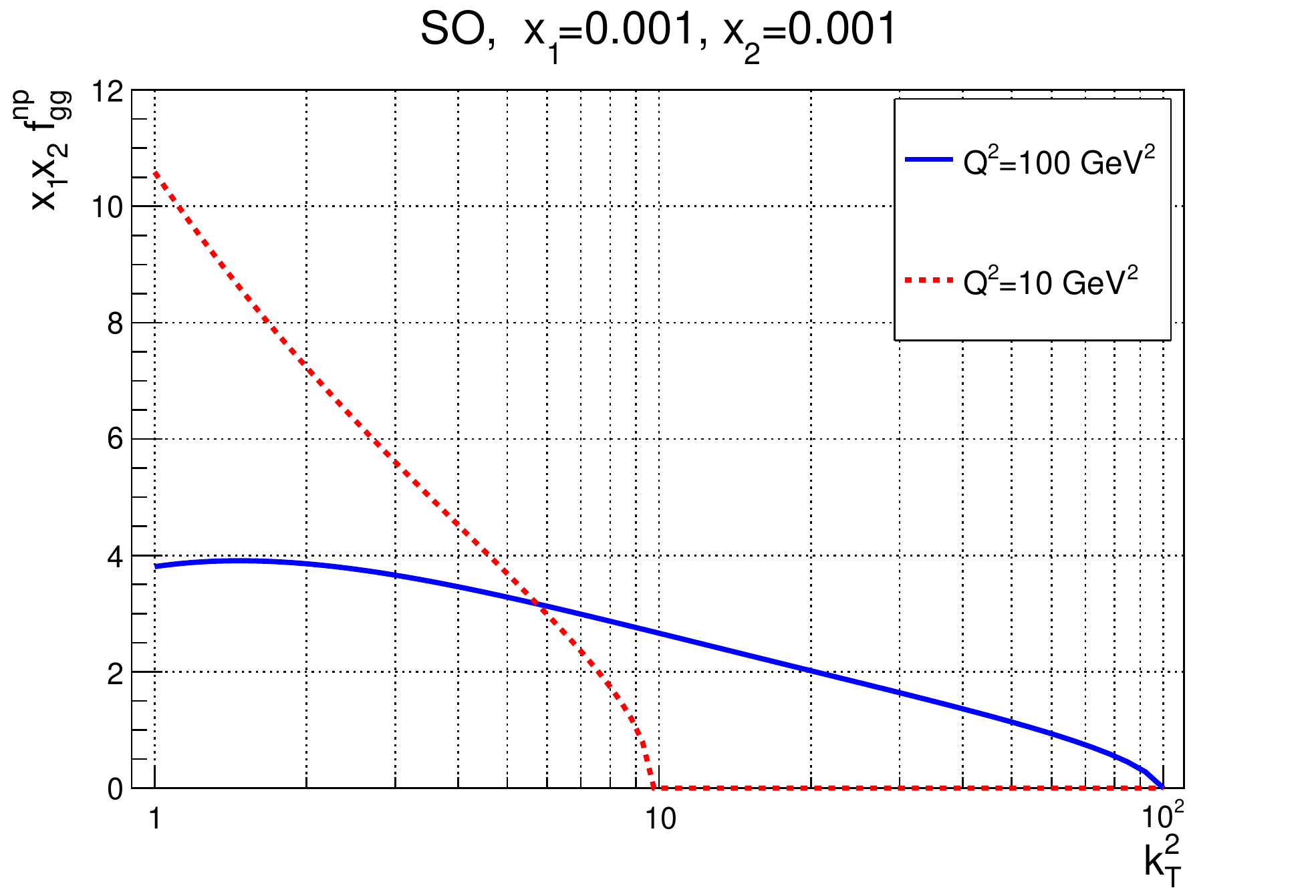}
\includegraphics[width=0.49\textwidth]{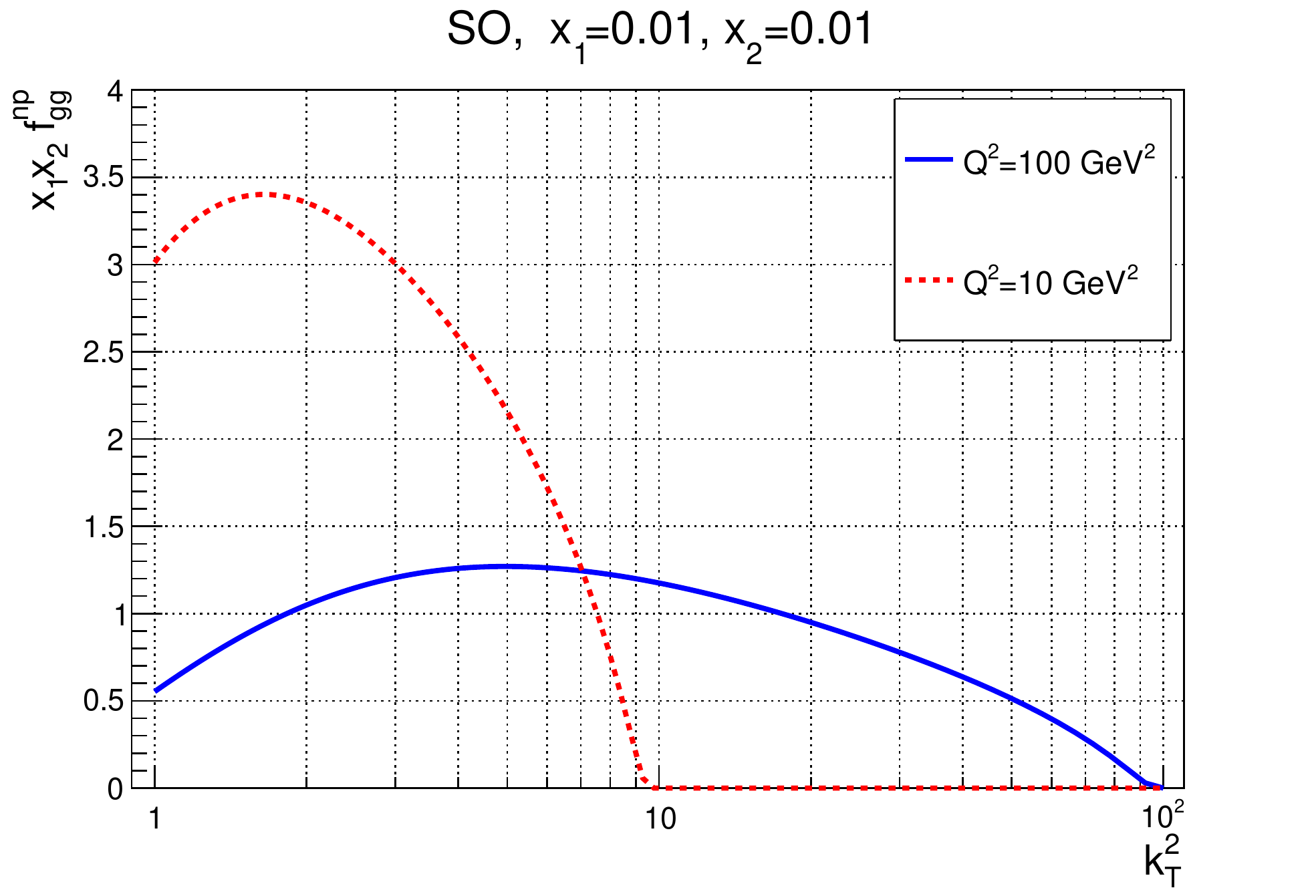}
\vspace*{1.5cm}
\includegraphics[width=0.49\textwidth]{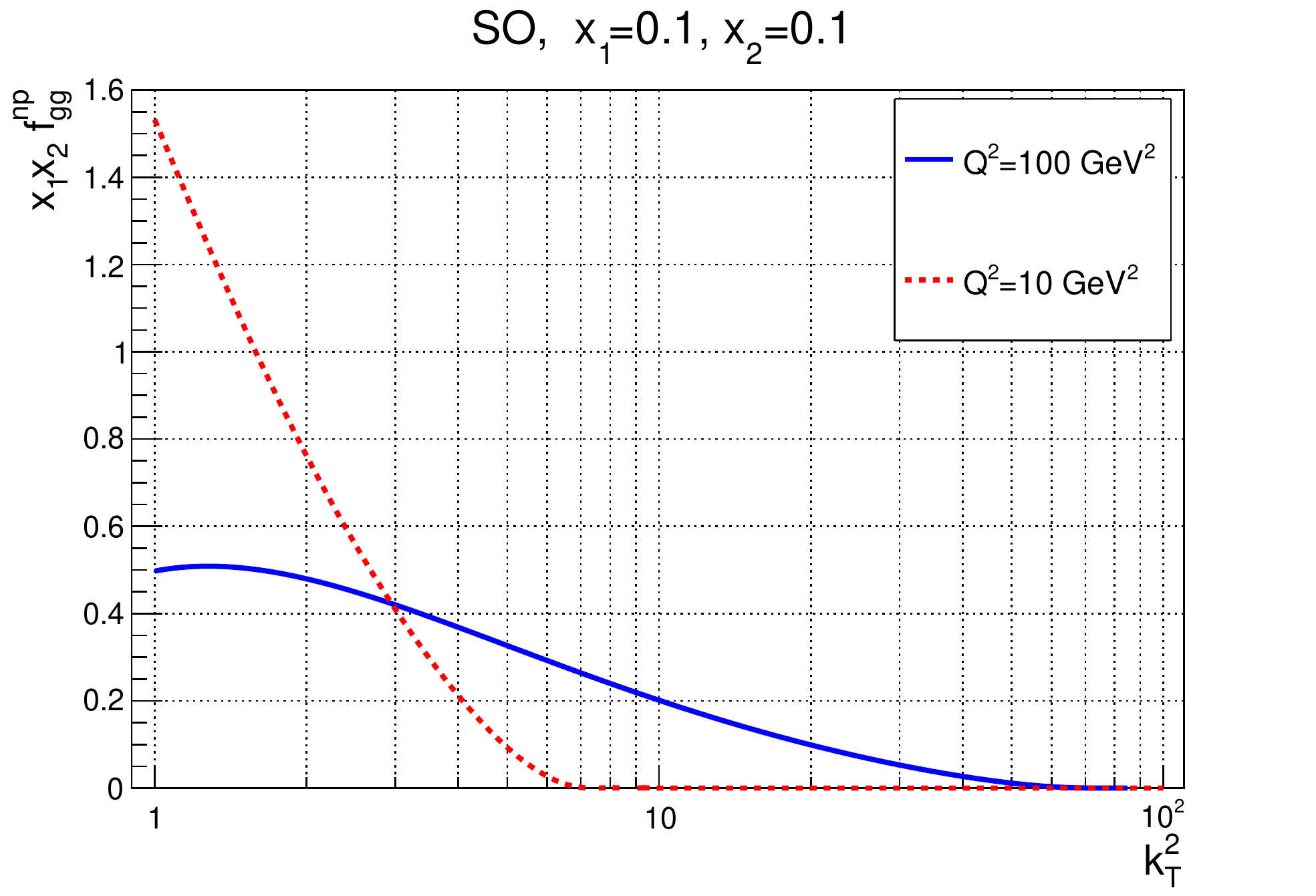}
\includegraphics[width=0.49\textwidth]{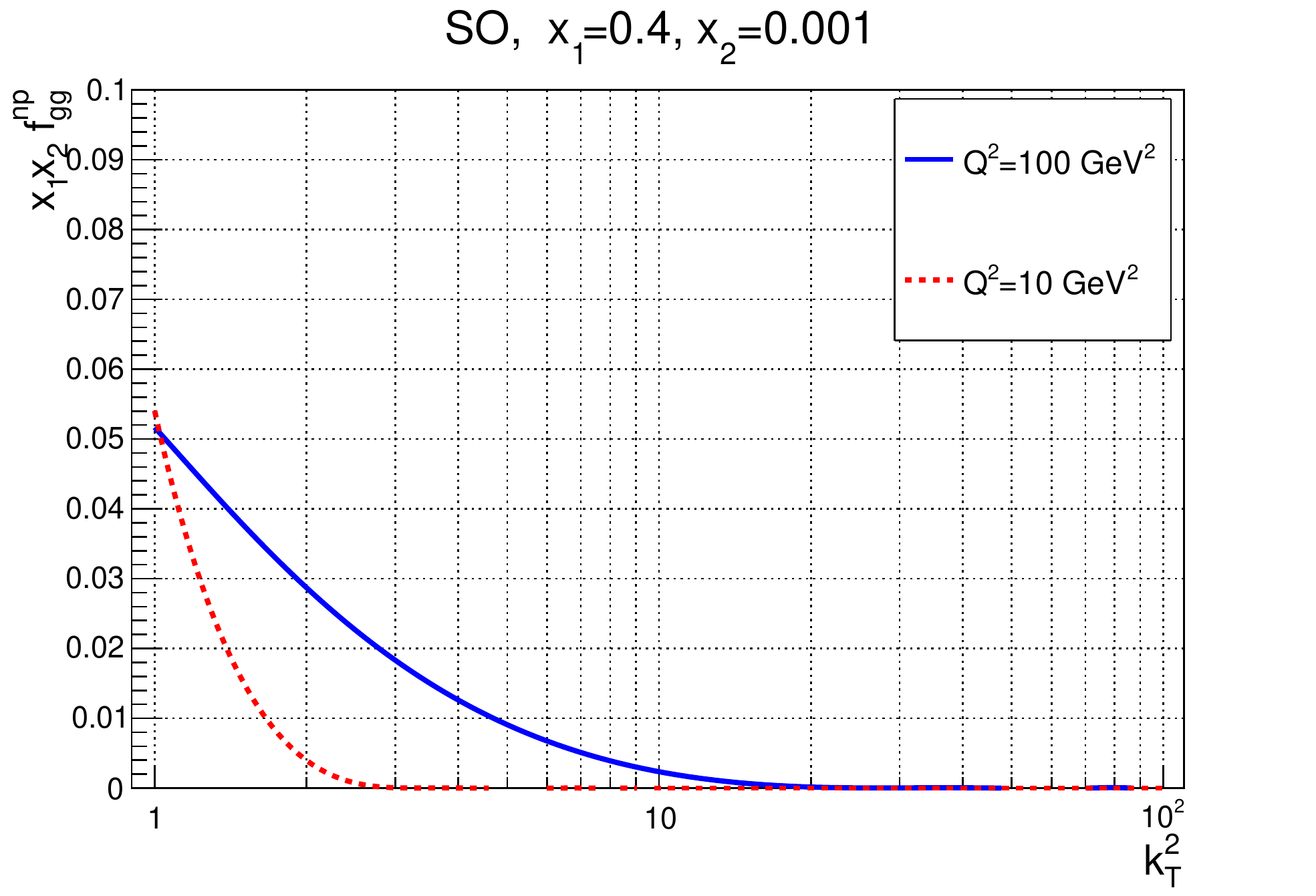}
\end{center}
\vskip -10mm
\caption{The non-perturbative distribution $x_1x_2 f_{gg}^{(h)}(x_1,x_2,\kperpone,Q,Q)$ from Eq.~\eqref{eq:3.7a}
as a function of   $k^2_\perp=\kperpone^2$ for $Q^2=10,100\,{\rm GeV}^2$ and
the  indicated values of $(x_1,x_2)$.
}
\label{fig:fig2as}
\end{figure}

In Fig.~\ref{fig:dpdf_kt_so2} we show the results for the homogeneous unintegrated double gluon distribution as a function of the transverse momentum 
$k^2_\perp\equiv\kperpone^2=\kperptwo^2$ for four pairs  of $(x_1,x_2)$ and  fixed hard scales $Q^2=Q_1^2=Q_2^2=100 \; {\rm GeV}^2$. We compare the evolution
which starts from two discussed initial conditions, given by Eqs.~\eqref{eq:dpdfinitial} and \eqref{eq:gauntinput}.
The calculations were performed for
 the cutoffs $\Delta_i$ which were taken according to the strong ordering (SO) scenario \eqref{eq:3.4}. 
We observe that the results for  the  different inputs are very similar to each other. This is somewhat surprising since the two inputs are rather different. 
The largest discrepancy is observed at the largest values of $x_1$. This is the region  where 
the effects due to the constraint from the momentum sum rule are most relevant. But even that difference is not large, and this is partially due to the fact that the GS input approximately satisfies the momentum sum rule, up to few percent for the relevant region in $x_1,x_2$.   We also see that the unintegrated double gluon distribution  vanishes at $\kperp=Q$,  as expected in the SO case, and the suppression at the lowest values of $k_T$ is due to the Sudakov form factor.

In Fig.~\ref{fig:dpdf_kt_q2depso}  we investigate the effect of different values of $Q^2$ on the unintegrated double gluon distribution for four fixed values  of  $(x_1,x_2)$ in the strong ordering case \eqref{eq:3.4}.
We observe that the distribution in $\kperp$ is shifted towards higher transverse momenta with increasing values of the  hard scale $Q^2$,
which leads to the greater spread in $\kperp$.
Also the peak of the distribution at fixed $Q^2$ is shifted towards larger values of transverse momenta with  decreasing $(x_1,x_2)$. This is due to the fact that as 
the gluon momentum fraction $x$ decreases, the probability of the gluon splitting increases.
The suppression at low values of $\kperp$ is increasing with $Q^2$  due to the Sudakov form factor. The visible sharp cutoff in transverse momenta at the value of the hard scale is consistent with the strong ordering cutoff.

In Fig.~\ref{fig:dpdf_kt_ao_gblsss} we compare the homogeneous unintegrated gluon distributions
in the strong ordering \eqref{eq:3.4} and the angular ordering  \eqref{eq:3.5} cases, plotted as a function of the transverse momentum
$k^2_\perp=\kperpone^2=\kperptwo^2$ for fixed hard scales $Q^2\equiv Q^2_1=Q^2_2=100 \; {\rm GeV}^2$. 
We should stress at this point that the angular ordering is not implemented in the strict sense since this condition is only incorporated in the last step of the evolution rather than in the entire cascade, like  in the original formulation for single gluon distribution \cite{Catani:1989sg,Catani:1989yc,Marchesini:1994wr}. Nevertheless, we  compare this model with the strong ordering case to illustrate the potential impact of the terms beyond the strong ordering scenario. As expected,  we see that the solution with the angular ordering exhibits tails  in transverse momentum  which extend well beyond the hard scale $Q$.  The tails are most prominent for very small values of $x$. This  can be  easily understood  since one needs to have $x<1-\Delta$ which for angular ordering leads to the condition $\kperp <Q(1/x-1)$ as compared to the condition $\kperp<Q(1-x)$ for the strong ordering. Therefore, at low values of $x$ and high $\kperp$, 
angular ordering allows for more phase space for emissions. At low values of $\kperp$, the calculation with angular ordering is typically lower than the calculation using strong ordering condition which stems from the fact that the suppression from the Sudakov form factor is   larger for the  angular ordering condition.

\subsection{Contributions from non-perturbative regions}

In Ref.~\cite{Golec-Biernat:2016vbt} we also provided formulae for the unintegrated double parton distributions in which
at least one of the parton transverse momentum is below the initial scale, i.e. for $k_{1\perp}>Q_0$
and $k_{2\perp} \le Q_0$,
\begin{align}\nonumber
\label{eq:3.7a}
f_{gg}^{(h)}(x_1,x_2,\kperpone, Q_1,Q_2) &=T_{g}(Q_1,\kperpone)\,T_{g}(Q_2,Q_0)\,\times
\\
&\times\,\int_{\frac{x_1}{1-x_2}}^{1-\Delta_1}\frac{dz_1}{z_1}\,
P_{gg}(z_1,\kperpone)\,D_{gg}^{(h)}\Big(\frac{x_1}{z_1},x_2,\kperpone,Q_0\Big)\,.
\end{align}
Notice that in such a case $f_{gg}^{(h)}$ only depends on  perturbative $\kperpone$ since
$\kperptwo$ is  integrated  out in the nonperturbative region below $Q_0$. This is
reflected in the dependence of $D_{gg}^{(h)}$ on $Q_0$.
In the opposite case, $\kperpone \le Q_0$ and $\kperptwo > Q_0$, we have
\begin{align}\nonumber
\label{eq:3.7b}
f_{gg}^{(h)}(x_1,x_2,\kperptwo, Q_1,Q_2) &=T_{g}(Q_1,Q_0)\,T_{g}(Q_2,\kperptwo)\,\times
\\
&\times\,\int_{\frac{x_2}{1-x_1}}^{1-\Delta_2}\frac{dz_2}{z_2}\,
P_{gg}(z_2,\kperptwo)\,
D_{gg}^{(h)}\Big(x_1,\frac{x_2}{z_2},Q_0,\kperptwo\Big)\,.
\end{align}
Finally, for the non-perturbative region, $\kperpone \le Q_0$ and $\kperptwo \le Q_0$,
\begin{align}
\label{eq:3.7c}
{f}_{gg}^{(h)}(x_1,x_2,Q_1,Q_2) =T_{g}(Q_1,Q_0)\,T_{g}(Q_2,Q_0)\,
D_{gg}^{(h)}(x_1,x_2,Q_0,Q_0)\,.
\end{align}
i.e. there is no transverse momentum dependence, since the transverse momenta have been integrated out below the initial scale $Q_0$ and
the function \eqref{eq:3.7c} is proportional to the integrated non-perturbative double parton distribution at the initial scales. 

In Fig.~\ref{fig:fig2as}, we show the non-perturbative gluon contributions given by Eq.~\eqref{eq:3.7a}, which   correspond to the configurations in which one gluon has the perturbative transverse momentum above the cutoff $Q_0$, while the other 
one is non-perturbative. In such a  case,  the double gluon distribution depends only on  transverse momentum of the first gluon.
We see that generally the  contributions from the non-perturbative regions are numerically much smaller than the perturbative ones, though they can become important for small scales.

\subsection{Non-homogeneous (parton splitting) contribution}

\begin{figure}[t]
\includegraphics[width=0.5\textwidth]{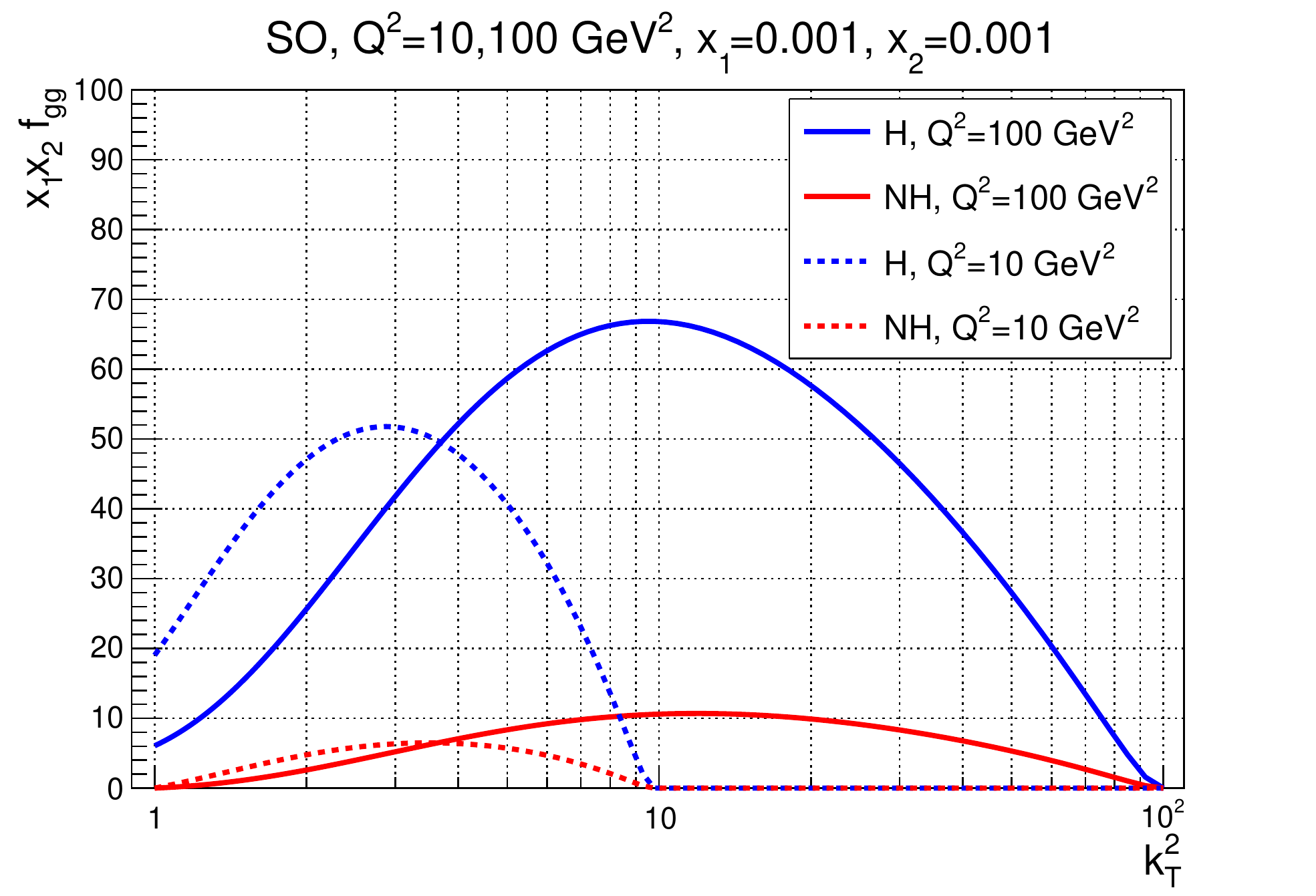}
\includegraphics[width=0.5\textwidth]{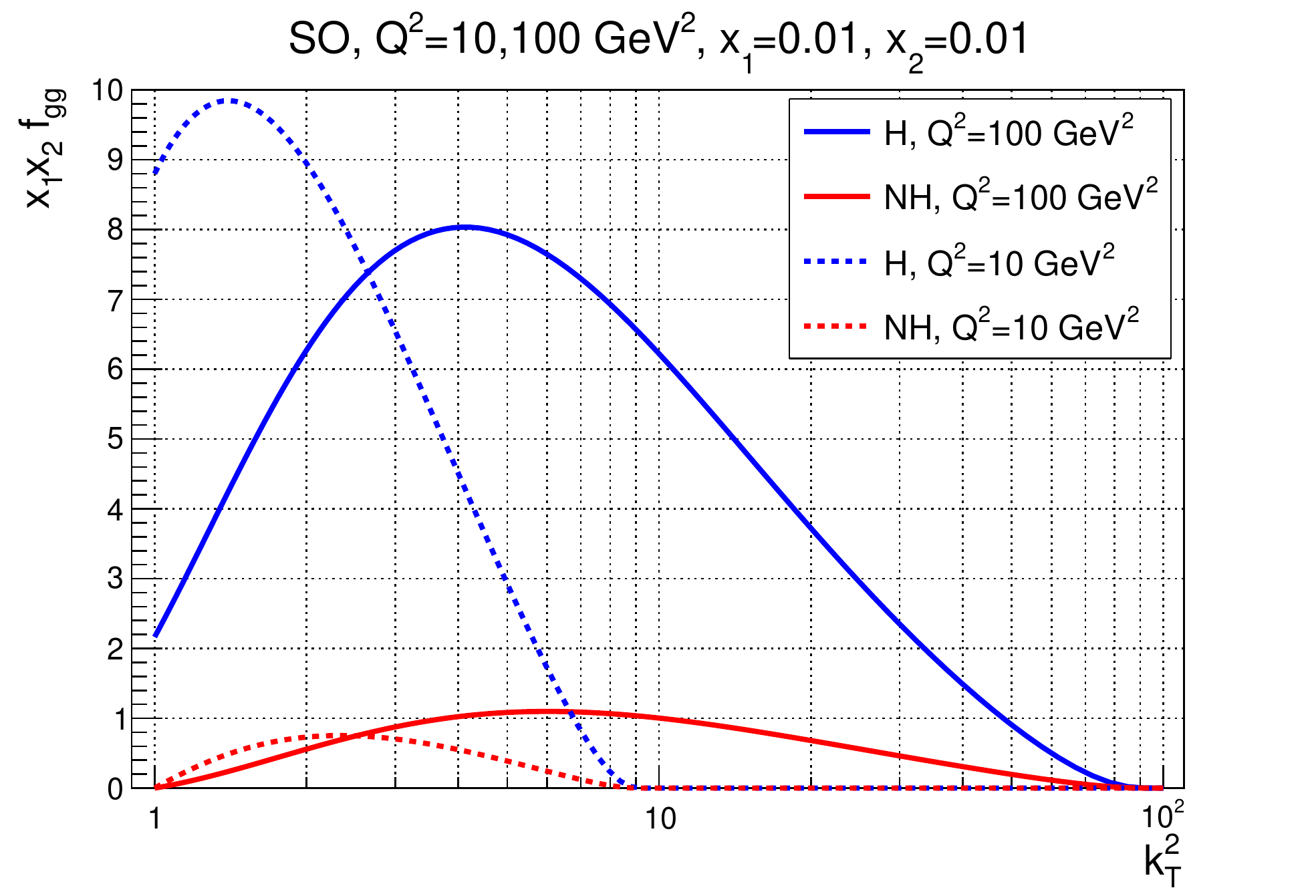}
\vspace*{1cm}
\includegraphics[width=0.5\textwidth]{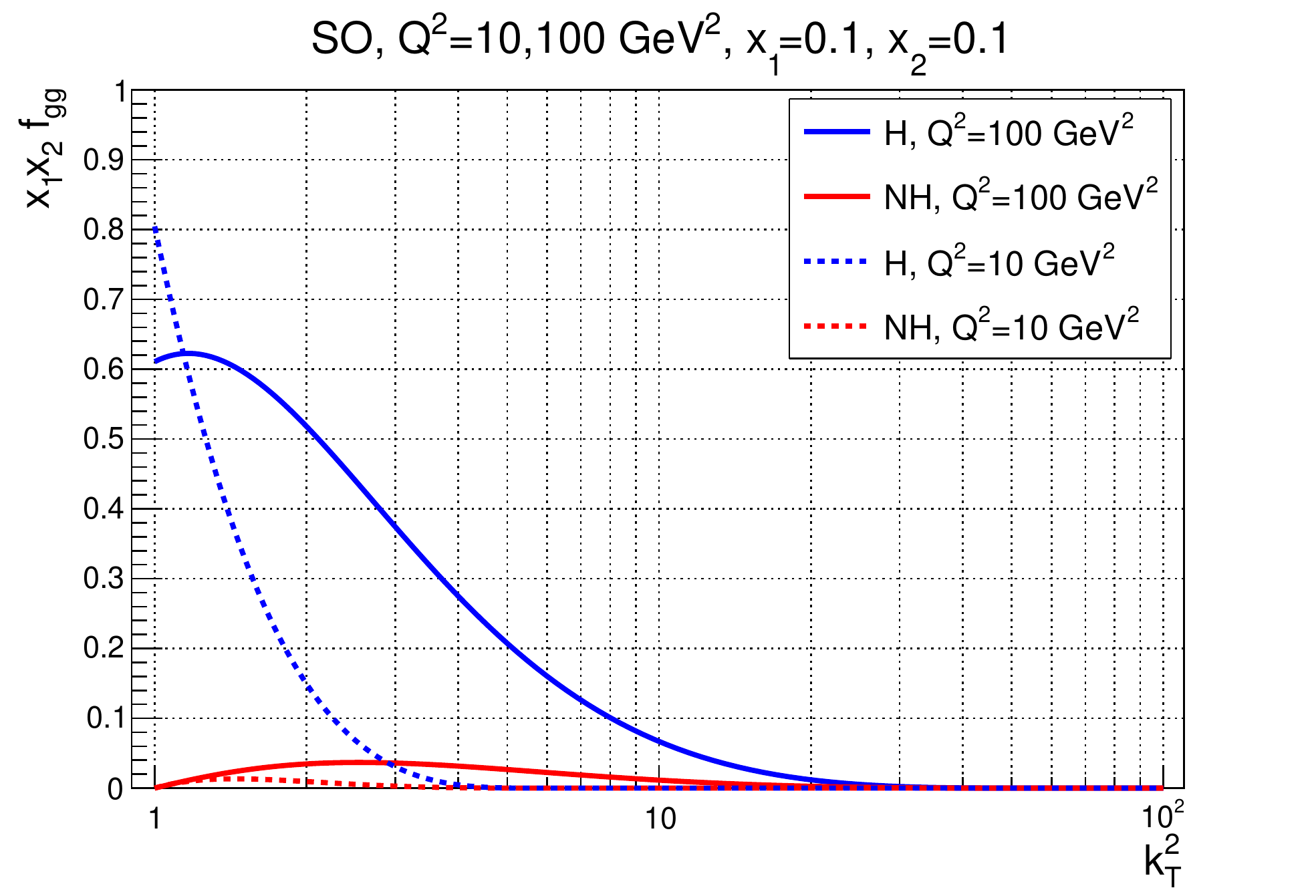}
\includegraphics[width=0.5\textwidth]{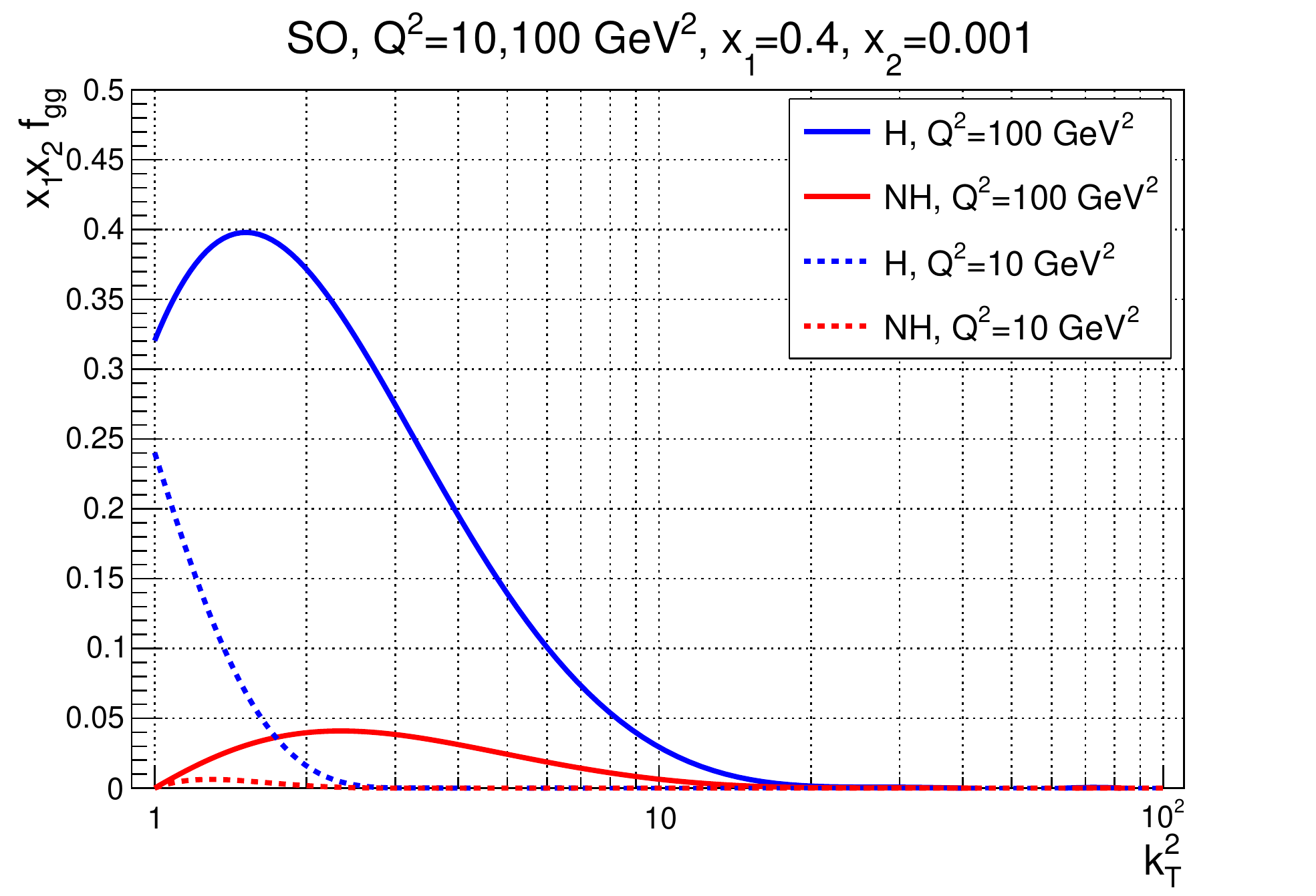}
\vskip -5mm
\caption{The homogeneous \eqref{eq:3.1}  (blue curves) 
and non-homogenous  \eqref{eq:4.3} (red curves)
components of the distribution $x_1x_2 f_{gg}(x_1,x_2,\kperp,\kperp,Q,Q)$ 
in the strong ordering case as a function of the  transverse momentum $k^2_\perp$ for  $Q^2=10\,{\rm GeV}^2$ (dashed curves) and $Q^2=100\,{\rm GeV}^2$ (solid curves).  The  values of $(x_1,x_2)$ are indicated and the ${\rm GBLS}^3$ input is used.}
\label{fig:figfactas10_100_smallx}
\end{figure}

\begin{figure}[t]
\centering\includegraphics[width=0.99\textwidth]{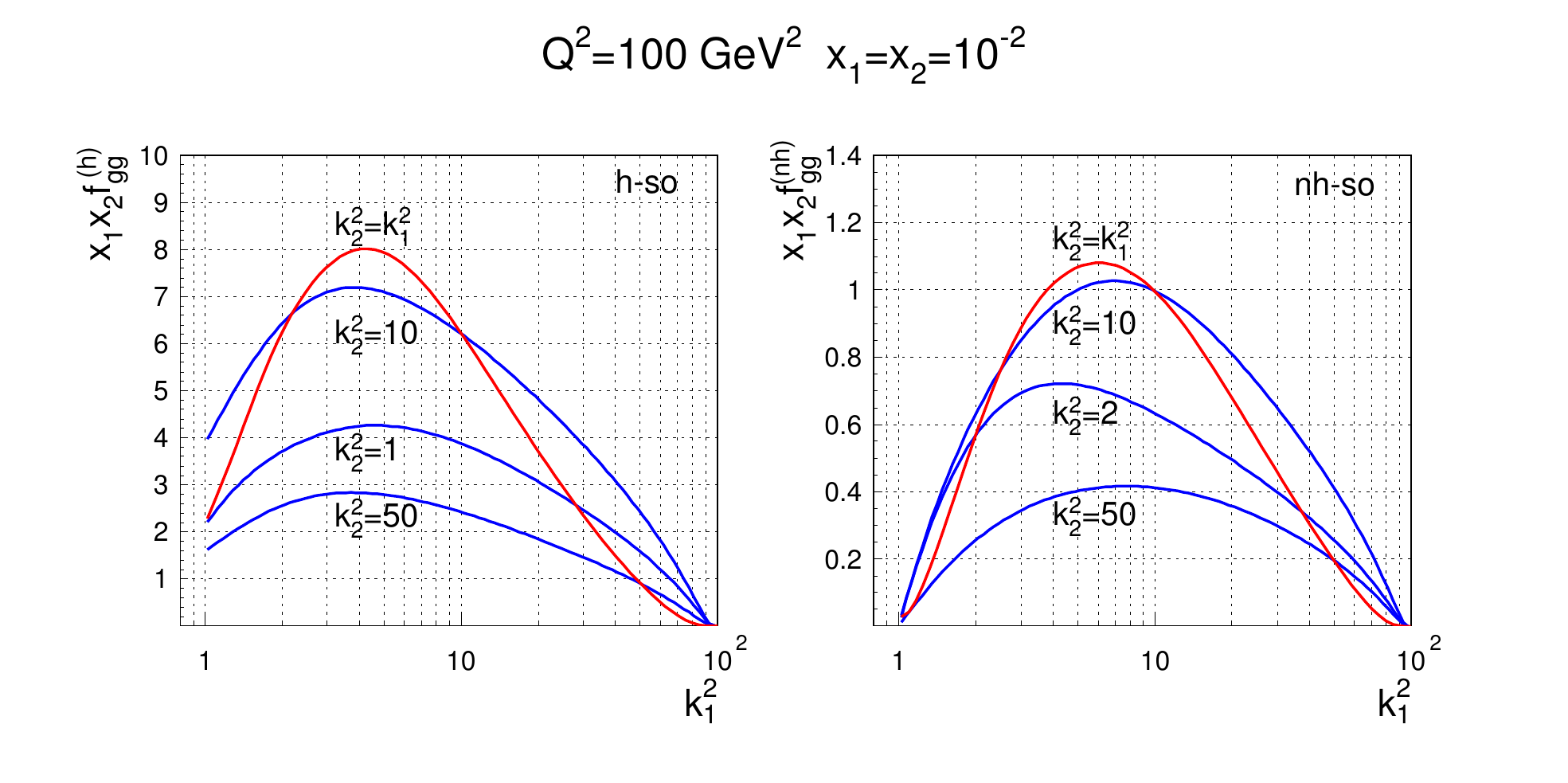}
\caption{The homogeneous \eqref{eq:3.1} (h-so)    and nonhomogeneous  \eqref{eq:4.3}  (nh-so)  distributions in the strong
ordering case  as a function of $\kperpone^2$ for the indicated fixed values of $\kperptwo^2$ (in ${\rm GeV}^2$),  and 
$x_1=x_2=10^{-2}$ and  $Q^2=100\,{\rm GeV}^2$.  The red lines correspond to the condition $\kperpone^2=\kperptwo^2$.}
\label{fig:fig2}
\end{figure}

\begin{figure}[t]
\begin{center}
\includegraphics[width=0.95\textwidth]{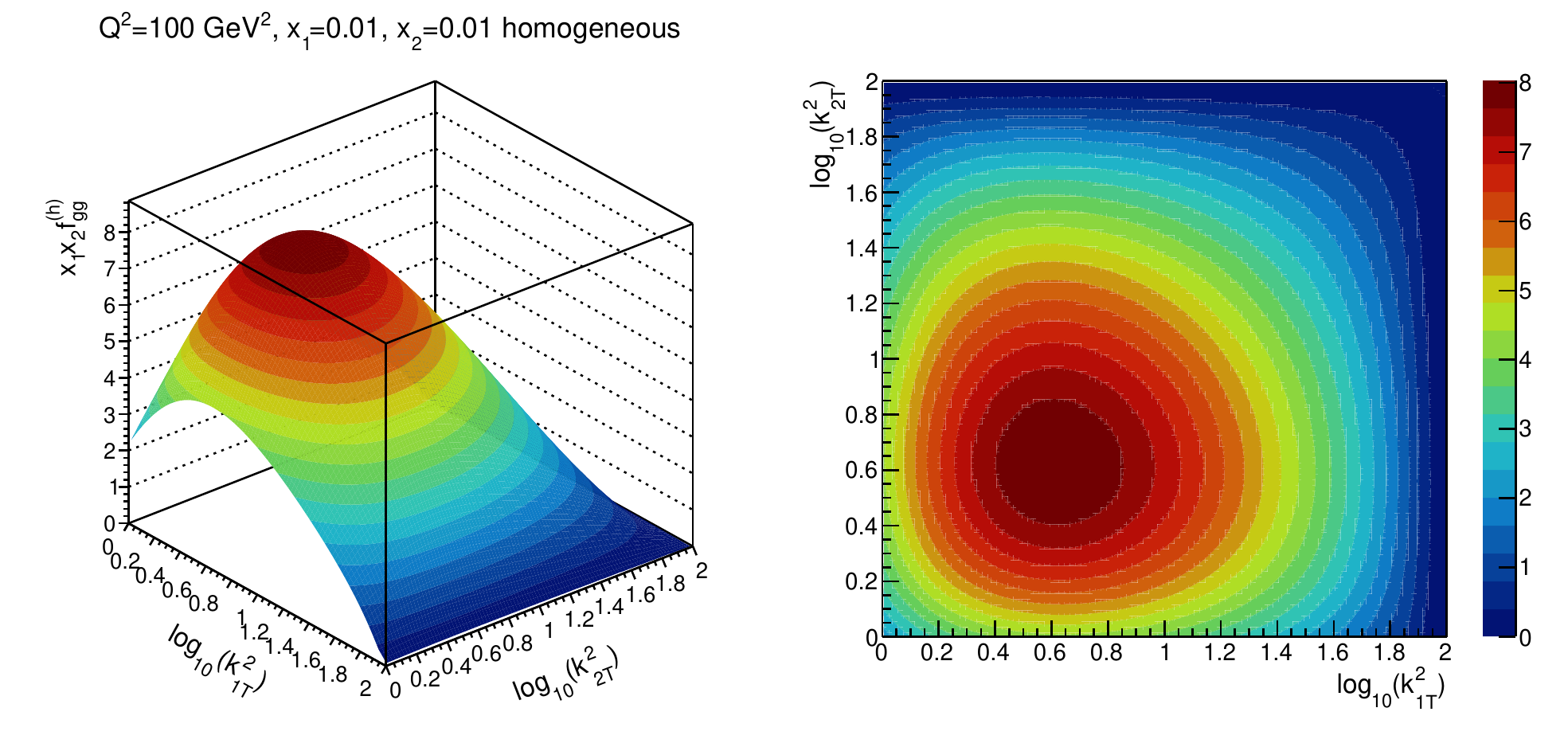}
\includegraphics[width=0.95\textwidth]{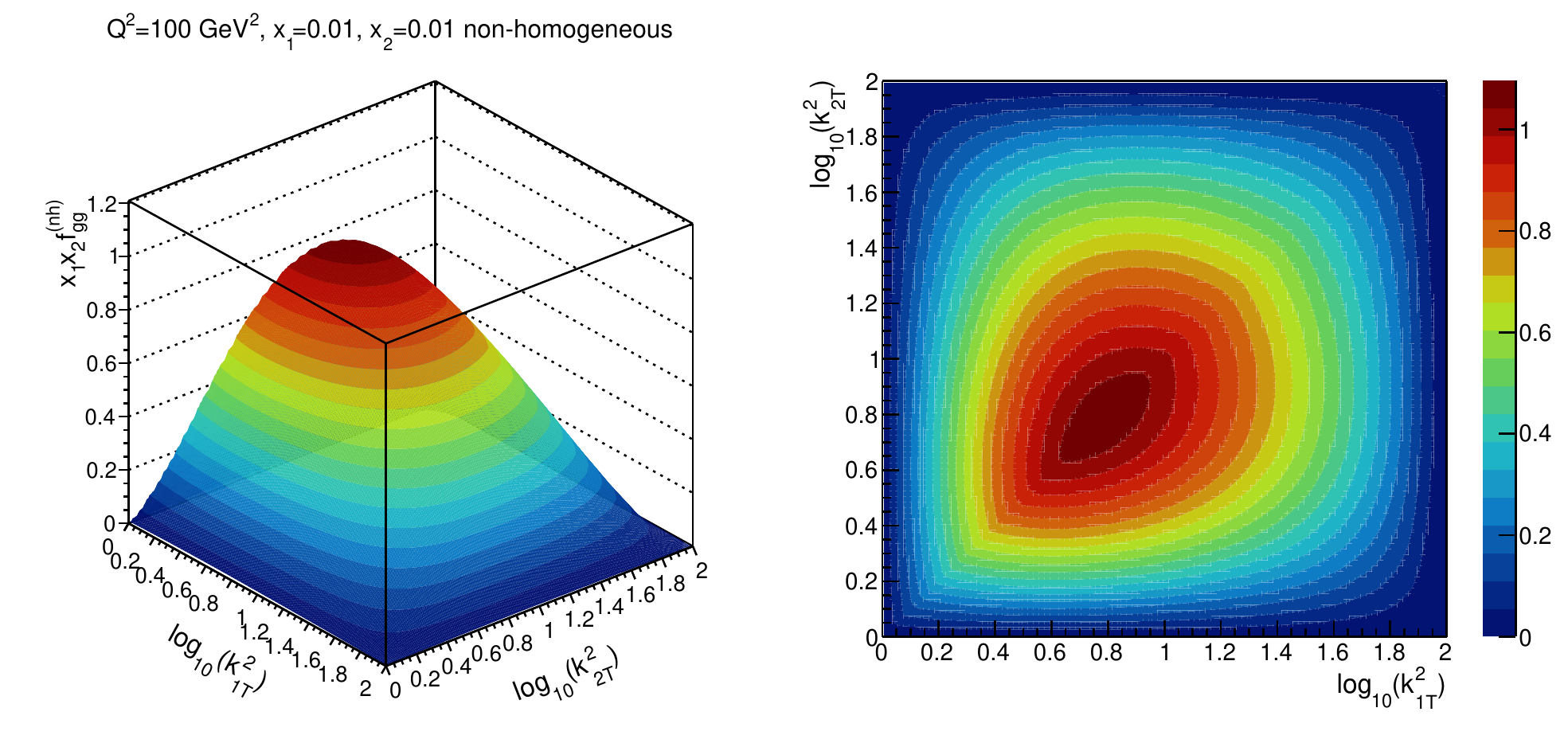}
\end{center}
\caption{The homogeneous \eqref{eq:3.1} (upper plots)  and non-homogeneous   \eqref{eq:4.3} (lower plots) distributions in the strong
ordering  case  as a function of  $(\kperpone^2,\kperptwo^2)$ for the indicated values of $(x_1,x_2)$ and $Q^2$. Contour projections  (with 20 equidistant contours) on the transverse momentum plane  are shown on the right.
}
\label{fig:fig3}
\end{figure}


In Ref.~\cite{Golec-Biernat:2016vbt} we also considered the contribution to the unintegrated double parton distributions originating from a splitting of a single
parton into a pair of partons, $f^{(nh)}_{f_1f_2}$. This contribution is generated by the special solution of the non-homogeneous evolution equations for the integrated double parton distributions, $D_{f_1f_2}^{(nh)}$. 

In the pure gluon case, the integrated double gluon distribution due to this contribution is given by
\begin{align}
\label{eq:4.1}
 D_{gg}^{(nh)}(x_1,x_2,Q_1,Q_2)= \int^{Q_{\min}^2}_{Q_0^2}\frac{dQ^2_s}{Q^2_s}\,{\cal D}^{(sp)}_{gg}(x_1,x_2,Q_1,Q_2,Q_s)\,,
\end{align}
where $Q_{\min}^2={\rm min}\{Q_1^2,Q_2^2\}$.
The distribution ${\cal D}^{(sp)}_{gg}$ under the integral, discussed at length in \cite{Golec-Biernat:2016vbt}, 
corresponds to the perturbative gluon splitting $g\to gg$ at the scale $Q_s$,
\begin{align}
&{\cal D}^{(sp)}_{gg}(x_1,x_2,Q_1,Q_2,Q_s) = 
\nonumber 
\\
&=\int_{x_1}^{1-x_2}\frac{dz_1}{z_1}
\int_{x_2}^{1-z_1}\frac{dz_2}{z_2}\,
E_{gg}\Big(\frac{x_1}{z_1},Q_1,Q_s\Big)
E_{gg}\Big(\frac{x_2}{z_2},Q_2,Q_s\Big)D_{gg}^{(sp)}(z_1,z_2,Q_s) \; ,
\label{eq:4.1a}
\end{align}
which 
evolves with the scales $Q_1$ and $Q_2$  from  the initial condition given by Eq.~(\ref{eq:10}).
The distribution ${\cal D}^{(sp)}_{gg}$ is used to define the unintegrated double gluon distribution from the gluon splitting,
\begin{align}
\nonumber
\label{eq:4.2}
f^{(nh)}_{gg}(x_1,x_2,&\kperpone,\kperptwo,Q_1,Q_2) =
T_{g}(Q_1,\kperpone)\,T_{g}(Q_2,\kperptwo)\,\times
\\\nonumber
&\times\,
\int_{\frac{x_1}{1-x_2}}^{1-\Delta_1}\frac{dz_1}{z_1}
\int_{\frac{x_2}{1-x_1/z_1}}^{1-\Delta_2}\frac{dz_2}{z_2}\,
P_{gg}(z_1,\kperpone)
P_{gg}(z_2,\kperptwo)\,\times
\\
&\times
\int_{Q_0^2}^{Q_{\min}^2}\frac{dQ_s^2}{Q_s^2}\,
\theta(k_{1\perp}^2-Q_s^2)\,\theta(k_{2\perp}^2-Q_s^2)\,
{\cal D}^{(sp)}_{gg}\Big(\frac{x_1}{z_1},\frac{x_2}{z_2},\kperpone,\kperptwo,Q_s\Big).
\end{align}


In the strong ordering case (\ref{eq:3.4}), the transverse momenta obey the condition $k_{i\perp}<Q_i$. Thus, for given $k_{1\perp}$ and $k_{2\perp}$, the upper limit in the integral over $Q_s^2$ in (\ref{eq:4.2}) is given by $Q_{\rm min}^2={\rm min}\{k^2_{1\perp},k^2_{2\perp}\}$. In such a case, the step functions in (\ref{eq:4.2}) are automatically satisfied,  which  allows to rewrite the above formula with the help of   the distribution (\ref{eq:4.1}),
\begin{align}
\nonumber
\label{eq:4.3}
f^{(nh)}_{gg}(&x_1,x_2,\kperpone,\kperptwo,Q_1,Q_2) =
T_{g}(Q_1,\kperpone)\,T_{g}(Q_2,\kperptwo)\,\times
\\
&\times
\int_{\frac{x_1}{1-x_2}}^{1-\Delta_1}\frac{dz_1}{z_1}
\int_{\frac{x_2}{1-x_1/z_1}}^{1-\Delta_2}\frac{dz_2}{z_2}\,
P_{gg}(z_1,\kperpone)
P_{gg}(z_2,\kperptwo)\, D_{gg}^{(nh)}\Big(\frac{x_1}{z_1},\frac{x_2}{z_2},k_1,k_2\Big)\,.
\end{align}
Therefore, $f^{(nh)}_{gg}$ is given by the formula analogous  to   Eq.~(\ref{eq:3.1}) in which the homogeneous  double gluon distribution, $D_{gg}^{(h)}$, is replaced by the non-homogeneous one, $D_{gg}^{(nh)}$.

In the angular ordering case (\ref{eq:3.5}), the transverse momenta are not bounded and
 we have to use formula (\ref{eq:4.2}) together with \eqref{eq:4.1a} in the numerical analysis. This would mean that  when $\kperpone,\kperptwo \ge Q_1,Q_2$,  the splitting could only occur at the scale provided by $Q^2_{\rm min}={\rm min}\{Q^2_1,Q^2_2\}$, and then the region up to $\kperpone,\kperptwo$ would evolve as two separate parton ladders. However, the parton splitting which generates two partonic ladders can occur at any point of the evolution, and therefore such a situation is unphysical. Since the transverse momentum dependence in the framework considered is anyway generated in the last splitting by convoluting with the splitting functions, therefore we shall argue that also in the case of the angular ordering  we should also use  
 $Q^2_{\rm min}={\rm min}\{k^2_{1\perp},k^2_{2\perp}\}$  in Eq.~(\ref{eq:4.2}).

In  Fig.~\ref{fig:figfactas10_100_smallx}, we show the relative size of the homogeneous \eqref{eq:3.1} and non-homogenous \eqref{eq:4.3} contributions 
by showing them as a function of the transverse momentum $k_\perp^2\equiv\kperpone=\kperptwo$ for $Q^2=10,100 \; {\rm GeV}^2$ 
and the  indicated values of $(x_1,x_2)$. 
We see that the non-homogeneous (splitting) contribution is generally much smaller than the homogenous one. As expected, the significance of the splitting contribution rises with increasing $Q^2$.

\begin{figure}[t]
\includegraphics[width=0.5\textwidth]{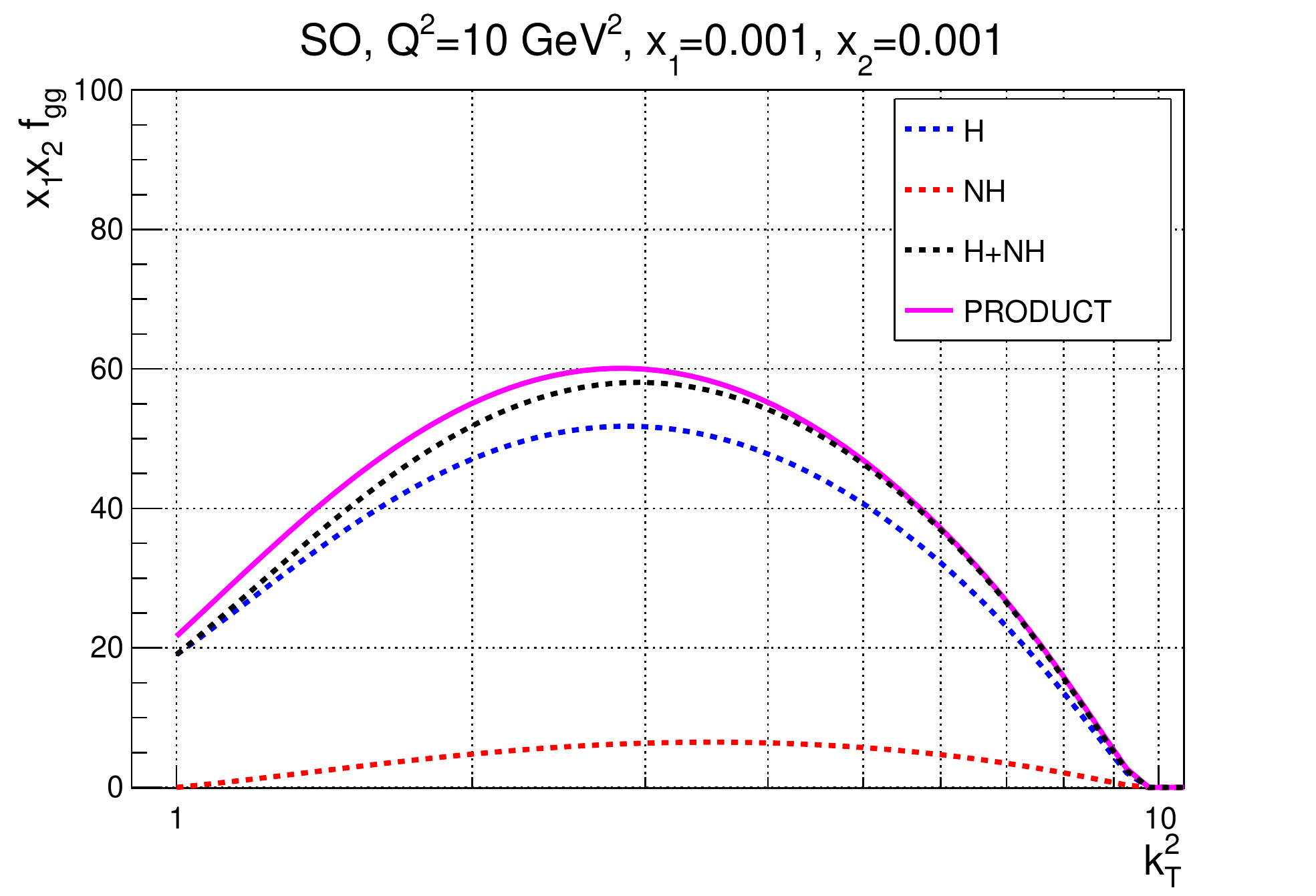}
\includegraphics[width=0.5\textwidth]{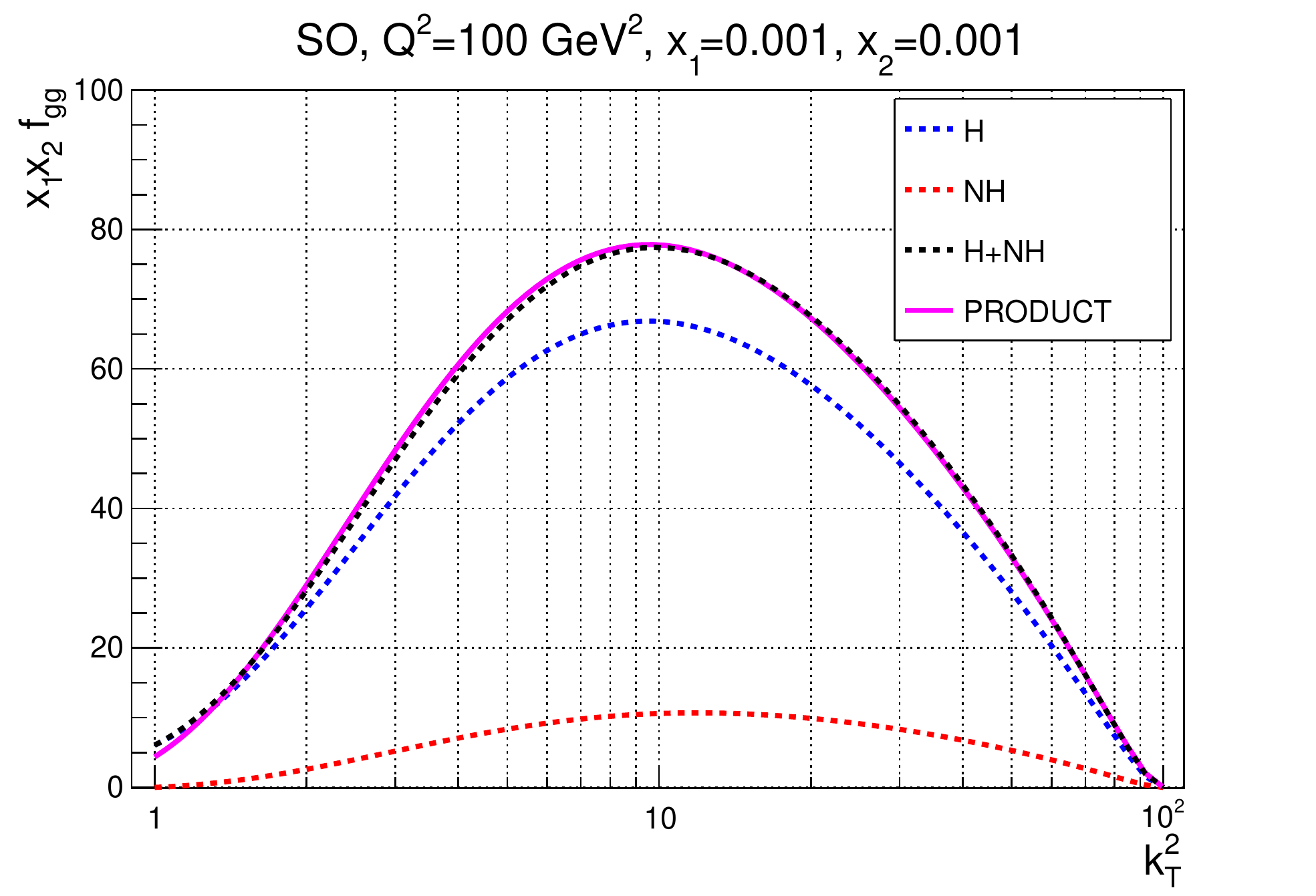}
\vspace*{1cm}
\includegraphics[width=0.5\textwidth]{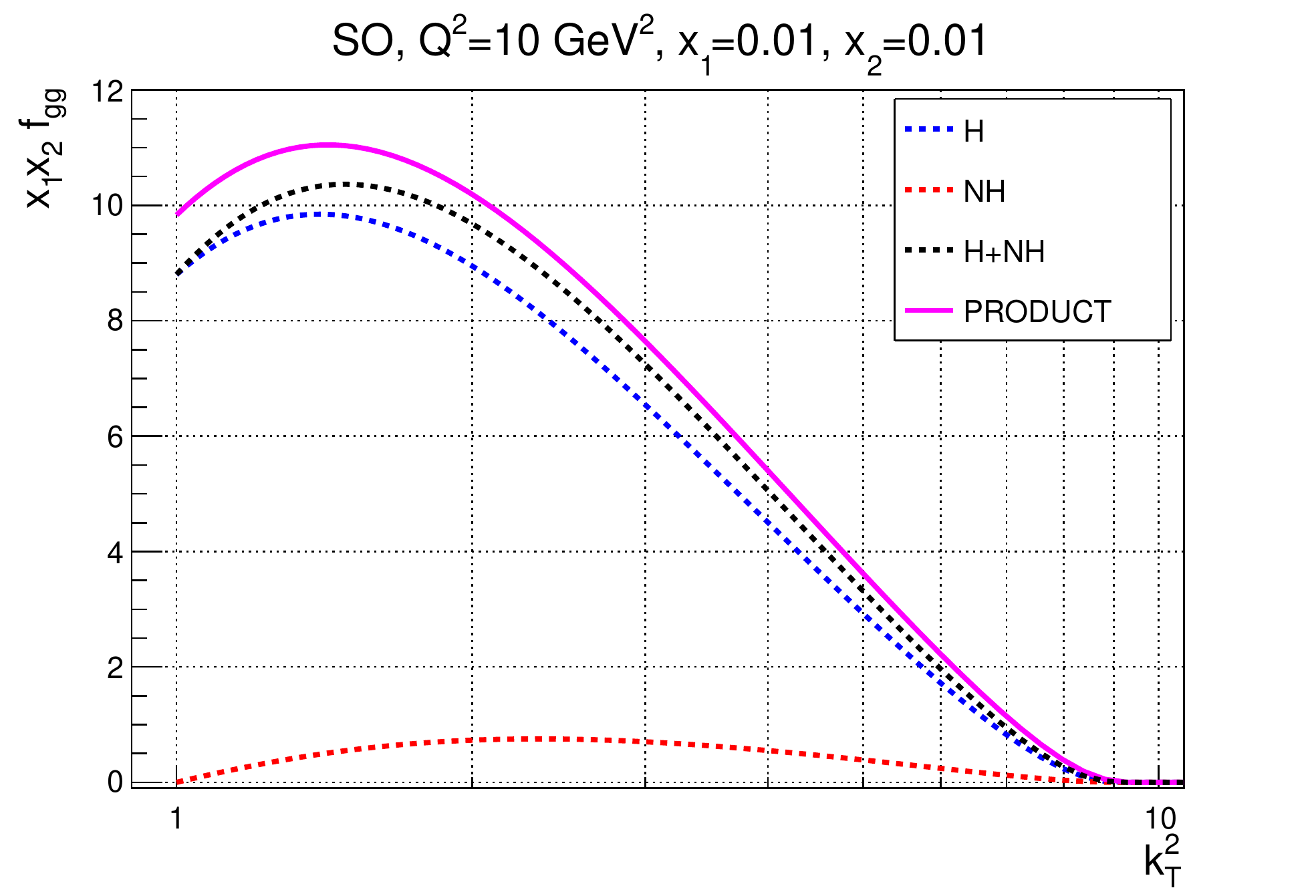}
\includegraphics[width=0.5\textwidth]{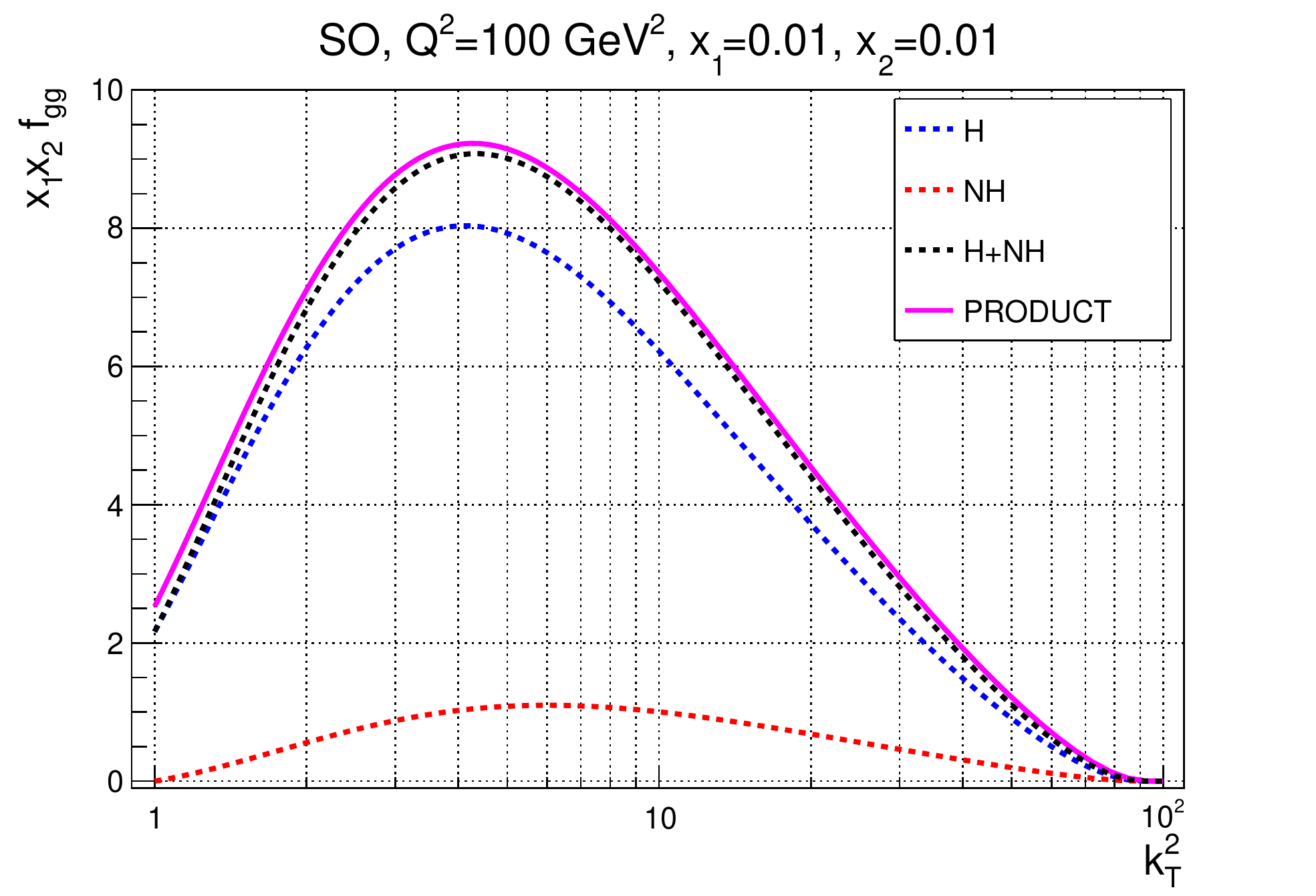}
\vskip -10mm
\caption{The distribution $x_1x_2 f_{gg}(x_1,x_2,\kperp,\kperp,Q,Q)$  (dashed black line)
against the product of the single distributions $x_1 f_g(x_1,\kperp,Q) x_2 f_g(x_2,\kperp,Q)$ (dashed magenta), plotted as a function of $k_\perp^2$ for the indicated small values of $(x_1,x_2)$ and $Q^2=10,100\,{\rm GeV}^2$, starting from the ${\rm GBLS}^3$ input. 
The breakdown  of  $f_{gg}$  into the homogeneous  \eqref{eq:3.1} (dashed blue) 
 and non-homogenous \eqref{eq:4.3}  (dashed red) components shows that the non-homogeneous component is essential for  factorization.
}
\label{fig:figfactas10_100_smallx1}
\end{figure}

We also present  the unintegrated double gluon distribution for the case with different  transverse momenta,  $\kperpone \neq \kperptwo$.  Slices of this distribution for fixed values of $\kperptwo$  are shown as a function of $\kperpone$ in Fig.~\ref{fig:fig2}.  The full two-dimensional distributions in $(\kperpone,\kperptwo)$ are shown in Fig.~\ref{fig:fig3} together with the contour plots.  From the  contour plots, it is evident that the distribution originating from the non-homogeneous contribution is more perturbative, i.e. the maximum of the distribution is  shifted towards higher transverse momenta and the region of very low transverse momenta is depleted.

\section{Factorization of  double gluon distributions}
\label{sec:5}

In this subsection we shall investigate to what extent the  factorization of the unintegrated double gluon distribution into a product of two single unintegrated gluon distributions, 
\be
\label{eq:5.1}
f_g(x,\kperp, Q) = 
T_g(Q,\kperp)\int_x^{1-\Delta} \frac{dz}{z}\,P_{gg}(z,\kperp)\, 
D_{g}\Big(\frac{x}{z},\kperp\Big) \;,
\ee
 is satisfied in different kinematic regimes. The  integrated single gluon distribution, $D_g$, on the rhs of the above equation is evolved using the  DGLAP evolution equation reduced to the gluon sector with the initial condition given by Eq.~(\ref{eq:2.2}).

\begin{figure}[t]
\includegraphics[width=0.5\textwidth]{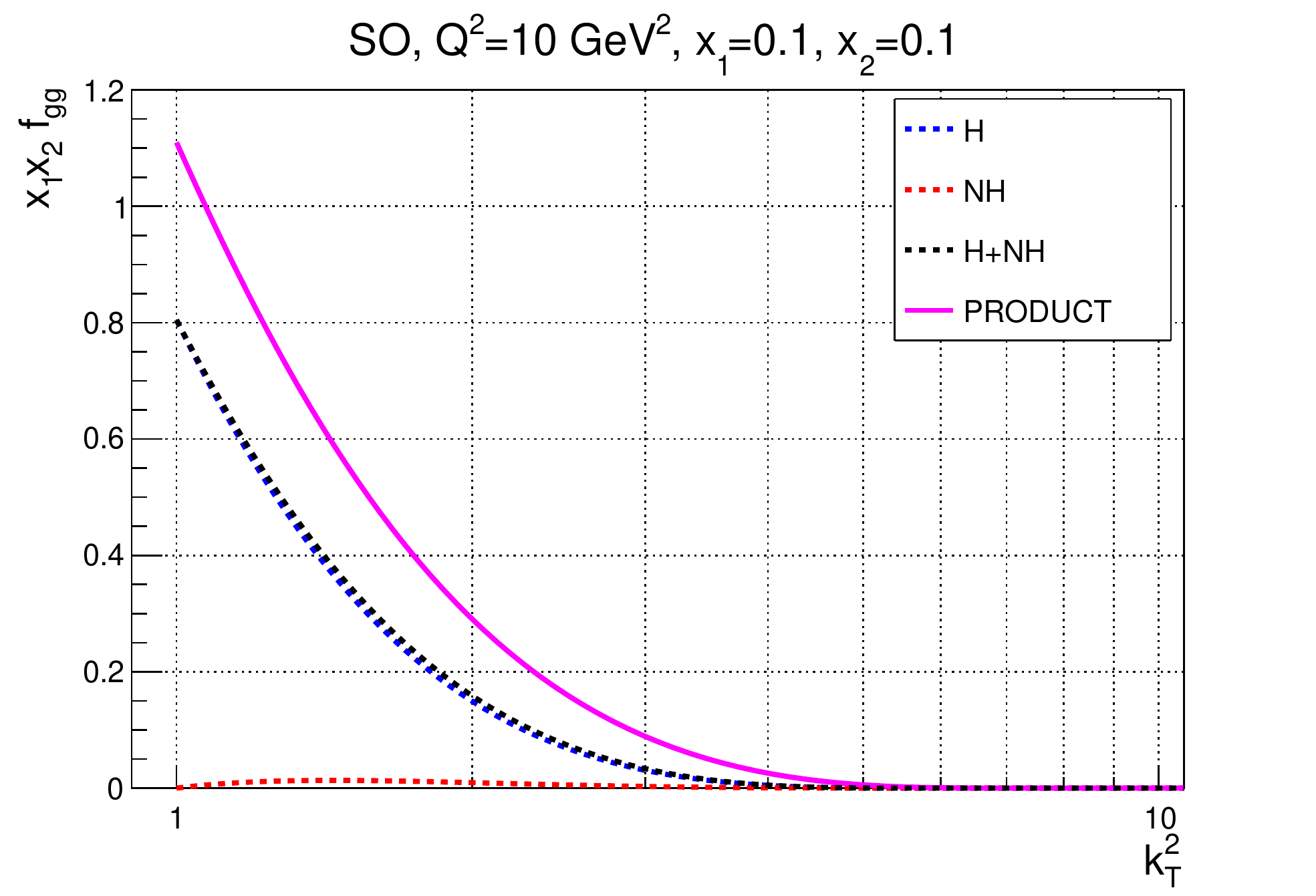}
\includegraphics[width=0.5\textwidth]{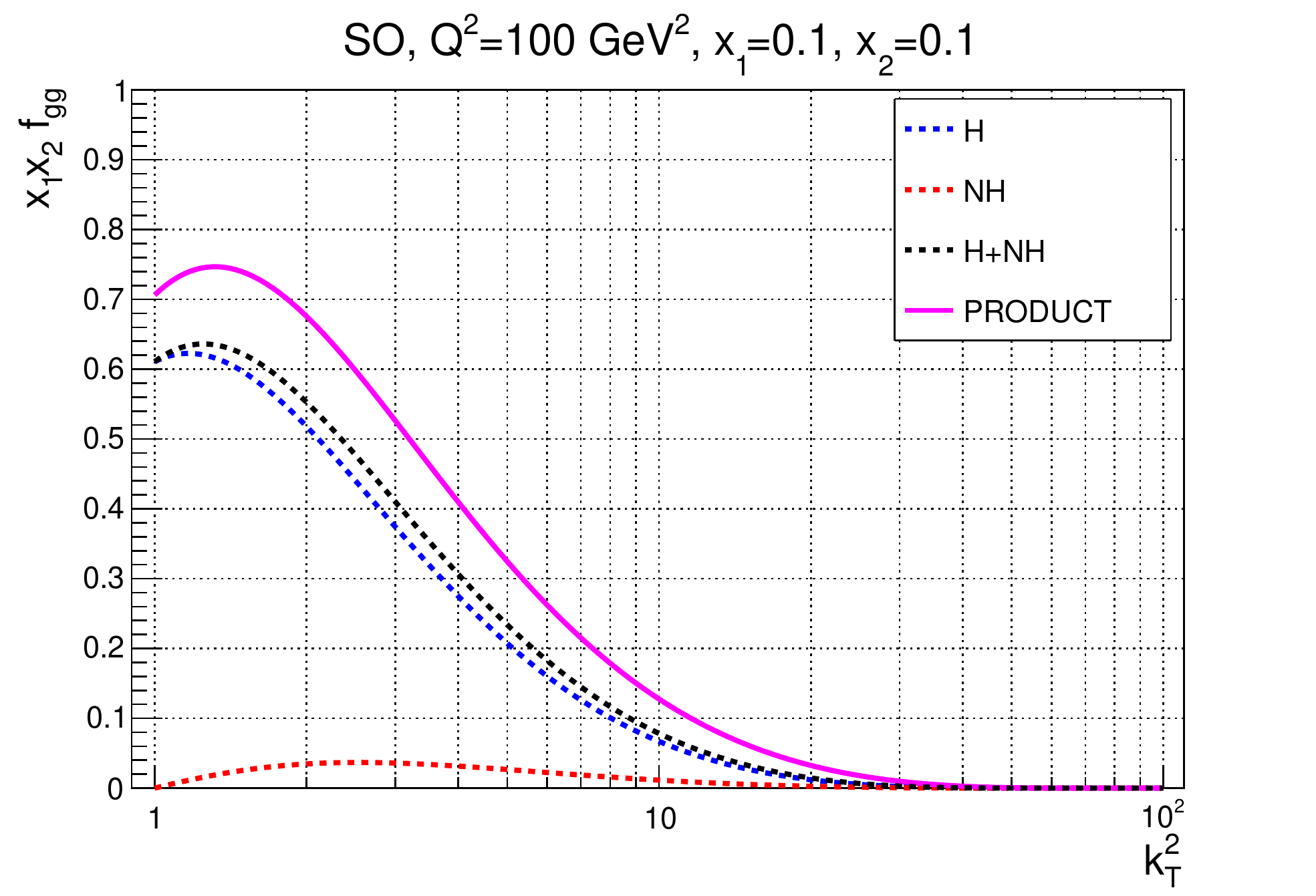}
\vspace*{1cm}
\includegraphics[width=0.5\textwidth]{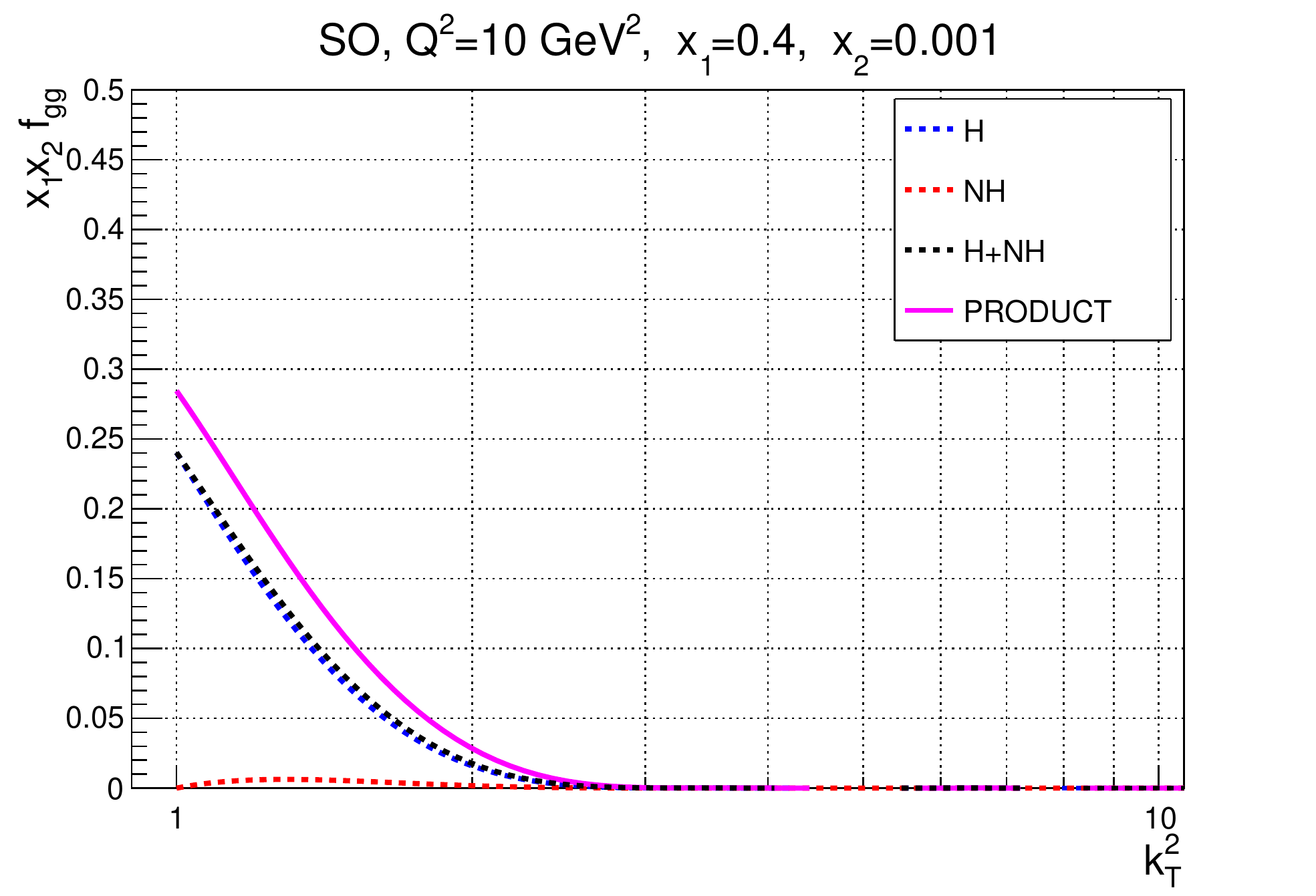}
\includegraphics[width=0.5\textwidth]{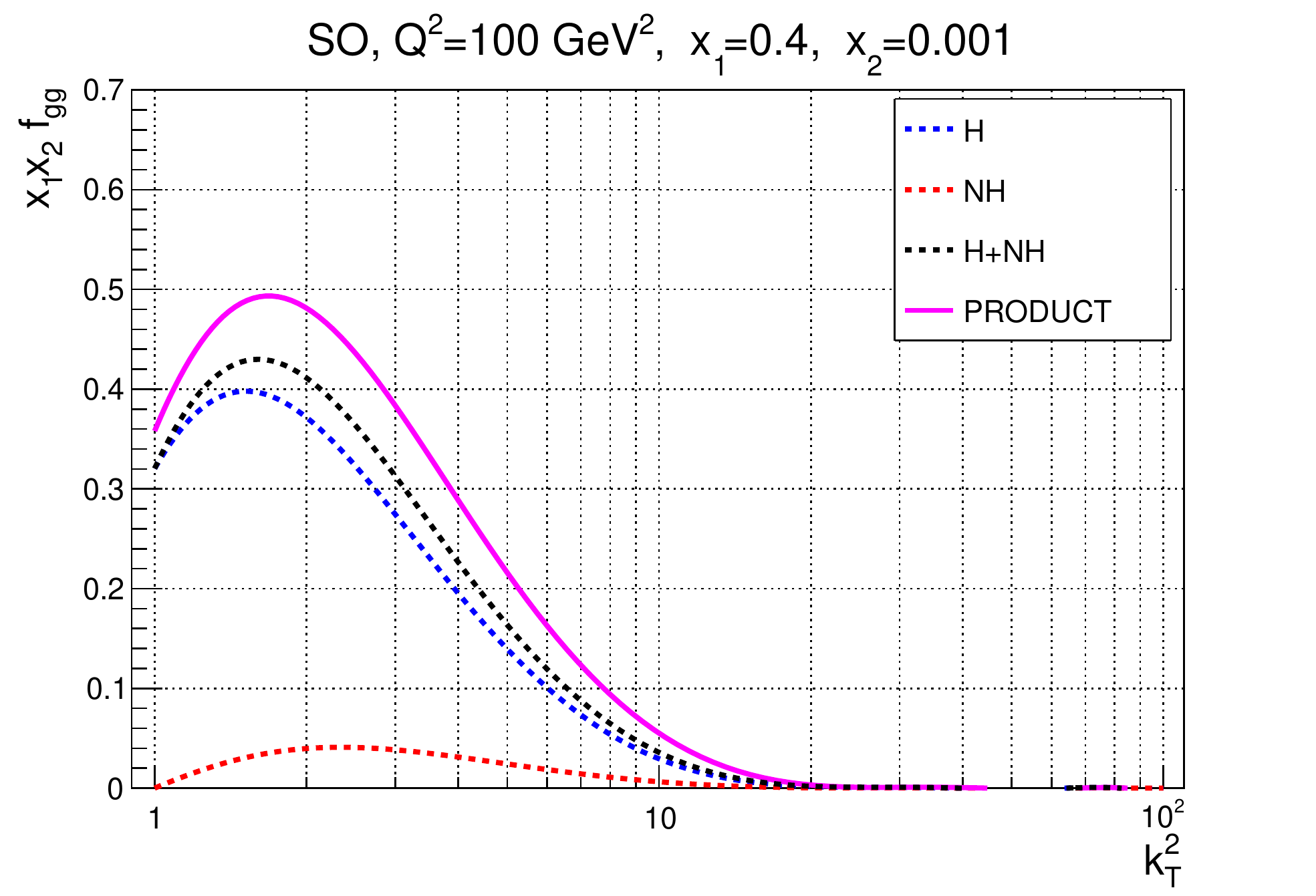}
\vskip -10mm
\caption{As in Figure\,\ref{fig:figfactas10_100_smallx1} but for large values of $x_1$ or $x_2$ when factorization does not hold.}
\label{fig:figfactas10_100_largex}
\end{figure}

\begin{figure}[t]
\begin{center}
\includegraphics[width=0.9\textwidth]{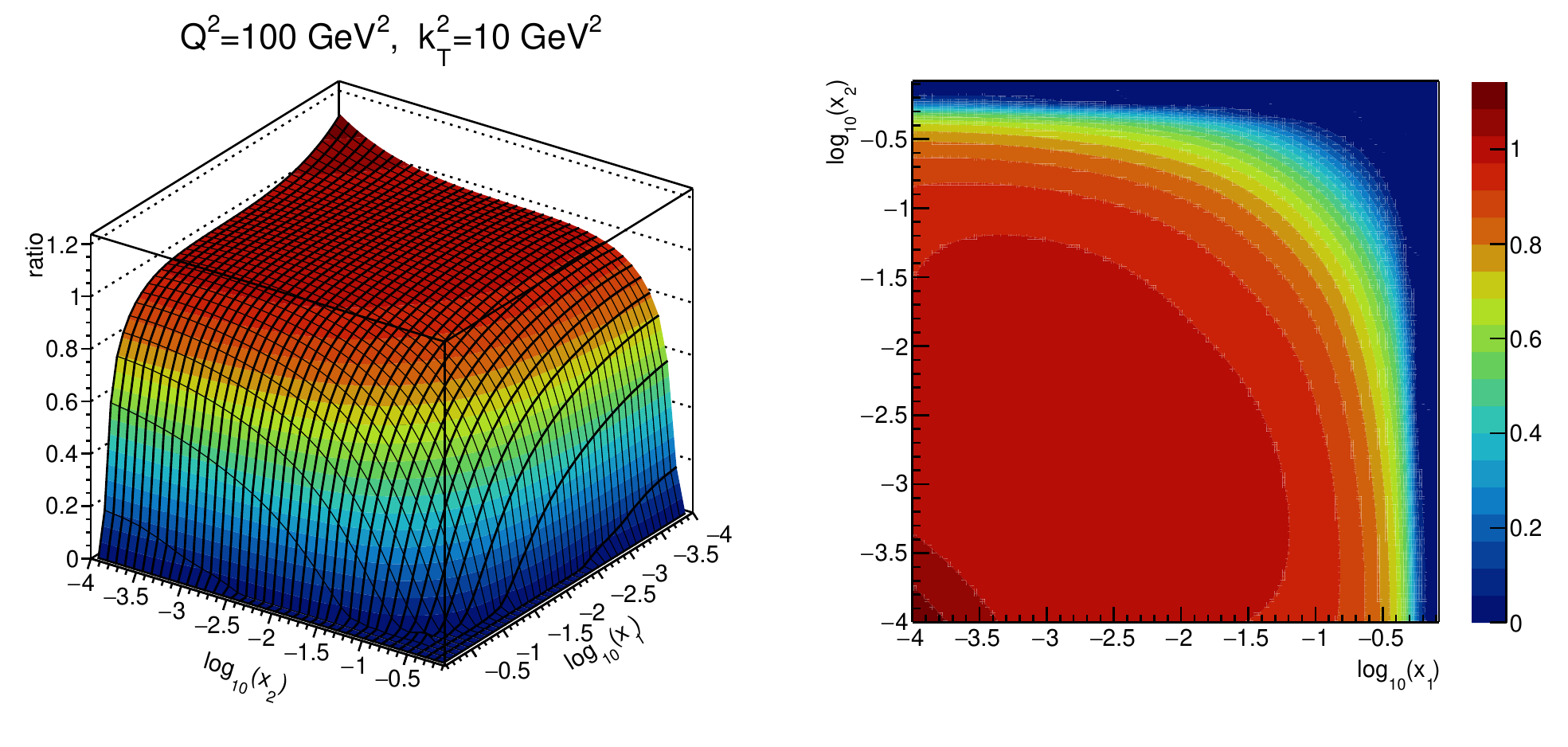}
\end{center}
\caption{The ratio  \eqref{eq:ratiofact}  as a function of the longitudinal momentum fractions $(x_1,x_2)$ for fixed values of transverse momenta,
 $\kperpone^2=\kperptwo^2=10 \; {\rm GeV}^2$, and $Q^2=100\,{\rm GeV}^2$, in the strong ordering case.
}
\label{fig:fig4a}
\end{figure}

In Figs.~\ref{fig:figfactas10_100_smallx1} and \ref{fig:figfactas10_100_largex}, we   compare the sum of the homogeneous and non-homogenous contributions with the  product of two single unintegrated gluon distributions, each given by Eq.~\eqref{eq:5.1}.   The results are shown as a function of $\kperp\equiv \kperpone=\kperptwo$ for fixed $Q^2$ and $(x_1,x_2)$. We observe in Fig.~\ref{fig:figfactas10_100_smallx1} that for small values of $(x_1,x_2)$, 
the sum of the homogeneous and non-homogeneous contribution   is   very close  to the product of two single gluon distributions. 
This shows that the factorization of the solution works very well in the small $x$ regime and for  large value of hard scale, $Q^2=100\,{\rm GeV}^2$. It has to be stressed that the factorization works for the sum of the two contributions, $f_{gg}=f^{(h)}_{gg}+f^{(nh)}_{gg}$,  but not for the homogeneous term only as  might be naively expected. We also observe   that the factorization is violated for large values of $x_1$ or $x_2$, see Fig.~\ref{fig:figfactas10_100_largex}.
This is also seen in Fig.~\ref{fig:fig4a}, where we plot the ratio 
\be
\label{eq:ratiofact}
{ r}=\frac{f_{gg}(x_1,x_2,\kperpone,\kperptwo,Q,Q)}{f_g(x_1,\kperpone,Q) f_g(x_2,\kperptwo,Q)}
\ee
as a function of $(x_1,x_2)$ for  fixed transverse momenta,  $\kperpone^2=\kperptwo^2=10\,{\rm GeV}^2$
and  $Q^2=100\,{\rm GeV}^2$,  in the strong ordering case. 

The factorization of the distribution $f_{gg}$ is an interesting  fact that deserves some explanation. One could naively expect that since the homogeneous term is  equivalent to the evolution of two disconnected ladder graphs, the factorization would occur with this term only. On the contrary, it is the  sum of the homogenous and non-homogenous contributions which is equal to the product of two single gluon distributions,
$f^{(h)}_{gg}+f^{(nh)}_{gg}\approx f_g\cdot f_g$.
To understand this feature, one needs to investigate the factorization  of the integrated double gluon distribution, $D_{gg}(x_1,x_2,Q_1,Q_2)$, since we expect that
the factorization at this level leads  to the factorization of the unintegrated double gluon distribution.

\begin{figure}[t]
\includegraphics[width=0.5\textwidth]{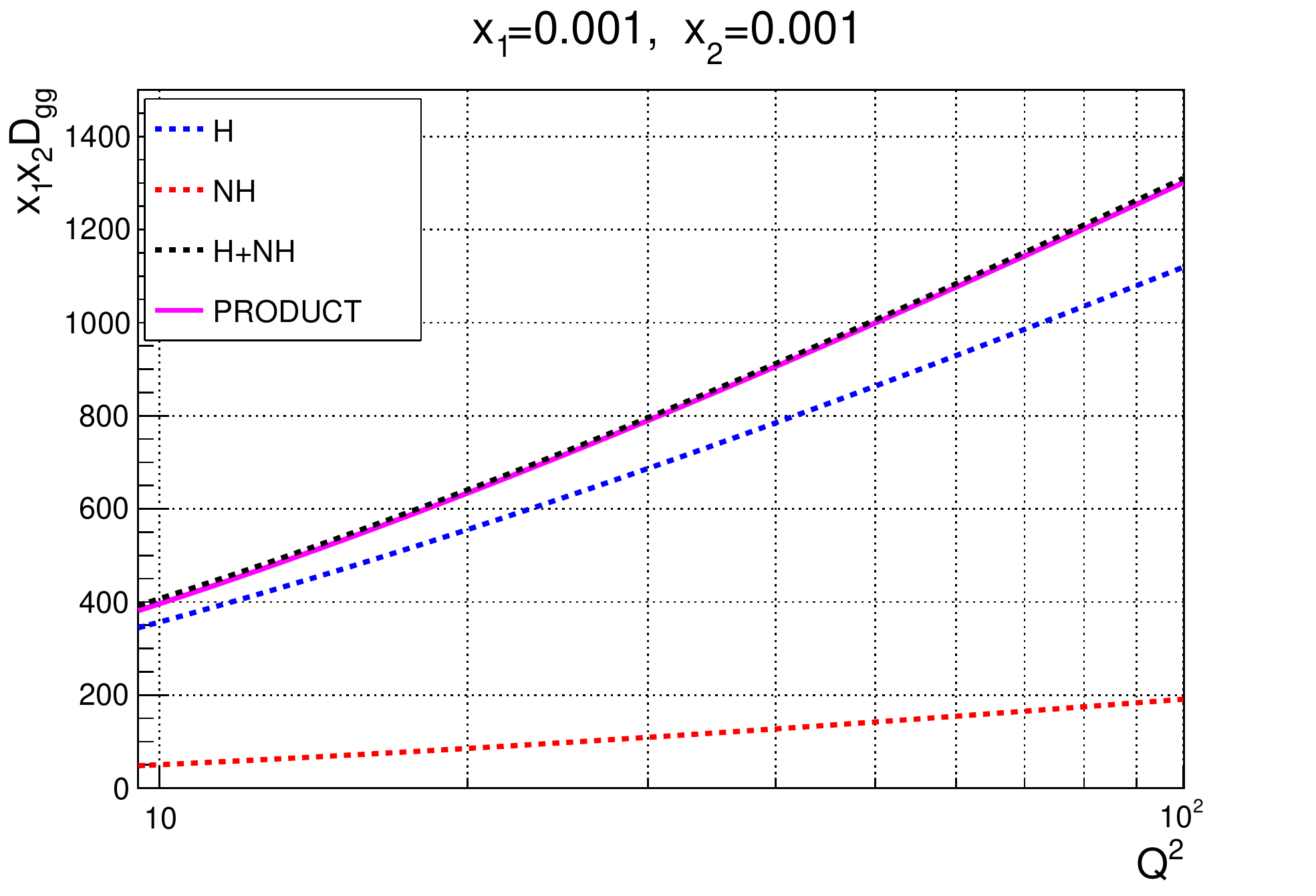}
\includegraphics[width=0.5\textwidth]{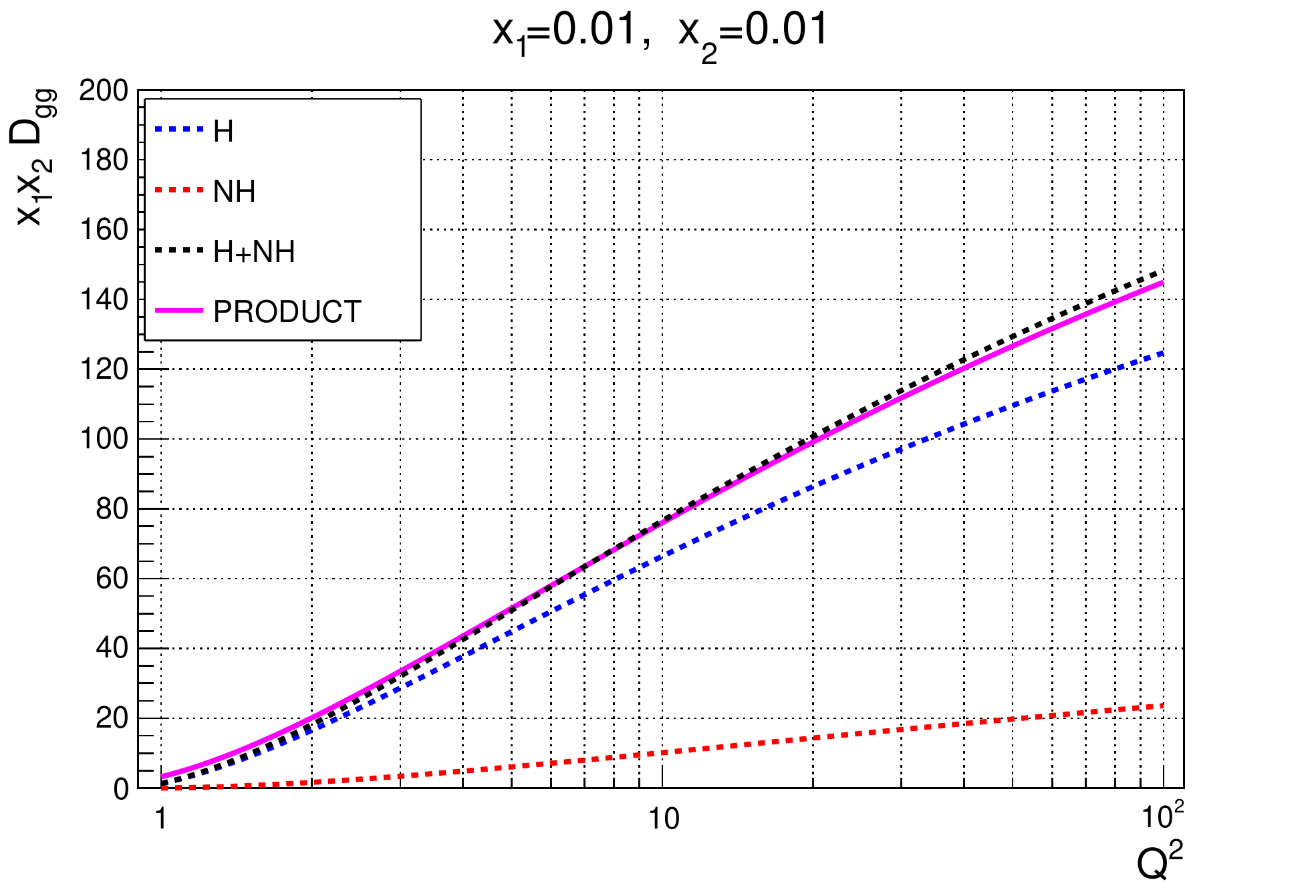}
\vspace*{0.5cm}
\includegraphics[width=0.5\textwidth]{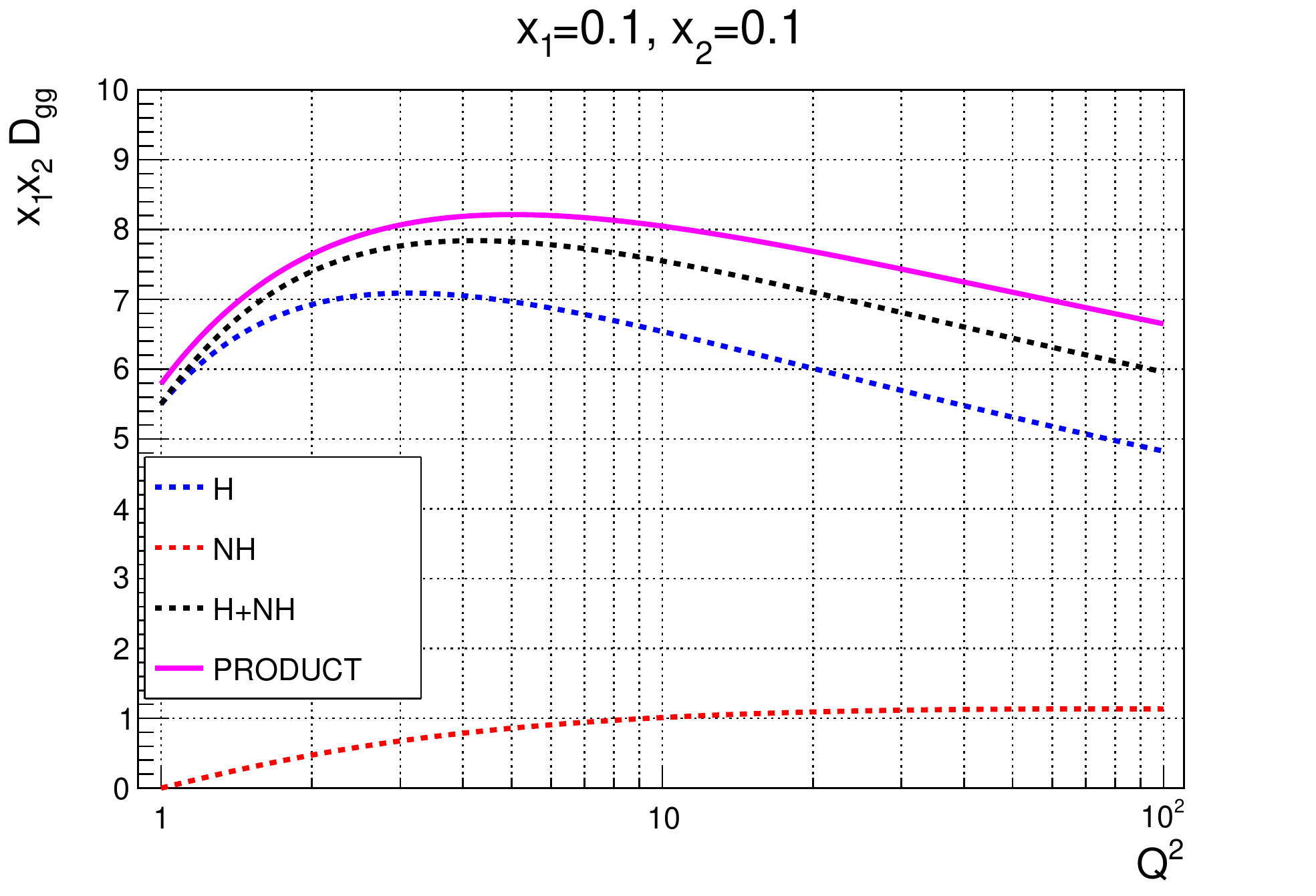}
\includegraphics[width=0.5\textwidth]{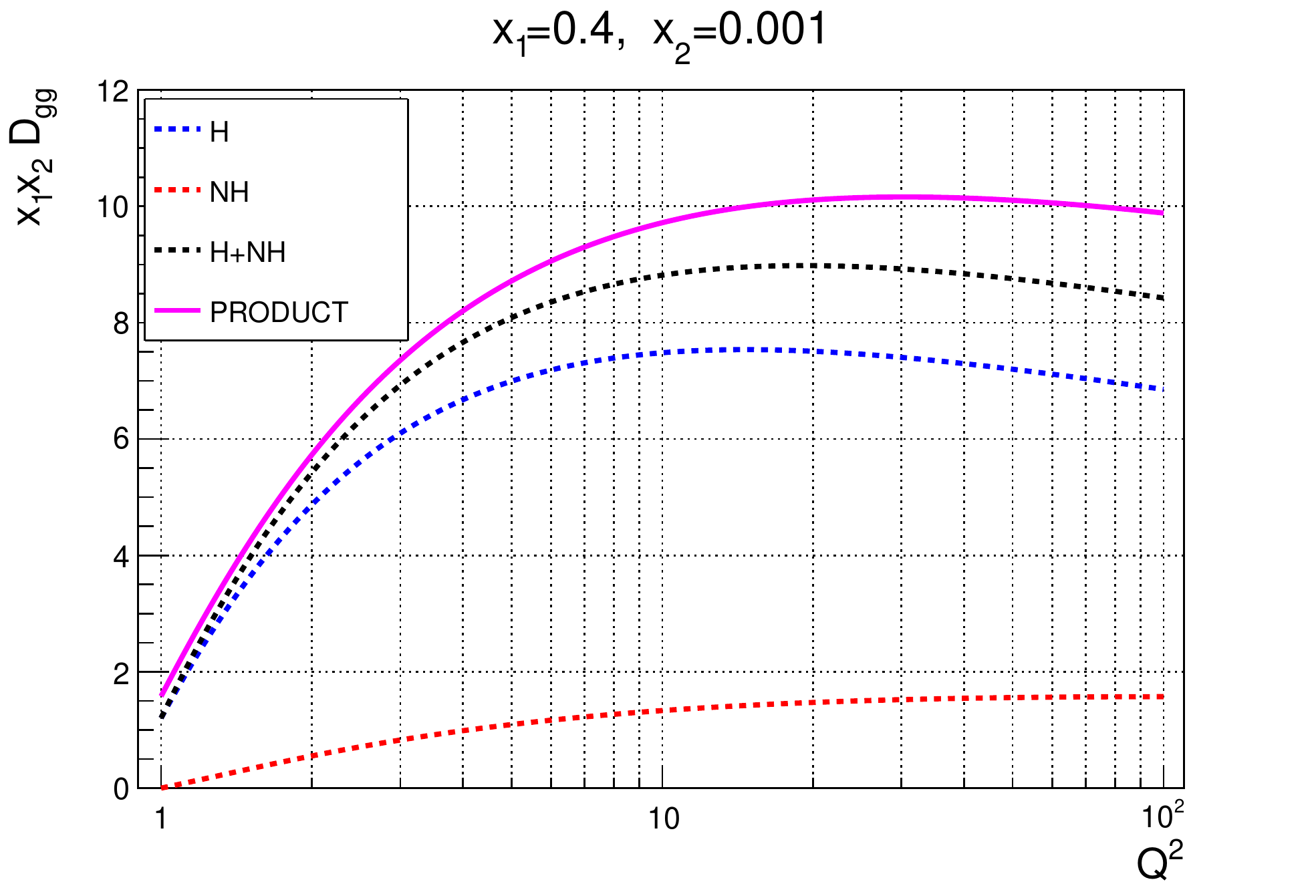}
\caption{The comparison of the integrated gluon distribution  $x_1 x_2 D_{gg}(x_1,x_2,Q,Q)$  (dashed black line)
with the product of two single gluon distributions, $x_1 D_g(x_1,Q) x_2 D_g(x_2,Q)$, (solid magenta) as a function of 
$Q^2$ for the indicated values of $(x_1,x_2)$ and the ${\rm GBLS}^3$ input.
The breakdown  of  $D_{gg}(x_1,x_2,Q,Q)$  into the homogeneous (dashed blue) and non-homogenous (dashed red) components, see Eq.~\eqref{eq:hnh}, 
shows that the non-homogeneous component is essential for the factorization.}
\vspace*{1cm}
\label{fig:figfactasipdf1}
\end{figure}

In Fig.~\ref{fig:figfactasipdf1}, we show the integrated  double gluon distribution $D_{gg}$ as a function of the hard scale $Q=Q_1=Q_2$ for  fixed values of $(x_1,x_2)$. We indeed observe that the sum of the homogenous and non-homogenous terms  nicely factorizes into a product of two single integrated gluon distributions,
provided $x_1$ and $x_2$ are sufficiently small and $Q^2$ is large. The latter condition is  related to the fact that at low $Q^2$, the solution is still close to the form of the initial condition which does not factorize in the case of the ${\rm GBLS}^3$ input.
So even though the initial condition does not factorize  at small $x$, after the evolution to large scales the solution becomes factorized for small $x$.

In order to further investigate the origin of the factorization,  we also performed calculations for the GS initial conditions. We found   that in this case the factorization at large scales holds to a lesser degree than for the ${\rm GBLS}^3$ input, though approximately it is still valid.  Finally, we  also tested initial conditions which violate the momentum sum rule more strongly than the GS input \cite{Gaunt:2009re}, and we found that in this case the factorization of the resulting solution is stronger violated. 

The crucial difference between the analyzed  inputs is the fact whether or not they satisfy the momentum sum rule. 
The best quality factorization at small $x$ occurs for the ${\rm GBLS}^3$ input  which satisfies the momentum sum rule exactly by construction. On the other hand,  the GS input does not satisfy the sum rule, but the violation is  of the order of a few percent (for $x<0.5$), which translates to a similar in size effect for the factorization. This suggests that whether or not the factorization at small $x$ and large $Q^2$ holds is related to the fulfillment of the momentum sum rule for the double PDFs. We stress, that this is true even if there is no factorization in the initial condition.  
However, more detailed analytic studies are necessary to quantify this observation, which we postpone for the future.

\section{Summary and outlook}
\label{sec:6}

We  performed detailed numerical analysis of the unintegrated double gluon distribution with the transverse momentum dependence. The construction of such distributions,   presented in  \cite{Golec-Biernat:2016vbt}, 
is based on the extension of a similar formalism for the construction of  the single unintegrated parton distributions \cite{Kimber:1999xc,Kimber:2001sc}. 
The unintegrated double  parton distributions in  \cite{Golec-Biernat:2016vbt}
were obtained as a convolution of the integrated parton distributions with the Altarelli-Parisi splitting functions and  the Sudakov form factors. 

As a general conclusion from the presented analysis, 
the transverse momentum dependent double gluon distribution is shifted towards larger values of transverse momenta with increasing values of the hard scales $Q_{1,2}$ and decreasing  values of the longitudinal momentum fractions $x_{1,2}$. 
The model with angular ordering in the last step of the evolution leads to a  distribution which extends further in transverse momenta than in the strong ordering scenario in which the transverse momenta are cut off by the hard scales. The homogeneous contribution dominates over the non-homogeneous one,  for all the values of transverse momenta, at least for the type of initial conditions considered in this work. We observe that the non-homogeneous contribution, which originates from the splitting, is more perturbative being shifted towards higher values of transverse momenta than  the homogeneous contribution. The sum of the homogeneous and non-homogeneous contribution factorizes into the product of two single unintegrated gluon distributions at small values of $x_{1,2}$ and high hard scales. We showed that this factorization stems from the factorization of the underlying integrated parton distributions and is contingent upon the validity of the momentum sum rule for the initial conditions.
  
There are number of possible avenues that could  be further explored. First of all, we have only considered the distributions for ${\bf q}_\perp=0$, i.e. integrated over the relative transverse positions of the two gluons.  Full dependence  on ${\bf q}_\perp$ should be included when one wants to use these distributions for the evaluation of the double parton scattering cross sections. In the simplest scenario, this could be done by including the appropriate form factors. However, it is  possible  that there will be large correlations between the transverse momenta of the partons and the momentum transfer ${\bf q}_\perp$, see 
\cite{Diehl:2017kgu, Buffing:2017mqm,Diehl:2017wew} for recent studies.
Second, the presented  analysis  was performed for the gluon sector only. The extension to include   quarks can be done, and in principle does not present any new technical difficulties. The biggest challenge is the appropriate choice of the initial conditions for the evolution which would satisfy both the momentum and the quark number sum rules simultaneously. So far this problem has not been fully solved. Finally, it would be also interesting to further explore the relation between the momentum sum rule and the factorization of the double distributions  for small values of the longitudinal momentum fractions.

\section*{Acknowledgments}
This work was supported by the Department of Energy  Grant No. DE-SC-0002145, 
the National Science Center, Poland, Grant No. 2015/17/B/ST2/01838 and
 by the Center
for Innovation and Transfer of Natural Sciences and Engineering Knowledge in Rzesz\'ow.

\newpage
\bibliographystyle{JHEP}
\bibliography{mybib}

\providecommand{\href}[2]{#2}\begingroup\raggedright\begin{thebibliography}{10}

\bibitem{Abt:1993cb}
{\scshape H1} collaboration, I.~Abt et~al., \emph{{Measurement of the proton
  structure function $F_2$(x, $Q^2$) in the low x region at HERA}},
  \href{http://dx.doi.org/10.1016/0550-3213(93)90090-C}{\emph{Nucl. Phys.} {\bf
  B407} (1993) 515--538}.

\bibitem{Derrick:1993fta}
{\scshape ZEUS} collaboration, M.~Derrick et~al., \emph{{Measurement of the
  proton structure function $F_2$ in e p scattering at HERA}},
  \href{http://dx.doi.org/10.1016/0370-2693(93)90347-K}{\emph{Phys. Lett.} {\bf
  B316} (1993) 412--426}.

\bibitem{Aid:1996au}
{\scshape H1} collaboration, S.~Aid et~al., \emph{{A Measurement and QCD
  analysis of the proton structure function $F_2$(x, $q^2$) at HERA}},
  \href{http://dx.doi.org/10.1016/0550-3213(96)00211-8}{\emph{Nucl. Phys.} {\bf
  B470} (1996) 3--40}, [\href{https://arxiv.org/abs/hep-ex/9603004}{{\tt
  hep-ex/9603004}}].

\bibitem{Derrick:1996hn}
{\scshape ZEUS} collaboration, M.~Derrick et~al., \emph{{Measurement of the F2
  structure function in deep inelastic e+ p scattering using 1994 data from the
  ZEUS detector at HERA}}, \href{http://dx.doi.org/10.1007/BF02909169,
  10.1007/s002880050260}{\emph{Z. Phys.} {\bf C72} (1996) 399--424},
  [\href{https://arxiv.org/abs/hep-ex/9607002}{{\tt hep-ex/9607002}}].

\bibitem{Akesson:1986iv}
{\scshape Axial Field Spectrometer Collaboration} collaboration, T.~Akesson
  et~al., \emph{{Double Parton Scattering in $p p$ Collisions at
  $\sqrt{s}=63$-{GeV}}},
  \href{http://dx.doi.org/10.1007/BF01566757}{\emph{Z.Phys.} {\bf C34} (1987)
  163}.

\bibitem{Abe:1997bp}
{\scshape CDF Collaboration} collaboration, F.~Abe et~al., \emph{{Measurement
  of double parton scattering in $\bar{p}p$ collisions at $\sqrt{s} = 1.8$
  TeV}},
  \href{http://dx.doi.org/10.1103/PhysRevLett.79.584}{\emph{Phys.Rev.Lett.}
  {\bf 79} (1997) 584--589}.

\bibitem{Abe:1997xk}
{\scshape CDF Collaboration} collaboration, F.~Abe et~al., \emph{{Double parton
  scattering in $\bar{p}p$ collisions at $\sqrt{s} = 1.8 $TeV}},
  \href{http://dx.doi.org/10.1103/PhysRevD.56.3811}{\emph{Phys.Rev.} {\bf D56}
  (1997) 3811--3832}.

\bibitem{Abazov:2009gc}
{\scshape D0 Collaboration} collaboration, V.~Abazov et~al., \emph{{Double
  parton interactions in photon+3 jet events in $p p^-$ bar collisions
  $\sqrt{s}=1.96$ TeV}},
  \href{http://dx.doi.org/10.1103/PhysRevD.81.052012}{\emph{Phys.Rev.} {\bf
  D81} (2010) 052012}, [\href{https://arxiv.org/abs/0912.5104}{{\tt
  0912.5104}}].

\bibitem{Aad:2013bjm}
{\scshape ATLAS Collaboration} collaboration, G.~Aad et~al., \emph{{Measurement
  of hard double-parton interactions in $W(\to l\nu)$+ 2 jet events at
  $\sqrt{s}$=7 TeV with the ATLAS detector}},
  \href{http://dx.doi.org/10.1088/1367-2630/15/3/033038}{\emph{New J.Phys.}
  {\bf 15} (2013) 033038}, [\href{https://arxiv.org/abs/1301.6872}{{\tt
  1301.6872}}].

\bibitem{Chatrchyan:2013xxa}
{\scshape CMS Collaboration} collaboration, S.~Chatrchyan et~al., \emph{{Study
  of double parton scattering using W + 2-jet events in proton-proton
  collisions at $\sqrt{s}$ = 7 TeV}},
  \href{http://dx.doi.org/10.1007/JHEP03(2014)032}{\emph{JHEP} {\bf 1403}
  (2014) 032}, [\href{https://arxiv.org/abs/1312.5729}{{\tt 1312.5729}}].

\bibitem{Aad:2014rua}
{\scshape ATLAS Collaboration} collaboration, G.~Aad et~al., \emph{{Measurement
  of the production cross section of prompt $J/\psi$ mesons in association with
  a $W^\pm$ boson in $pp$ collisions at $\sqrt{s} =$ 7 TeV with the ATLAS
  detector}}, \href{http://dx.doi.org/10.1007/JHEP04(2014)172}{\emph{JHEP} {\bf
  1404} (2014) 172}, [\href{https://arxiv.org/abs/1401.2831}{{\tt 1401.2831}}].

\bibitem{Paver:1982yp}
N.~Paver and D.~Treleani, \emph{{Multi - Quark Scattering and Large $p_T$ Jet
  Production in Hadronic Collisions}},
  \href{http://dx.doi.org/10.1007/BF02814035}{\emph{Nuovo Cim.} {\bf A70}
  (1982) 215}.

\bibitem{Diehl:2011yj}
M.~Diehl, D.~Ostermeier and A.~Schafer, \emph{{Elements of a theory for
  multiparton interactions in QCD}},
  \href{http://dx.doi.org/10.1007/JHEP03(2012)089}{\emph{JHEP} {\bf 1203}
  (2012) 089}, [\href{https://arxiv.org/abs/1111.0910}{{\tt 1111.0910}}].

\bibitem{Treleani:2017zzl}
D.~Treleani and G.~Calucci, \emph{{Multiple Parton Interactions, inclusive and
  exclusive cross sections, sum rules}},
  \href{https://arxiv.org/abs/1707.00271}{{\tt 1707.00271}}.

\bibitem{Diehl:2017wew}
M.~Diehl and J.~R. Gaunt, \emph{{Double parton scattering theory overview}},
  \href{https://arxiv.org/abs/1710.04408}{{\tt 1710.04408}}.

\bibitem{Shelest:1982dg}
V.~Shelest, A.~Snigirev and G.~Zinovev, \emph{{The Multiparton Distribution
  Equations in {QCD}}},
  \href{http://dx.doi.org/10.1016/0370-2693(82)90049-1}{\emph{Phys.Lett.} {\bf
  B113} (1982) 325}.

\bibitem{Zinovev:1982be}
G.~Zinovev, A.~Snigirev and V.~Shelest, \emph{{Equations for many parton
  distributions in quantum chromodynamics}},
  \href{http://dx.doi.org/10.1007/BF01017270}{\emph{Theor.Math.Phys.} {\bf 51}
  (1982) 523--528}.

\bibitem{Ellis:1982cd}
R.~K. Ellis, W.~Furmanski and R.~Petronzio, \emph{{Unraveling Higher Twists}},
  \href{http://dx.doi.org/10.1016/0550-3213(83)90597-7}{\emph{Nucl.Phys.} {\bf
  B212} (1983) 29}.

\bibitem{Bukhvostov:1985rn}
A.~Bukhvostov, G.~Frolov, L.~Lipatov and E.~Kuraev, \emph{{Evolution Equations
  for Quasi-Partonic Operators}},
  \href{http://dx.doi.org/10.1016/0550-3213(85)90628-5}{\emph{Nucl.Phys.} {\bf
  B258} (1985) 601--646}.

\bibitem{Snigirev:2003cq}
A.~M. Snigirev, \emph{{Double parton distributions in the leading logarithm
  approximation of perturbative QCD}},
  \href{http://dx.doi.org/10.1103/PhysRevD.68.114012}{\emph{Phys. Rev.} {\bf
  D68} (2003) 114012}, [\href{https://arxiv.org/abs/hep-ph/0304172}{{\tt
  hep-ph/0304172}}].

\bibitem{Korotkikh:2004bz}
V.~L. Korotkikh and A.~M. Snigirev, \emph{{Double parton correlations versus
  factorized distributions}},
  \href{http://dx.doi.org/10.1016/j.physletb.2004.05.012}{\emph{Phys. Lett.}
  {\bf B594} (2004) 171--176},
  [\href{https://arxiv.org/abs/hep-ph/0404155}{{\tt hep-ph/0404155}}].

\bibitem{Gaunt:2009re}
J.~R. Gaunt and W.~J. Stirling, \emph{{Double Parton Distributions
  Incorporating Perturbative QCD Evolution and Momentum and Quark Number Sum
  Rules}}, \href{http://dx.doi.org/10.1007/JHEP03(2010)005}{\emph{JHEP} {\bf
  03} (2010) 005}, [\href{https://arxiv.org/abs/0910.4347}{{\tt 0910.4347}}].

\bibitem{Blok:2010ge}
B.~Blok, Y.~Dokshitzer, L.~Frankfurt and M.~Strikman, \emph{{The Four jet
  production at LHC and Tevatron in QCD}},
  \href{http://dx.doi.org/10.1103/PhysRevD.83.071501}{\emph{Phys.Rev.} {\bf
  D83} (2011) 071501}, [\href{https://arxiv.org/abs/1009.2714}{{\tt
  1009.2714}}].

\bibitem{Ceccopieri:2010kg}
F.~A. Ceccopieri, \emph{{An update on the evolution of double parton
  distributions}},
  \href{http://dx.doi.org/10.1016/j.physletb.2011.02.047}{\emph{Phys. Lett.}
  {\bf B697} (2011) 482--487}, [\href{https://arxiv.org/abs/1011.6586}{{\tt
  1011.6586}}].

\bibitem{Diehl:2011tt}
M.~Diehl and A.~Schafer, \emph{{Theoretical considerations on multiparton
  interactions in QCD}},
  \href{http://dx.doi.org/10.1016/j.physletb.2011.03.024}{\emph{Phys. Lett.}
  {\bf B698} (2011) 389--402}, [\href{https://arxiv.org/abs/1102.3081}{{\tt
  1102.3081}}].

\bibitem{Gaunt:2011xd}
J.~R. Gaunt and W.~J. Stirling, \emph{{Double Parton Scattering Singularity in
  One-Loop Integrals}},
  \href{http://dx.doi.org/10.1007/JHEP06(2011)048}{\emph{JHEP} {\bf 1106}
  (2011) 048}, [\href{https://arxiv.org/abs/1103.1888}{{\tt 1103.1888}}].

\bibitem{Ryskin:2011kk}
M.~Ryskin and A.~Snigirev, \emph{{A Fresh look at double parton scattering}},
  \href{http://dx.doi.org/10.1103/PhysRevD.83.114047}{\emph{Phys.Rev.} {\bf
  D83} (2011) 114047}, [\href{https://arxiv.org/abs/1103.3495}{{\tt
  1103.3495}}].

\bibitem{Bartels:2011qi}
J.~Bartels and M.~G. Ryskin, \emph{{Recombination within multi-chain
  contributions in pp scattering}},
  \href{https://arxiv.org/abs/1105.1638}{{\tt 1105.1638}}.

\bibitem{Blok:2011bu}
B.~Blok, Y.~Dokshitzer, L.~Frankfurt and M.~Strikman, \emph{{pQCD physics of
  multiparton interactions}},
  \href{http://dx.doi.org/10.1140/epjc/s10052-012-1963-8}{\emph{Eur.Phys.J.}
  {\bf C72} (2012) 1963}, [\href{https://arxiv.org/abs/1106.5533}{{\tt
  1106.5533}}].

\bibitem{Luszczak:2011zp}
M.~Luszczak, R.~Maciula and A.~Szczurek, \emph{{Production of two $c \bar c$
  pairs in double-parton scattering}},
  \href{http://dx.doi.org/10.1103/PhysRevD.85.094034}{\emph{Phys. Rev.} {\bf
  D85} (2012) 094034}, [\href{https://arxiv.org/abs/1111.3255}{{\tt
  1111.3255}}].

\bibitem{Manohar:2012jr}
A.~V. Manohar and W.~J. Waalewijn, \emph{{A QCD Analysis of Double Parton
  Scattering: Color Correlations, Interference Effects and Evolution}},
  \href{http://dx.doi.org/10.1103/PhysRevD.85.114009}{\emph{Phys.Rev.} {\bf
  D85} (2012) 114009}, [\href{https://arxiv.org/abs/1202.3794}{{\tt
  1202.3794}}].

\bibitem{Ryskin:2012qx}
M.~Ryskin and A.~Snigirev, \emph{{Double parton scattering in double logarithm
  approximation of perturbative QCD}},
  \href{http://dx.doi.org/10.1103/PhysRevD.86.014018}{\emph{Phys.Rev.} {\bf
  D86} (2012) 014018}, [\href{https://arxiv.org/abs/1203.2330}{{\tt
  1203.2330}}].

\bibitem{Gaunt:2012dd}
J.~R. Gaunt, \emph{{Single Perturbative Splitting Diagrams in Double Parton
  Scattering}}, \href{http://dx.doi.org/10.1007/JHEP01(2013)042}{\emph{JHEP}
  {\bf 1301} (2013) 042}, [\href{https://arxiv.org/abs/1207.0480}{{\tt
  1207.0480}}].

\bibitem{Blok:2013bpa}
B.~Blok, Y.~Dokshitzer, L.~Frankfurt and M.~Strikman, \emph{{Perturbative QCD
  correlations in multi-parton collisions}},
  \href{http://dx.doi.org/10.1140/epjc/s10052-014-2926-z}{\emph{Eur.Phys.J.}
  {\bf C74} (2014) 2926}, [\href{https://arxiv.org/abs/1306.3763}{{\tt
  1306.3763}}].

\bibitem{vanHameren:2014ava}
A.~van Hameren, R.~Maciula and A.~Szczurek, \emph{{Single-parton scattering
  versus double-parton scattering in the production of two $c \bar c$ pairs and
  charmed meson correlations at the LHC}},
  \href{http://dx.doi.org/10.1103/PhysRevD.89.094019}{\emph{Phys. Rev.} {\bf
  D89} (2014) 094019}, [\href{https://arxiv.org/abs/1402.6972}{{\tt
  1402.6972}}].

\bibitem{Maciula:2014pla}
R.~Maciula and A.~Szczurek, \emph{{Double-parton scattering contribution to
  production of jet pairs with large rapidity separation at the LHC}},
  \href{http://dx.doi.org/10.1103/PhysRevD.90.014022}{\emph{Phys. Rev.} {\bf
  D90} (2014) 014022}, [\href{https://arxiv.org/abs/1403.2595}{{\tt
  1403.2595}}].

\bibitem{Snigirev:2014eua}
A.~Snigirev, N.~Snigireva and G.~Zinovjev, \emph{{Perturbative and
  nonperturbative correlations in double parton distributions}},
  \href{http://dx.doi.org/10.1103/PhysRevD.90.014015}{\emph{Phys.Rev.} {\bf
  D90} (2014) 014015}, [\href{https://arxiv.org/abs/1403.6947}{{\tt
  1403.6947}}].

\bibitem{Golec-Biernat:2014nsa}
K.~Golec-Biernat and E.~Lewandowska, \emph{{Electroweak boson production in
  double parton scattering}},
  \href{http://dx.doi.org/10.1103/PhysRevD.90.094032}{\emph{Phys.Rev.} {\bf
  D90} (2014) 094032}, [\href{https://arxiv.org/abs/1407.4038}{{\tt
  1407.4038}}].

\bibitem{Gaunt:2014rua}
J.~R. Gaunt, R.~Maciula and A.~Szczurek, \emph{{Conventional versus
  single-ladder-splitting contributions to double parton scattering production
  of two quarkonia, two Higgs bosons and $c \bar c c \bar c$}},
  \href{http://dx.doi.org/10.1103/PhysRevD.90.054017}{\emph{Phys. Rev.} {\bf
  D90} (2014) 054017}, [\href{https://arxiv.org/abs/1407.5821}{{\tt
  1407.5821}}].

\bibitem{Harland-Lang:2014efa}
L.~A. Harland-Lang, V.~A. Khoze and M.~G. Ryskin, \emph{{Exclusive production
  of double J/$\Psi$ mesons in hadronic collisions}},
  \href{http://dx.doi.org/10.1088/0954-3899/42/5/055001}{\emph{J. Phys.} {\bf
  G42} (2015) 055001}, [\href{https://arxiv.org/abs/1409.4785}{{\tt
  1409.4785}}].

\bibitem{Blok:2014rza}
B.~Blok and M.~Strikman, \emph{{Double parton interactions in $\gamma p, \gamma
  A$ collisions in the direct photon kinematics}},
  \href{http://dx.doi.org/10.1140/epjc/s10052-014-3214-7}{\emph{Eur. Phys. J.}
  {\bf C74} (2014) 3214}, [\href{https://arxiv.org/abs/1410.5064}{{\tt
  1410.5064}}].

\bibitem{Maciula:2015vza}
R.~Maciula and A.~Szczurek, \emph{{Searching for and exploring double-parton
  scattering effects in four-jet production at the LHC}},
  \href{http://dx.doi.org/10.1016/j.physletb.2015.07.035}{\emph{Phys. Lett.}
  {\bf B749} (2015) 57--62}, [\href{https://arxiv.org/abs/1503.08022}{{\tt
  1503.08022}}].

\bibitem{Kirschner:1979im}
R.~Kirschner, \emph{{Generalized {Lipatov-Altarelli-Parisi} Equations and Jet
  Calculus Rules}},
  \href{http://dx.doi.org/10.1016/0370-2693(79)90300-9}{\emph{Phys.Lett.} {\bf
  B84} (1979) 266}.

\bibitem{Konishi:1978yx}
K.~Konishi, A.~Ukawa and G.~Veneziano, \emph{{A Simple Algorithm for QCD
  Jets}}, \href{http://dx.doi.org/10.1016/0370-2693(78)90015-1}{\emph{Phys.
  Lett.} {\bf B78} (1978) 243--248}.

\bibitem{Konishi:1979cb}
K.~Konishi, A.~Ukawa and G.~Veneziano, \emph{{Jet Calculus: A Simple Algorithm
  for Resolving QCD Jets}},
  \href{http://dx.doi.org/10.1016/0550-3213(79)90053-1}{\emph{Nucl. Phys.} {\bf
  B157} (1979) 45--107}.

\bibitem{Diehl:2017kgu}
M.~Diehl, J.~R. Gaunt and K.~Schoenwald, \emph{{Double hard scattering without
  double counting}},
  \href{http://dx.doi.org/10.1007/JHEP06(2017)083}{\emph{JHEP} {\bf 06} (2017)
  083}, [\href{https://arxiv.org/abs/1702.06486}{{\tt 1702.06486}}].

\bibitem{Collins:2005uv}
J.~Collins and H.~Jung, \emph{{Need for fully unintegrated parton densities}},
  in \emph{{HERA and the LHC: A Workshop on the implications of HERA for LHC
  physics. Proceedings, Part B}}, 2005.
\newblock \href{https://arxiv.org/abs/hep-ph/0508280}{{\tt hep-ph/0508280}}.

\bibitem{Collins:2007ph}
J.~C. Collins, T.~C. Rogers and A.~M. Stasto, \emph{{Fully unintegrated parton
  correlation functions and factorization in lowest-order hard scattering}},
  \href{http://dx.doi.org/10.1103/PhysRevD.77.085009}{\emph{Phys. Rev.} {\bf
  D77} (2008) 085009}, [\href{https://arxiv.org/abs/0708.2833}{{\tt
  0708.2833}}].

\bibitem{Collins:2011zzd}
{John Collins}, \emph{{Foundations of perturbative QCD}}, vol.~{32}.
\newblock {Cambridge Univ. Press}, {2011}.

\bibitem{Collins:2011ca}
J.~Collins, \emph{{New definition of TMD parton densities}},
  \href{http://dx.doi.org/10.1142/S2010194511001590}{\emph{Int. J. Mod. Phys.
  Conf. Ser.} {\bf 4} (2011) 85--96},
  [\href{https://arxiv.org/abs/1107.4123}{{\tt 1107.4123}}].

\bibitem{Catani:1990eg}
S.~Catani, M.~Ciafaloni and F.~Hautmann, \emph{{High-energy factorization and
  small x heavy flavor production}},
  \href{http://dx.doi.org/10.1016/0550-3213(91)90055-3}{\emph{Nucl. Phys.} {\bf
  B366} (1991) 135--188}.

\bibitem{Buffing:2016qql}
M.~G.~A. Buffing, M.~Diehl and T.~Kasemets, \emph{{Double parton scattering for
  perturbative transverse momenta}},  2016.
\newblock \href{https://arxiv.org/abs/1611.00178}{{\tt 1611.00178}}.

\bibitem{Buffing:2017mqm}
M.~G.~A. Buffing, M.~Diehl and T.~Kasemets, \emph{{Transverse momentum in
  double parton scattering: factorisation, evolution and matching}},
  \href{https://arxiv.org/abs/1708.03528}{{\tt 1708.03528}}.

\bibitem{Golec-Biernat:2016vbt}
K.~Golec-Biernat and A.~M. Stasto, \emph{{Unintegrated double parton
  distributions}},
  \href{http://dx.doi.org/10.1103/PhysRevD.95.034033}{\emph{Phys. Rev.} {\bf
  D95} (2017) 034033}, [\href{https://arxiv.org/abs/1611.02033}{{\tt
  1611.02033}}].

\bibitem{Kimber:1999xc}
M.~A. Kimber, A.~D. Martin and M.~G. Ryskin, \emph{{Unintegrated parton
  distributions and prompt photon hadroproduction}},
  \href{http://dx.doi.org/10.1007/s100520000326}{\emph{Eur. Phys. J.} {\bf C12}
  (2000) 655--661}, [\href{https://arxiv.org/abs/hep-ph/9911379}{{\tt
  hep-ph/9911379}}].

\bibitem{Kimber:2001sc}
M.~A. Kimber, A.~D. Martin and M.~G. Ryskin, \emph{{Unintegrated parton
  distributions}},
  \href{http://dx.doi.org/10.1103/PhysRevD.63.114027}{\emph{Phys. Rev.} {\bf
  D63} (2001) 114027}, [\href{https://arxiv.org/abs/hep-ph/0101348}{{\tt
  hep-ph/0101348}}].

\bibitem{Kimber:2000bg}
M.~A. Kimber, J.~Kwiecinski, A.~D. Martin and A.~M. Stasto, \emph{{The
  Unintegrated gluon distribution from the CCFM equation}},
  \href{http://dx.doi.org/10.1103/PhysRevD.62.094006}{\emph{Phys. Rev.} {\bf
  D62} (2000) 094006}, [\href{https://arxiv.org/abs/hep-ph/0006184}{{\tt
  hep-ph/0006184}}].

\bibitem{Blok:2017alw}
B.~Blok and M.~Strikman, \emph{{Multiparton pp and pA collisions - from
  geometry to parton - parton correlations}},
  \href{https://arxiv.org/abs/1709.00334}{{\tt 1709.00334}}.

\bibitem{Strikman:2011cx}
M.~Strikman, \emph{{Transverse Nucleon Structure and Multiparton
  Interactions}}, \href{http://dx.doi.org/10.5506/APhysPolB.42.2607}{\emph{Acta
  Phys. Polon.} {\bf B42} (2011) 2607--2630},
  [\href{https://arxiv.org/abs/1112.3834}{{\tt 1112.3834}}].

\bibitem{Rogers:2009ke}
T.~C. Rogers and M.~Strikman, \emph{{Multiple Hard Partonic Collisions with
  Correlations in Proton-Proton Scattering}},
  \href{http://dx.doi.org/10.1103/PhysRevD.81.016013}{\emph{Phys. Rev.} {\bf
  D81} (2010) 016013}, [\href{https://arxiv.org/abs/0908.0251}{{\tt
  0908.0251}}].

\bibitem{Rogers:2008ua}
T.~C. Rogers, A.~M. Stasto and M.~I. Strikman, \emph{{Unitarity Constraints on
  Semi-hard Jet Production in Impact Parameter Space}},
  \href{http://dx.doi.org/10.1103/PhysRevD.77.114009}{\emph{Phys. Rev.} {\bf
  D77} (2008) 114009}, [\href{https://arxiv.org/abs/0801.0303}{{\tt
  0801.0303}}].

\bibitem{Broniowski:2013xba}
W.~Broniowski and E.~Ruiz~Arriola, \emph{{Valence double parton distributions
  of the nucleon in a simple model}},
  \href{http://dx.doi.org/10.1007/s00601-014-0840-4}{\emph{Few Body Syst.} {\bf
  55} (2014) 381--387}, [\href{https://arxiv.org/abs/1310.8419}{{\tt
  1310.8419}}].

\bibitem{Golec-Biernat:2014bva}
K.~Golec-Biernat and E.~Lewandowska, \emph{{How to impose initial conditions
  for QCD evolution of double parton distributions?}},
  \href{http://dx.doi.org/10.1103/PhysRevD.90.014032}{\emph{Phys.Rev.} {\bf
  D90} (2014) 014032}, [\href{https://arxiv.org/abs/1402.4079}{{\tt
  1402.4079}}].

\bibitem{Golec-Biernat:2015aza}
K.~Golec-Biernat, E.~Lewandowska, M.~Serino, Z.~Snyder and A.~M. Stasto,
  \emph{{Constraining the double gluon distribution by the single gluon
  distribution}},
  \href{http://dx.doi.org/10.1016/j.physletb.2015.09.067}{\emph{Phys. Lett.}
  {\bf B750} (2015) 559--564}, [\href{https://arxiv.org/abs/1507.08583}{{\tt
  1507.08583}}].

\bibitem{Martin:2009iq}
A.~Martin, W.~Stirling, R.~Thorne and G.~Watt, \emph{{Parton distributions for
  the LHC}},
  \href{http://dx.doi.org/10.1140/epjc/s10052-009-1072-5}{\emph{Eur.Phys.J.}
  {\bf C63} (2009) 189--285}, [\href{https://arxiv.org/abs/0901.0002}{{\tt
  0901.0002}}].

\bibitem{Ciafaloni:1987ur}
M.~Ciafaloni, \emph{{Coherence Effects in Initial Jets at Small $q^2$/s}},
  \href{http://dx.doi.org/10.1016/0550-3213(88)90380-X}{\emph{Nucl. Phys.} {\bf
  B296} (1988) 49--74}.

\bibitem{Catani:1989sg}
S.~Catani, F.~Fiorani and G.~Marchesini, \emph{{Small x Behavior of Initial
  State Radiation in Perturbative QCD}},
  \href{http://dx.doi.org/10.1016/0550-3213(90)90342-B}{\emph{Nucl. Phys.} {\bf
  B336} (1990) 18--85}.

\bibitem{Catani:1989yc}
S.~Catani, F.~Fiorani and G.~Marchesini, \emph{{QCD Coherence in Initial State
  Radiation}},
  \href{http://dx.doi.org/10.1016/0370-2693(90)91938-8}{\emph{Phys. Lett.} {\bf
  B234} (1990) 339--345}.

\bibitem{Marchesini:1994wr}
G.~Marchesini, \emph{{QCD coherence in the structure function and associated
  distributions at small $x$}},
  \href{http://dx.doi.org/10.1016/0550-3213(95)00149-M}{\emph{Nucl. Phys.} {\bf
  B445} (1995) 49--80}, [\href{https://arxiv.org/abs/hep-ph/9412327}{{\tt
  hep-ph/9412327}}].

\end{thebibliography}\endgroup

\end{document}